\documentclass{jfm}

\ifCUPmtlplainloaded \else
  \checkfont{eurm10}
  \iffontfound
    \IfFileExists{upmath.sty}
      {\typeout{^^JFound AMS Euler Roman fonts on the system,
                   using the 'upmath' package.^^J}%
       \usepackage{upmath}}
      {\typeout{^^JFound AMS Euler Roman fonts on the system, but you
                   dont seem to have the}%
       \typeout{'upmath' package installed. JFM.cls can take advantage
                 of these fonts,^^Jif you use 'upmath' package.^^J}%
      }
  \else
  \fi
\fi

\ifCUPmtlplainloaded \else
  \checkfont{msam10}
  \iffontfound
    \IfFileExists{amssymb.sty}
      {\typeout{^^JFound AMS Symbol fonts on the system, using the
                'amssymb' package.^^J}%
       \usepackage{amssymb}%

      }{}
  \fi
\fi

\ifCUPmtlplainloaded \else
  \IfFileExists{amsbsy.sty}
    {\typeout{^^JFound the 'amsbsy' package on the system, using it.^^J}%
     \usepackage{amsbsy}}
    {}
\fi

\usepackage{amsmath, amssymb}
\usepackage{amsfonts}
\usepackage{graphicx, psfrag}
\usepackage{subfigure}
\usepackage{bm}
\usepackage[authoryear,round]{natbib}
\newcommand{\sliver}{{\hskip 0.6pt}}
\newcommand{\antisliver}{{\hskip -0.6pt}}

\newcommand{\Sliver}{{\hskip 1.2pt}}
\newcommand{\Antisliver}{{\hskip -1.2pt}}

\newcommand{\seventh}{{\tfrac{1}{\raisebox{-1pt}{\scriptsize7}}}}
\newcommand{\shalf}
             {{\!\raisebox{1.3pt}{$\genfrac{}{}{}{2}{1}{2}$}}}

\newcommand{\pottemp}{\vartheta}
\DeclareRobustCommand{\alphat}{\alpha_{\raisebox{-1pt}{\scriptsize$\vartheta$}}}
\newcommand{\alphamu}{\alpha_\mu}

\newcommand{\bbbt}{b_{\pottemp}}
\newcommand{\aaat}{a_{\pottemp}}
\newcommand{\rt}{r_{\pottemp}}

\newcommand{\bbbmu}{b_\mu}
\newcommand{\aaamu}{a_\mu}
\newcommand{\rmu}{r_{\Antisliver\mu}}
\newcommand{\rperiph}{r_{\rm d}}

\newcommand{\Eq}{Equation~}  
\newcommand{\Eqs}{Equations~}
\newcommand{\eq}{equation~}
\newcommand{\eqs}{equations~}
\newcommand{\Fig}{Figure~}  
\newcommand{\Figs}{Figures~}
\newcommand{\fig}{figure~}
\newcommand{\figs}{figures~}

\newcommand{\definedas}{=}

\newcommand{\yr}{{\rm yr}}

\newcommand{\nlb}{\protect\nolinebreak}
\newcommand{\degg}{$^\circ$}
\newcommand{\Mm}{\nlb\Sliver Mm}
\newcommand{\G}{\nlb\Sliver\Sliver gauss}
\newcommand{\cms}{\nlb\sliver cm\,s$^{-1}$}
\newcommand{\cmms}{\nlb\sliver cm$^2$s$^{-1}$}

\newcommand{\gcmmm}{\nlb\sliver g\,cm$^{-3}$}

\newcommand{\sm}{\nlb\sliver s$^{-1}$}
\newcommand{\nHz}{\nlb\sliver nHz}

\newcommand{\san}{semi-analytical}
\newcommand{\muchoke}{\mbox{mu-choke}}
\newcommand{\solid}{solid}
\newcommand{\extrapolar}{extra-polar}
\newcommand{\mfp}{mag\-net\-ic flux pump\-ing}

\newcommand{\mhd}{mag\-net\-o\-hy\-dro\-dy\-nam\-ic}

\newcommand{\base}{bottom}

\newcommand{\tsss}{therm\-al strat\-if\-i\-ca\-tion sur\-faces}
\newcommand{\cz}{con\-vec\-tion zone}

\newcommand{\tc}{ta\-cho\-cline}
\newcommand{\HSL}{hel\-ium sett\-ling lay\-er}
\newcommand{\hsl}{hel\-ium sub\-lay\-er}
\newcommand{\hhsl}{hel\-ium-sub\-lay\-er}

\newcommand{\cl}{con\-fine\-ment layer}
\newcommand{\chl}{con\-fine\-ment-layer}

\newcommand{\hsy}{helio\-seis\-mol\-ogy}

\newcommand{\hsc}{helio\-seis\-mic}

\newcommand{\te}{thought ex\-periment}
\newcommand{\mb}{mag\-ne\-to\-stroph\-ic bal\-ance}

\newcommand{\ie}{i.e.}
\newcommand{\eg}{{e.g.}}

\newcommand{\ii}{{\rm i}}

\newcommand{\dd}{{\rm d}}
\newcommand{\DD}{{\rm D}}

\newcommand{\twoandabit}{\gamma}

\newcommand{\smallparam}{\epsilon}
\newcommand{\CLthick}{\delta}
\newcommand{\SLthick}{\delta_{\chi}}
\newcommand{\subtailthick}{\delta_{\rm \ell}}
\newcommand{\Ekthick}{\delta_{\rm Ek}}

\newcommand{\Ro}{{\rm Ro}}
\newcommand{\Ek}{{\rm Ek}}

\newcommand{\pp}{p'}
\newcommand{\PP}{p}

\newcommand{\OmegaI}{\Omega_{\rm i}}
\newcommand{\APHI}{A}
\newcommand\bu{\mathbf{u}}
\newcommand\B{\mathbf{B}}
\newcommand\BI{\mathbf{B}_\ii}
\newcommand\BZO{\mathsf{B}}

\newcommand{\dr}{{\Delta r}}
\newcommand{\dz}{{\Delta z}}
\newcommand{\dt}{{\Delta t}}

\newcommand{\LHS}{left-hand side}
\newcommand{\RHS}{right-hand side}
\newcommand{\st}{subtail}

\begin{document}

\title[Confinement of the Sun's interior magnetic field]
{Polar
       confinement of the Sun's interior magnetic field
       by laminar magnetostrophic flow}

\author[T. S. Wood and M. E. McIntyre]
{T\ls O\ls B\ls Y\ns
S.\ns
W\ls O\ls O\ls D
\and
M\ls I\ls C\ls H\ls A\ls E\ls L\ns
E.\ns
M\ls {\small C}\ls I\ls N\ls T\ls Y\ls R\ls E}
\affiliation{Department of Applied Mathematics and Theoretical Physics,
University of Cambridge}
\date{28 May 2010,
            accepted 24 January 2011}

\maketitle

\begin{abstract}

The global-scale interior
magnetic field $\BI$ needed to account
for the Sun's observed differential rotation
can be effective only if confined
below the \cz\ in all latitudes
including, most critically, the polar caps.
Axisymmetric solutions are obtained to the
nonlinear
magnetohydrodynamic equations
showing that such polar
confinement can be brought about by
a very weak downwelling flow
$U\sim10^{-5}$\cms\
over each pole.  Such downwelling is
consistent with the \hsc\ evidence.
All three components of the magnetic field $\B$
decay exponentially with altitude across
a thin,
laminar
``magnetic \cl'' located at the bottom of the \tc\
and permeated by spiralling field lines.
With realistic parameter values, the thickness of the \cl\
$\sim10^{-3}$ of the Sun's radius,
the thickness scale being the magnetic advection--diffusion
scale $\CLthick=\eta/U$ where the
magnetic
(ohmic)
diffusivity $\eta\approx 4.1\times10^{2}$\cmms.
Alongside
baroclinic effects and stable
thermal stratification,
the solutions take into account
the stable
compositional stratification
of the \HSL,
if present as in today's Sun,
and
the small diffusivity of helium through hydrogen,
$\chi\approx0.9\times10^{1}$\cmms.
The
small value of $\chi$
relative to $\eta$
produces a
double boundary-layer structure in which a
``\hsl''
of smaller vertical scale
$(\chi/\eta)^{1/2}\CLthick$
is sandwiched
between the
top of the \HSL\
and the
rest of the \cl.
Solutions are obtained using both \san\
and purely numerical,
finite-difference techniques.
The \chl\ flows are
magnetostrophic to excellent approximation.
More precisely,
the principal force balances
are between Lorentz,
Coriolis, pressure-gradient
and buoyancy forces, with relative accelerations negligible
to excellent approximation.
Viscous forces are also negligible,
even in the \hsl\
where shears are greatest.
This is
despite the kinematic viscosity
being somewhat greater
than
$\chi$. \
We discuss how the \cl s at each pole might fit into a global
dynamical picture of the solar \tc.  That picture, in turn,
suggests a new insight into the early Sun and into
the longstanding enigma of
solar lithium depletion.

\end{abstract}

\hrule

\section{Introduction}
\label{sec:intro}

\begin{figure}
  \centering
\subfigure[][]{
  \label{fig:helioseismology}
  \psfrag{cz}{\begin{tabular}{c}convection\\[-0.1em]zone\end{tabular}}
  \psfrag{rz}{\begin{tabular}{c}radiative\\[-0.1em]interior\end{tabular}}
  \psfrag{tc}{tachocline}
  \includegraphics[clip=true,bb=118 -12 614 510,height=6cm]{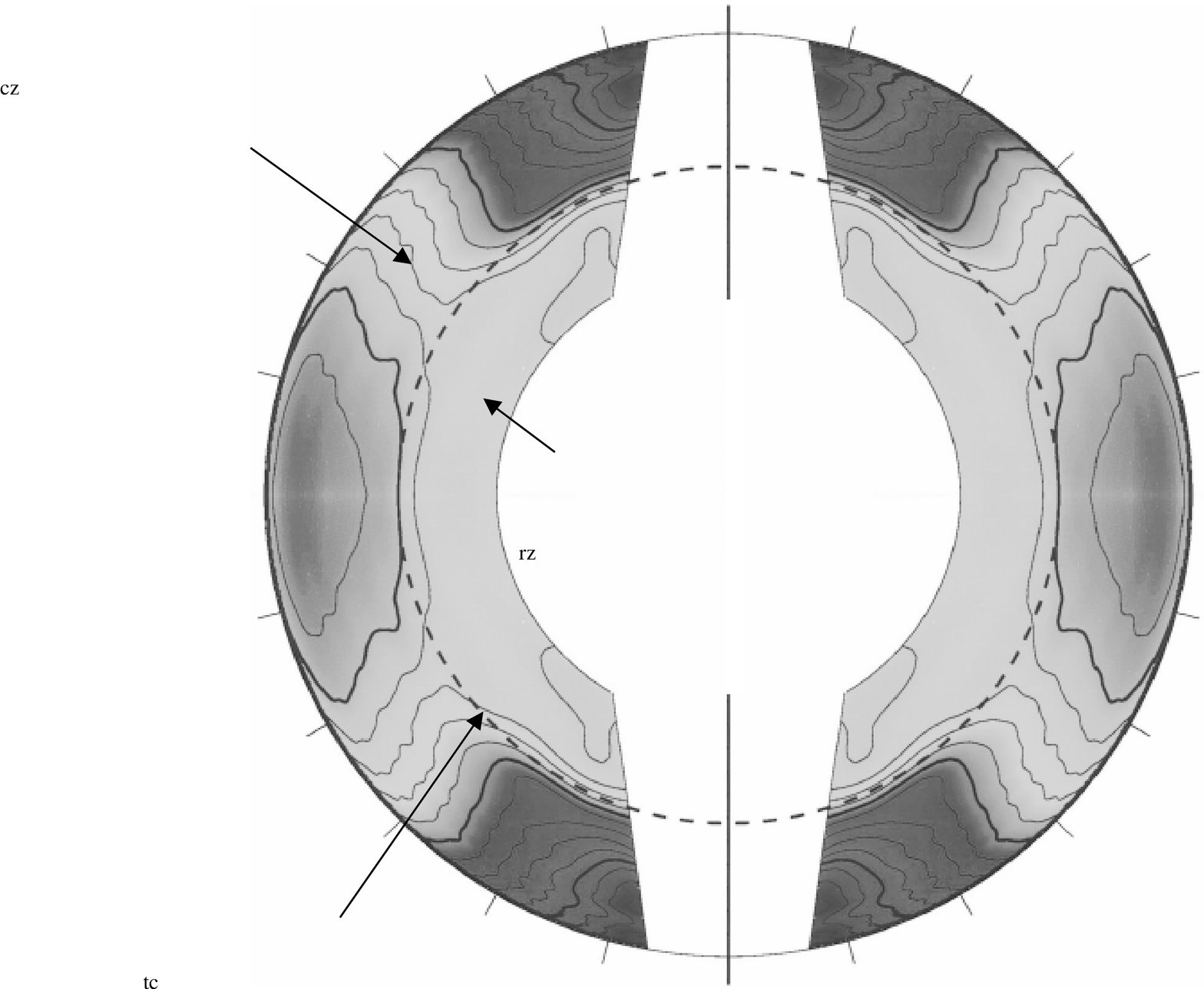}
}
\hspace{1cm}
\subfigure[][]{
  \label{fig:sphere}
  \includegraphics[clip=true,bb=14 -152 878 1048,height=6cm]{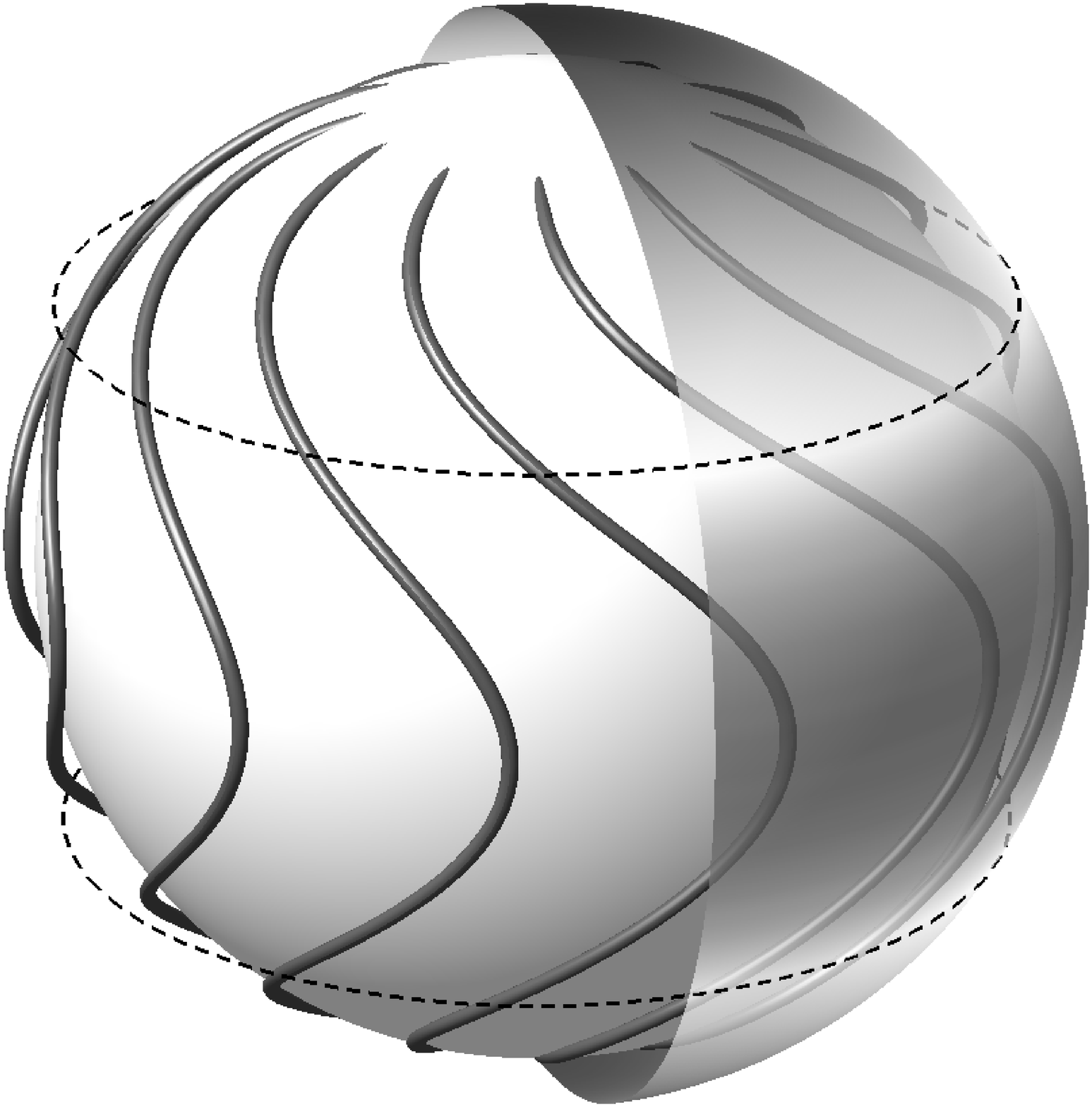}
}
  \caption{\footnotesize
    (a) The Sun's differential rotation
    deduced from helioseismic data using inverse methods
    \citep[adapted from][]{Schou-etal98}.
    The radiative interior rotates approximately \solid ly
    with angular velocity $\OmegaI = 2.7\times10^{-6}$\sm,
    or 435\,\nHz.
    Within the \cz, the angular velocity
    increases with
    colatitude
    through 350, 400, 450\,\nHz\ (heavy contours)
    to a maximum
    just under
    470\,\nHz\ at the equator.
\newline
  (b)
Schematic illustration
showing the top of the radiative interior
(inner sphere)
and the time-averaged magnetic field threading the \tc\ just above.
The cutaway outer sphere indicates the top of the \tc, whose
depth has been exaggerated.
  Poloidal magnetic field lines emerge
  from the interior
  in high latitudes
  and are wound up
  into their curved shapes
  by
  the
  \tc's
  differential rotation,
acting against turbulent
eddy diffusion.
A prograde torque is transmitted
  from low to high latitudes
  along these field lines.
  The slow polar and fast equatorial rotation are indicated
  by the darker shadings
  of the outer sphere.
  The dashed lines indicate the latitudes at which the
  rotation of the
  \cz\ matches that of the interior.
  }
\end{figure}

This paper analyses a new family of
laminar
magnetostrophic
flows that
may be important for confining
the
interior
magnetic
field
$\BI$
needed to explain the Sun's
differential rotation.
As illustrated in \fig\ref{fig:helioseismology},
the differential rotation
observed
within
the \cz\
goes over into near-\solid\
rotation within the
radiative, stably stratified
interior,
via a thin shear layer called the ``\tc''
much of which is also stably stratified.
The need for
the interior
field $\BI$
has been argued elsewhere
\citep[][hereafter GM98]{McIntyre94,
RudigerKitchatinov97,
GM98};
the main
arguments are briefly recalled below.
The observational evidence together with many
ideas about
the \tc\ are reviewed and further referenced in a recent major
compendium, \emph{The Solar Tachocline}
\citep{Hughesetal07},
and further discussed in the second edition of
Mestel's \emph{Stellar Magnetism}
\citep{Mestel11}.

By a ``confined'' $\BI$ we mean a field
most if not all of
whose lines are
contained
beneath the \cz, and held there against
magnetic (ohmic)
diffusion.
Such confinement is well known to be
necessary in order for
the field to help enforce
\solid\ rotation in the interior
\citep[\eg][]{Ferraro37,MestelWeiss87,
CharbonneauMacGregor93,
MacGregorCharbonneau99},
and thereby keep the \tc\ thin.
Confinement against magnetic
diffusion
requires fluid motion.
So, besides magnetic effects,
a realistic theory
of
confinement
must take account of
Coriolis effects,
stable stratification,
baroclinicity,
and thermal relaxation.
Without these effects we cannot
correctly describe, for instance, the
overall torque balance,
which
necessarily involves
mean meridional circulations (MMCs)
as well as Maxwell stresses.

The first attempt at a tachocline theory was that of
\citet{SpiegelZahn92}.
It
included all
the above effects except $\BI$. \
\citet{RudigerKitchatinov97}
included $\BI$ but
omitted the other effects.  The first attempt to include
all of them was that of
GM98,
in
a line of investigation
further developed by
\citet{GaraudGaraud08}.
Meanwhile, the dynamical importance of compositional as
well as thermal stratification
\citep[\eg][]{Mestel53}
was
suggested
for tachocline theories
\citep{McIntyre07}.
In particular, the helium settling layer
beneath the
tachocline is
nearly
impermeable to MMCs
because of
the small diffusivity of helium through hydrogen.
This near-impermeability of compositionally stratified regions
has been called the
``\muchoke''
\citep{MestelMoss86}.
The reality of the
Sun's
\HSL\
is strongly indicated
both by standard solar-evolution models
and by helioseismology
\citep[\eg][]{
JCD-etal93,
Ciacio-etal97,
ElliottGough99,
JCDThompson07}.\footnote{
Also Christensen-Dalsgaard \& Gough 2011,
in preparation.
}
As will be seen, this combination of circumstances gives rise to some
new and interesting fluid dynamics.

The need for the interior field $\BI$ arises from
a well known difficulty with
non-magnetic theories.
They
tend to spread
the strong differential rotation of
the convection zone down
into the radiative interior.
Although sometimes disputed,
this is a robust and well-understood
consequence of thermal relaxation,
interacting with Coriolis effects and
gyroscopically-pumped MMCs
\citep{Haynes-etal91,SpiegelZahn92,Elliott97,McIntyre07,GaraudBrummell08}.
As shown
by
\citeauthor{SpiegelZahn92}
and confirmed by
\citeauthor{Elliott97},
this downward spreading or burrowing
would have produced a \tc\ far thicker
than observed.
The accompanying MMC,
acting throughout the Sun's
lifetime,
would also have prevented the
\HSL\ from forming.

To counter
the burrowing
tendency
and
to
allow the interior to rotate \solid ly,
angular momentum has
to be transported somehow from the low-latitude tachocline
to
the high-latitude tachocline.  The
non-magnetic
horizontal eddy viscosity
proposed for this purpose by
\citeauthor{SpiegelZahn92}
is inconsistent with the properties of
non-magnetic
stratified turbulence
known from many studies of the terrestrial atmosphere
\citep[][\& refs.]{McIntyre94,McIntyre03}.
Angular-momentum
transport by
internal gravity waves is a physically possible
alternative
\citep[\eg][\& refs.]{Schatzman93,Zahn-etal97,
RogersGlatzmaier06,CharbonnelTalon07}.
However,
it is
highly improbable
as the main mechanism
because, by itself,
it has no natural
tendency to produce
\solid\
rotation
at all latitudes and depths
\citep[\eg][]{PlumbMcEwan78}.

A suitably-shaped magnetic field can, by contrast, naturally
produce the required angular momentum transport,
via the
Alfv\'enic elasticity of the field lines.
A suitable shape is one in which
the field lines
link low latitudes to high latitudes within the \tc.
The simplest such shape
--- simplest by virtue of its axisymmetry ---
is that suggested schematically
in \fig\ref{fig:sphere},
in which the linkage
is via a time-averaged field whose lines thread the \tc,
forming the superficial part of
a global-scale interior dipole
stabilized by a deep toroidal field
\citep[\eg][]{BraithwaiteSpruit04}.
Such an interior dipole has a diffusive lifetime
somewhat greater than
the Sun's lifetime
of around
$4.5\times10^9$\yr.
The dipole
imposes an Alfv\'enic
``Ferraro constraint''
on the interior.  It is this constraint
that helps to enforce the
interior's \solid\ rotation
\citep[\eg][]{Ferraro37,MestelWeiss87,
  MacGregorCharbonneau99}.

The
field lines shown in
\fig\ref{fig:sphere}
emerge
from
the interior (light-grey sphere)
near the north pole and,
after threading their way through the \tc,
re-enter the interior near the south
pole.
They
return northward
through an interior ``apple-core'' region, not shown,
surrounding the rotation axis.
It is crucial that
the field lines emerging
from the interior
bend over
toward the horizontal
as they enter the \tc.
They must be prevented from extending upward
through the polar cap, as
occurs
when
magnetic diffusion dominates
\citep[\eg][]{BraithwaiteSpruit04,BrunZahn06}.
The
curved
shapes of the
field lines
in \fig\ref{fig:sphere}
are evidently such as to transport angular momentum from
low to high latitudes by means of
persistent Alfv\'enic torques,
exactly as required to
prevent the \tc's MMCs
from burrowing into the interior and thickening the \tc.

The time averaging
envisaged in \fig\ref{fig:sphere}
conceals
a plethora of
fast processes,
including the 22-year dynamo cycle, convective
overshoot, and other turbulent processes
arising from
various instabilities
in the \tc.
All these are fast relative to the timescales on which the
mean structure of the \tc\ is maintained,
$\sim10^5$\yr\ or more.
We presume that
the fast processes have two important consequences.  The first
is to produce
a turbulent magnetic diffusivity
that stops the field
lines
being wound up arbitrarily tightly by the
shear in the \tc,
keeping
the curved
shapes
shown.\footnote{Of
course the persistent
angular momentum transport from the curved $\B$ lines
could be supplemented by equally persistent
contributions
coming directly from
MHD-turbulent stresses
\citep[\eg][]{Spruit02,GilmanCally07,ParfreyMenou07}.
}

The second important consequence is that,
away from the poles,
the field
lines are held down, and
held approximately horizontal, by
turbulent
``\mfp''
from
the convective
overshoot layer.
The effectiveness of such flux pumping
can be strongly argued from
several lines of evidence,
including three-dimensional direct numerical simulations,
with varying emphasis on the role
of turbulent anisotropy and of
vertical gradients of density and turbulent intensity
\citep[\eg][\& refs.]{Tobias-etal01,
KitchatinovRudiger06}
\citep[see also \S3 of][for a historical review]{Weiss-etal04}.

Near the poles it is less clear that \mfp\
will
be effective
in confining the field.
At least its effectiveness for
near-vertical
magnetic
fields
has not, to our knowledge, been convincingly demonstrated.
However, as argued for instance in GM98,
there are
good reasons in any case
(\S\ref{sec:downwelling}
below) to expect the \tc's
MMC
near the poles
to take the form of weak
but persistent downwelling.
This suggests
that the field can,
in any case,
be confined in the polar caps through an
advection--diffusion balance,
the kind of balance
argued
for
heuristically
in GM98.
The purpose of this paper is to
show in detail, by solving
an appropriate set of
nonlinear
magnetohydrodynamic equations,
that such
polar
confinement
by downwelling
is indeed
possible in a physically realistic model,
applicable to the Sun both
today and early in its main-sequence lifetime.

A large family of
axisymmetric
nonlinear
solutions
showing polar
confinement
has been obtained using two different techniques.
The first technique is
\san\ in a sense to be explained, and the second is numerical
on a \mbox{2-dimensional} grid.
The solutions are to be regarded as candidate solutions for
possible flows in the real Sun, all
showing
confinement
in the sense that
the total
magnetic
field strength $|\B|$ dies off exponentially
with altitude,
thanks to downward advection acting against upward diffusion.
In this sense
the poloidal and toroidal field components are both
well confined.
We call these flows ``\cl s''.
They are not to be confused with the \tc\ itself.
Rather, they occupy relatively thin regions
at the bottom of the \tc\ and are
much more weakly sheared,
with relatively long
timescales
$\sim 10^5$\yr.

The detailed
dynamics
involves not only magnetic advection, stretching,
twisting and diffusion but also a near-perfect balance between
Lorentz, Coriolis, pressure-gradient and buoyancy forces
(\S\S\ref{sec:model-equations}ff.).
Thus the \chl\
flows are magnetostrophic, like certain flows that
have been studied
in connection with
models of
the Earth's liquid core
\citep[\eg][\& refs.]{Kleeorinetal97},
though different in most other respects.
For instance the
latter flows
are viscous but unstratified: buoyancy forces
and thermal diffusion are absent.
In the \chl\ flows studied here,
by contrast,
viscosity turns out to be
wholly
unimportant
while
buoyancy and thermal diffusion are crucial,
along with magnetic diffusion.

\begin{figure}
  \centering
  \psfrag{U}[l][]{$\mathbf{u}$}
  \psfrag{B}[tl][B]{$\mathbf{B}$}
  \psfrag{W}[][t]{$_{\phantom{\ii}}\OmegaI$}
  \psfrag{TC}[][]{\sc \tc}
  \psfrag{cl}[][]{\begin{tabular}{c}confinement\\[-0.1em]layer\end{tabular}}
  \psfrag{sl}[][]{\begin{tabular}{c}helium\\[-0.1em]sublayer\end{tabular}}
  \psfrag{HSL}[][]{\sc\begin{tabular}{c}helium\\[-0.1em]settling\\[-0.1em]layer\end{tabular}}
  \includegraphics[clip=true,bb=-205 150 818 675,width=10cm]{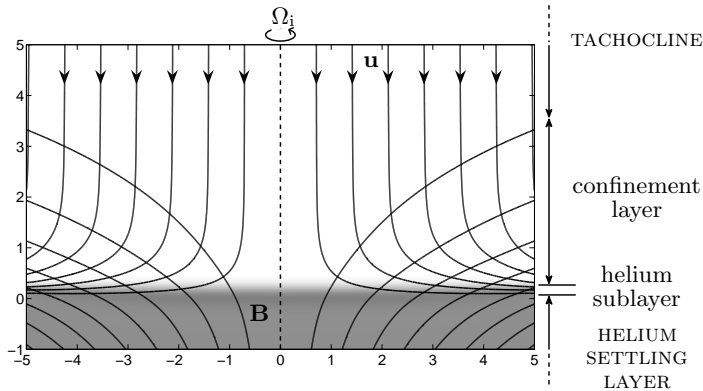}
  \caption{\footnotesize The magnetic confinement layer
  near the
  north pole
  in a model for today's Sun.
  The
  field strength $|\B|$
  falls off
  exponentially with
  altitude
  $z$. \
  The toroidal components of $\B$ and velocity $\bu$
  are not shown.
  The streamlines with arrows
  show the downwelling responsible for the confinement.
  If the downwelling were switched off, the field
  near the pole would diffuse and become nearly vertical, as
  illustrated for instance in 
  \citet{BrunZahn06}.
  Compositional stratification
  is indicated by shading.
  The plot is from a numerical solution;
  the corresponding
  \san\ solution looks
  almost
  identical.
  The horizontal and vertical axes are colatitudinal distance
  $r$ and altitude $z$
in units of $\CLthick$,
the advection--diffusion scale
defined in
(\ref{eq:CLthickdef}).
 With typical parameter values, the scale $\CLthick$
 is of the order of a fraction of a megametre,
$\sim10^{-3}$ of the Sun's radius.
}
\label{fig:section}
\end{figure}

\Fig\ref{fig:section}
gives a preview of a typical \chl\ flow, seen in
vertical cross-section.
It shows
the poloidal velocity and magnetic field
components
from
a
numerical solution.
The emerging magnetic
field lines are bent over
within the \cl,
as required
to fit into the global picture
sketched
in \fig\ref{fig:sphere}.
The magnetic field $\B$
has a
toroidal component,
not shown in the figure,
imparting spiral shapes to the three-dimensional field lines
and providing
the prograde Alfv\'enic
torque
demanded by the global picture,
in balance with a retrograde Coriolis torque
on the equatorward flow.

The vertical and colatitudinal distances
in \fig\ref{fig:section}
are shown in units of the magnetic advection--diffusion scale 
\vspace{-0.15cm}
\begin{equation}
\CLthick \definedas \eta/U
\,,
\label{eq:CLthickdef}
\end{equation}
say,
where $U$ is the magnitude of the downwelling and $\eta$
is the magnetic diffusivity.
Throughout this paper,
we assume that the \chl\ flow is
laminar,
and therefore use
molecular or microscopic diffusivity
values
(\S\ref{sec:model-equations}).
Issues of
stability or instability
lie beyond the scope of this paper but,
close to the pole at least,
there appears to be a
strong case for stability, to be argued in
a future paper,
arising from the smallness of the scale $\CLthick$.
Under reasonable assumptions,
$\CLthick$ is only a fraction of a megametre,
far smaller than the thickness of the overlying \tc\
which latter, by contrast, is probably unstable and
indeed turbulent,
as already mentioned
\citep[\eg][]{Spruit02,GilmanCally07,ParfreyMenou07}.

In the
present-day
Sun's \HSL, the top of which corresponds to
the shaded
region in \fig\ref{fig:section},
a downward gradient of helium
concentration
reinforces
the
stable stratification due to the
sub-adiabatic temperature
gradient.
Because the diffusivity
of helium through hydrogen,
$\chi\approx0.9\times10^{1}$\cmms,
is much less
than the magnetic diffusivity
$\eta\approx4.1\times10^2$\cmms,
the \HSL\ is nearly impermeable to
the \chl\
flow.
Helium advection and diffusion are comparable
only in
the extremely thin
``\hsl''
marked in \fig\ref{fig:section}.
In this
and other respects,
all the solutions in the present paper supersede
those described
in a first report on this work
\citep[][hereafter WM07]{WM07}.
For instance,
in WM07
we took $\chi$
to be zero, implying a \hsl\ of vanishing thickness.
We also took $\nu$, the
kinematic viscosity,
to be zero
and allowed a finite slip velocity at the top of the \HSL,
assuming that this slip velocity
would in reality be resolved into a weak Ekman layer.
However, the solutions presented here show that, on the
contrary, no Ekman layer forms.
The slip discontinuity is replaced by a
smooth velocity profile across
the \hsl\ and,
as
will be
shown in \S\ref{sec:helium},
the flow stays essentially inviscid.

The plan of the paper is as follows.
In \S\ref{sec:downwelling} we
summarize the
reasons for expecting
persistent
downwelling
over the poles.
In \S\ref{sec:model-equations}
we
present the model equations and in
\S\ref{sec:self-similar}
the \san\ solutions.  Those solutions
rely on assuming
a self-similar horizontal
structure
that is asymptotically
valid in the limit of strong stable stratification.
The same limit was taken
in WM07.

The validity of the strong-stratification limit
is assessed in \S\S\ref{sec:CL-scalings}
and~\ref{sec:helium},
which take a
thorough
look at the dynamical balances
and scalings
in the \cl\ and \hsl\ respectively.
Strong stable stratification
means that the
thermal and compositional stratification surfaces
are
``flat'', meaning gravitationally
horizontal, 
to sufficient approximation in some region surrounding
the poles,
which for reasonable parameter values
can be quite large in horizontal extent, up to tens
of degrees of colatitude.
Within the \hsl,
the low magnetic Reynolds number and flat geometry
cause
the momentum balance to take on the character of
flow in a porous medium, as
fluid pushes horizontally
past the field
lines.  As already indicated, true viscous effects are
negligible everywhere, even in the
sublayer.

Boundary conditions for the numerical solutions
are discussed in
\S\ref{sec:boundary-conditions}.
The numerical solutions
themselves are presented and discussed in
\S\ref{sec:num-solutions}.
They
provide cross-checks with the \san\ solutions
plus
additional insights.
In particular, they
directly demonstrate the
flatness of the stratification surfaces
by
solving the
full equations,
for finite stratification.
The solutions
allow the stratification surfaces to tilt as they may,
but
confirm that the
departures from flatness
are
indeed small when the stratification
is realistically strong.
In
\fig\ref{fig:section},
for instance,
the departures from flatness are barely visible.

In \S\ref{sec:comparison} we discuss
a subtlety
that arises when comparing
the \san\ and numerical solutions
in the upper part of the flow.
The dynamical balances aloft
become delicate
as the Lorentz and Coriolis forces become
vanishingly small.
The effects of
truncation error
and other small effects
thus complicate the comparison.
However, this is
something of an academic point because
of our expectation
that, in reality, the \chl\ solutions will need to be matched to
a turbulent \tc\ aloft, a task that remains a challenge for
the future.

In \S\ref{sec:noHSL}
we show that
the presence of
the \HSL\ is
not crucial
to our
\chl\ model.
The interior field
$\BI$
is sufficient by itself to turn
the flow equatorward, and the field remains confined in much
the same way.
That result
has relevance to
the Sun's early main-sequence evolution.
It explains
for instance how the
burrowing tendency could have been held
in check
from the start,
allowing the \HSL\ to form.
In the concluding discussion,
\S\ref{sec:conclusions},
we consider the implications for
early
solar evolution and lithium depletion.

\section{Downwelling in the polar tachocline}
\label{sec:downwelling}

Our polar-confinement scenario
relies
on the MMC pattern in the
stably-stratified
polar \tc\ being robustly and persistently downward
above the \cl,
after
averaging out any fast fluctuations due to waves and
turbulence.
As recognized in GM98,
\hsy\ provides a compelling
reason to expect polar
downwelling rather than upwelling,
at least in today's \tc.
A further reason
is that a downward MMC over the pole
is
a robust consequence
of the
gyroscopic pumping
already mentioned,
which, in the absence of the interior field $\BI$,
would mediate the
downward spreading or burrowing
of the \cz's slow polar rotation.
The distinction between gyroscopically-pumped MMCs and
MMCs driven in other ways
is reviewed in
\citet[][\S\S8.1\nlb--\nlb8.2]{McIntyre07},
confirming also the robustness
of the burrowing tendency itself,
despite recent controversy.
The upshot
is that
we expect persistent polar downwelling
to be present
not only in today's Sun, but also
throughout the Sun's main-sequence lifetime.

The
argument from \hsy\
is as follows.
As is well known, the pressure, density and angular velocity
fields, averaged with respect to time and longitude,
satisfy hydrostatic and cyclostrophic balance
to excellent approximation.
Departures from
such balance must take the form of fast oscillations
such as \mbox{$p$-modes} and
\mbox{$g$-modes},
or turbulent
fluctuations.
From the curl of the momentum equation,
taking its azimuthal component,
we may show
in the standard way that
balance implies the so-called
``thermal-wind relation''.
In
cylindrical polar coordinates $(z,r,\phi)$ centered on the
axis of rotation,
with the axial coordinate $z$ directed
vertically upward at the north pole,
the thermal-wind
relation can be expressed as
\vspace{-0.15cm}
\begin{equation}
  r\rho^2\frac{\partial|\bm{\Omega}|^2}{\partial z}
  =
  (\bm{\nabla} \PP \times \bm{\nabla} \rho) \cdot \mathbf{e}_\phi
  \label{eq:thermal-wind}
\end{equation}
where $\bm{\Omega}$ is the absolute angular velocity of the Sun's
differential rotation,
$\rho$ is the density,
  and
$\PP$ is the total pressure.
The
unit vector $\mathbf{e}_\phi$ is directed azimuthally,
while
$\bm{\nabla} \PP$, being
dominated by its hydrostatic part, is
   very
close to being vertically
downward.
On the assumption that the observed negative sign of
$\partial|\bm{\Omega}|^2\Antisliver/\partial z$
persists into the region near the
pole invisible to helioseismology
--- Occam's razor
makes this a reasonable assumption ---
we must have a minimum in $\rho$,
and hence a maximum in temperature $T$, on each isobaric surface
at the pole.

The stably-stratified radiative envelope is a thermally relaxing
system.  Local temperature anomalies, defined as departures of
$T$ from local radiative equilibrium,
will tend to relax
back toward zero.  To hold $T$ above radiative equilibrium
near the pole, there has to be persistent adiabatic
compression by downwelling,
with compensating upwelling and negative $T$ anomalies
in lower latitudes.

The strength $U$
of the polar downwelling
is difficult to estimate precisely.
Among other things it
depends
on the \tc\ thickness, which is not
well constrained by helioseismology.
The thickness scale enters both via \eq(\ref{eq:thermal-wind})
and, more sensitively, via
the rate of
diffusive thermal relaxation
within the \tc.
GM98 estimated
$U\sim10^{-5}$\cms, using
the rather small
\tc\
thickness estimate, 13\Mm, derived by \citet{ElliottGough99}.
The value of $U$
thus estimated
is inversely proportional to the cube
of the \tc\ thickness, and so
a similar estimate using a deeper \tc\ would
yield a
much smaller value of $U$.

However, GM98 assumed that the bulk of the \tc\ is
laminar.
\citet{McIntyre07} considered an alternative scenario
in which
\mhd\ turbulent stresses
within a deeper \tc\
dominate the angular-momentum transport
from the overlying \cz,
except near the bottom of the \tc.
The turbulent stresses were estimated
by assuming a particular prescription
for the turbulence,
following \citet{Spruit02}.
The stresses diverge in a thin layer
near the bottom of the \tc, just above
the \cl,
where they
gyroscopically
pump a downwelling
of the order of
$U\sim 4\times10^{-5}$\cms\
or greater.

Fortunately,
our \chl\ solutions can accommodate a wide range of
uncertainty
over the value of $U$.
They will show that polar field confinement
by downwelling
is robust
over a
range of $U$ values
at least as wide as
$10^{-6}$\cms\ to $10^{-4}$\cms.
From here on
we use GM98's value
$U\sim10^{-5}$\cms\ for illustrative purposes.

\section{The model equations}
\label{sec:model-equations}

Consider
the magnetic confinement layer
near
the north pole.
As already noted,
the magnetic advection--diffusion
thickness scale
$\CLthick \definedas \eta/U$
is to be evaluated with the microscopic
magnetic diffusivity $\eta$, whose value
in the neighbourhood of the \tc\
is carefully
estimated by \citet{Gough07} to be
$\eta \approx 4.1\times10^2$\cmms.
This gives
the value
$\CLthick\approx 0.4$\Mm\
if $U\approx10^{-5}$\cms.

We work in a frame rotating
with the same angular velocity
as the interior,
$\OmegaI = 2.7\times10^{-6}$\sm, and
seek axisymmetric
solutions of the Boussinesq MHD equations within a
domain consisting of a
cylindrical volume $V$
surrounding the pole.
Cylindrical coordinates  $(z,\,r,\,\phi)$ centred on the
rotation axis will be used,
with corresponding unit vectors
$(\mathbf{e}_z,\,
 \mathbf{e}_r,\,
 \mathbf{e}_\phi
)$.
The use of cylindrical coordinates will lead to
significant mathematical simplifications.
We may regard them as
slightly-distorted spherical coordinates, with
$r$ representing approximate colatitudinal distance
and $z$ locally vertical.
The \chl\ flows to be studied are thin-shell flows,
with $z\sim\CLthick$ and
$r\gg\CLthick$, and so
the coordinate distortions should
be qualitatively unimportant
out to colatitudes as far as
20\degg\ or so.

The Boussinesq
framework should
itself
be highly accurate because
typical flow and Alfv\'en velocities are
tiny fractions
of the local sound speed, and
because
$\CLthick$
values
of the order of a fraction of a megametre are far
smaller
than the pressure scale height,
$\approx 60$\Mm.
Conveniently, Boussinesq dynamics permits us to
measure the strength of the magnetic field $\B$
in terms of the corresponding Alfv\'en speed,
with 1\G\ corresponding to 0.6\cms\
at a (constant) \tc\ density of 0.2\gcmmm\
\citep{Gough07}.

We impose
uniform
downwelling of magnitude $U$
aloft
and
a simple axial
dipolar,
fully-diffused
poloidal
magnetic field structure beneath,
to represent the interior magnetic field
$\BI = (B_{\ii z},\,B_{\ii r},\,B_{\ii\phi})$,
on
to which the field $\B$ in the \cl\ is to be matched.
This interior
dipolar field has
$B_{\ii r}/r$
constant
and
$B_{\ii z}$
a linear function of $z$.
It is possible to have
$B_{\ii\phi}\ne0$ with
$B_{\ii\phi}/r$ constant; however,
for reasons
to be
explained at the end of the section,
the main focus will be on
the purely poloidal case
$B_{\ii\phi}=0$.  This
implies the vanishing of the Alfv\'enic torque beneath.
For the \san\ solutions, the condition
$B_{\ii\phi}=0$ is imposed directly.
For the numerical solutions a less direct procedure is necessary,
to be
explained in \S\ref{sec:num-solutions}.

It is convenient to
nondimensionalize the
equations using
$\CLthick$ as the
lengthscale
in
the horizontal ($r$)
as well as
in the vertical ($z$) direction.
Thus the thin-shell nature of the flow is expressed by the
dimensionless relation $r\gg 1$.
We take
$U$ as
the scale for the velocity field $\bu$,
and $\CLthick/U$ as the timescale,
$\sim 10^5$\yr\ if
   $U\sim   10^{-5}$\cms.
This is the advection timescale for the flow through the \cl\
and, by construction, is also the timescale on which $\B$ diffuses
across the \cl.
Since this timescale far exceeds the typical turnover time of the
turbulent eddies in the overlying layers,
we may
neglect any fluctuations in the downwelling aloft.
We therefore take $U$ to be
steady as well as uniform,
representing the
time-averaged downwelling
that is
gyroscopically pumped by
turbulence in the overlying layers,
whether those layers consist of the \cz\ or the \tc\
or both.
Fluctuations in $U$
may well be present,
but should not greatly
influence the structure
of the confinement layer on timescales
$\sim 10^5$\yr\
or more.
Even if the downwelling comes in pulses, the cumulative effect
will arguably be much the same as if it were
steady.

We nondimensionalize
the magnetic field $\B$
(expressed as Alfv\'en velocity)
with respect to
a different
velocity scale
$(2\OmegaI\eta)^{1/2} \approx 0.05$\cms.
The significance of this last
choice
will emerge
in \S\ref{sec:CL-scalings}.
It will simplify
the scaling relations
(\ref{eq:Lambda-scalings-a})--(\ref{eq:Lambda-scalings-f}).
We suppose that the thermal
and compositional stratifications are
approximately
uniform within the \HSL,
shaded in \fig\ref{fig:section},
since the \cl's MMCs do not penetrate into that region.
Writing
$\hat{\pottemp}$ and $\hat{\mu}$ for the fractional
(therefore dimensionless)
perturbations of Boussinesq
potential
temperature and
mean molecular weight,
we
therefore
impose that
the corresponding buoyancy frequencies
are
exactly constant
at the bottom of the domain.
That is, we impose,
with $z$ now the dimensionless
vertical coordinate,
\begin{align}
  \left.\frac{\partial\hat{\pottemp}}{\partial z}\right|_{\rm bottom}
&\;\definedas\;
\frac{N_\pottemp^2\CLthick}{g}
~\;\;\definedas
\mbox{~const.,}
\label{eq:NTdef}
\\
  \left.\frac{\partial\hat{\mu}}{\partial z}\right|_{\rm bottom}
&\;=\;
-\frac{N_\mu^2\CLthick}{g}
\;=\;
\mbox{const.,}
\label{eq:Nmudef}
\end{align}
the dimensional buoyancy frequencies
$N_\pottemp$ and $N_\mu$
being constant by definition.
For today's Sun we have
$N_\pottemp\approx0.8\times10^{-3}$\sm\
\citep[][and Gough 2010, personal communication]{Gough07},
and
$N_\mu\approx0.5\times10^{-3}$\sm\
\citep[\eg][]{JCDThompson07}\footnote{Estimates
  of $N_\mu$ vary \citep[\eg][]{JCD-etal93}.
  The value $N_\mu\approx0.5\times10^{-3}$\sm\
  was computed in
  \citet[][\S8.5]{McIntyre07}
  from information given in \citet{JCDThompson07}.
  However, the results
  presented in this paper
  are
  not critically dependent on
  the value of $N_\mu$;
see \S\ref{sec:helium}.
}
corresponding to a
total buoyancy frequency
$N = (N_\pottemp^2 + N_\mu^2)^{1/2}\approx0.94\times10^{-3}$\sm\
representative of the stratification
just inside the \HSL.
In place of
$\hat{\pottemp}$ and $\hat{\mu}$
it proves convenient to define
rescaled quantities
$\pottemp$ and $\mu$, also dimensionless,
by
\begin{align}
  \frac{N_\pottemp^2\CLthick}{g}\pottemp
&\definedas
\hat{\pottemp}
\,,
\label{eq:Tdef}
\\
  \frac{N_\mu^2\CLthick}{g}\mu
&\definedas
\hat{\mu}
\,,
\label{eq:mudef}
\end{align}
so that the dimensionless stratifications
inside the \HSL\ become simply
$\partial\pottemp/\partial z = 1$ and
$\partial\mu/\partial z = -1$.

Finally, we nondimensionalize the pressure anomaly $\pp$
by $2\OmegaI\eta\rho$ where $\rho\approx0.2$\gcmmm,
the constant Boussinesq density,
and thus
arrive at the following dimensionless
equations,
\begin{flalign}
  \Ro\frac{\DD\bu}{\DD t} + \mathbf{e}_z
\times\bu\,
=& \; -\bm{\nabla} \pp
  + \alphat\sliver
\pottemp\mathbf{e}_z
  - \alphamu\sliver
\mu\sliver\mathbf{e}_z \nonumber \\
&\phantom{\Big|}
\:+\;(\bm{\nabla}\times\B)
 \times\B
  + \Ek\nabla^2\bu
\label{eq:momentum} \\
  0
=& \;\bm{\nabla}\cdot\bu\phantom{\raisebox{-0.cm}{|}}
\label{eq:divu} \\
  \frac{\partial\B}{\partial t}
=& \;\bm{\nabla}\times(\bu\times\B)
  + \nabla^2\B
\label{eq:B-induc} \\
  0
=& \;\bm{\nabla}\cdot\B
\label{eq:divB} \\
  \frac{\DD \pottemp}{\DD t}
=& \;\frac{\kappa}{\eta}\nabla^2\pottemp
  \label{eq:T} \\
  \frac{\DD\mu}{\DD t}
=& \;\frac{\chi}{\eta}\nabla^2\mu
\label{eq:mu}
\end{flalign}
where
$\DD/\DD t\definedas\partial/\partial t + \bu\cdot\bm{\nabla}$,
the
material derivative,
and where
$\mathbf{e}_z$ is the
vertical
unit vector.
The thermal and compositional diffusivities
$\kappa$ and $\chi$ take numerical values
$\kappa\approx1.4\times10^7$
\cmms\
and
$\chi\approx0.9\times10^1$
\cmms\
in the neighbourhood of the \tc\
\citep{Gough07}.
In (\ref{eq:mu}) we have neglected
gravitational settling,
an excellent approximation
in virtue of
the short timescale $\sim10^5$\yr\ of the \chl\
dynamics relative to the Sun's lifetime,
$\gtrsim10^9$\yr.
We have defined four
dimensionless
constants
\begin{equation}
\alphat \definedas \frac{N_\pottemp^2\CLthick^2}{2\OmegaI\eta}
\:,
\hspace{0.65cm}
\alphamu \definedas \frac{N_\mu^2\CLthick^2}{2\OmegaI\eta} 
\;,
\end{equation}
\vspace{-0.3cm}
\begin{equation}
  \Ro \definedas \frac{U}{2\OmegaI\CLthick}
  = \frac{\eta}{2\OmegaI\CLthick^2}
\;,
\hspace{0.45cm} \mbox{and} \hspace{0.55cm}
  \Ek \definedas \frac{\nu}{2\OmegaI\CLthick^2} = \frac{\nu}{\eta}\Ro
\;
  \label{eq:EkRo}
\vspace{0.1cm}
\end{equation}
where
$\nu$ is the kinematic viscosity,
the sum of
molecular and radiative contributions,
$\approx 2.7\times10^1$\cmms\
in the neighbourhood of the \tc\
\citep{Gough07}. \
The Rossby and Ekman numbers, $\Ro$ and $\Ek$,
quantify how far magnetic flux and
fluid momentum diffuse across the \cl\ during one solar rotation.
For $U \sim 10^{-5}$\cms\
the Rossby number
is tiny,
$\Ro\sim0.5\times10^{-7}$,
and the Ekman number is smaller still because
$\nu/\eta \approx 0.7\times10^{-1}$.
To excellent approximation,
therefore,
the flows under consideration will
be
magnetostrophic.
That is,
in (\ref{eq:momentum})
the Coriolis force
will be balanced against the combined pressure-gradient,
buoyancy, and Lorentz forces:
\begin{align}
  \mathbf{e}_z\times\bu
\;=\; -\bm{\nabla} \pp
  + \alphat\sliver
& \pottemp\mathbf{e}_z
  - \alphamu\sliver
  \mu\sliver\mathbf{e}_z
+ (\bm{\nabla}\times\B)\times\B
  \;,
\label{eq:Tmubalance}
\end{align}
and this will be verified independently from the
numerical solutions, from which
\mb\ emerges rather than
being imposed.

We are concerned here only with axisymmetric steady states.
Then
the azimuthal components of (\ref{eq:Tmubalance}) and its curl
are respectively
\begin{flalign}
  u_r
&~=~
  \frac{1}{r}\B\cdot\bm{\nabla}(rB_\phi)
\;,
\phantom{\raisebox{-0.3cm}{j}}
\label{eq:torque}
\\
  \frac{\partial u_\phi}{\partial z}
& \:=~
\alphat\sliver
\frac{\partial \pottemp}{\partial r}
-
\alphamu\sliver
\frac{\partial\mu}{\partial r}
+ \frac{1}{r}\frac{\partial}{\partial z}(B_\phi^2)
  - r\Sliver\B\cdot\bm{\nabla}\left(
         \frac{[\bm{\nabla}\times\B]_\phi}{r}
       \right)
\label{eq:T-W}
\end{flalign}
where
$r$ is the
dimensionless
perpendicular distance from the
rotation axis,
and
where suffixes $z,\,r,\,\phi$
denote
vector components.

\Eq(\ref{eq:torque})
represents the
local
torque balance about the rotation axis,
after multiplication by $r$.
It
describes how
the retrograde Coriolis torque
from the
equatorward flow
is balanced by
the prograde Lorentz torque
from
the confined magnetic
field.\footnote{A
referee reminds us that this torque balance is related to the
well-known
Taylor constraint for magnetostrophic flow
\citep{Taylor63}.
If (\ref{eq:torque}) is integrated vertically between two
hypothetical
impermeable boundaries then,
because $\bm{\nabla}\cdot\bu=0$,
the integral on the
left and hence that on the right
must vanish.
In the \chl\ problem, however,
this constraint
is broken
by the presence of downwelling aloft.
}
\Eq(\ref{eq:T-W}) represents
thermal-wind balance
generalized to include
compositional gradients and
the Lorentz force-curl.
In the
upper part of \fig\ref{fig:section},
where the magnetic field and compositional stratification
are both negligible,
this balance becomes the standard thermal-wind
balance
\vspace{-0.2cm}
\begin{equation}
  \frac{\partial u_\phi}{\partial z}
  =
  \alphat\sliver
  \frac{\partial \pottemp}{\partial r}~,
  \label{eq:justT-W}
\vspace{0.1cm}
\end{equation}
which is the Boussinesq, low-Rossby-number limit
of (\ref{eq:thermal-wind})
in dimensionless units.

The overall torque balance
for the \cl\ can be
expressed
by integrating $r$ times (\ref{eq:torque}) over
the
volume $V$ of the
cylindrical domain,
then using the divergence theorem
and the fact that
$\bm{\nabla}\cdot\bu=0$.
The result is
\begin{equation}
        \int_{\partial V}
     \tfrac{1}{2}r^2
        \bu\cdot\dd\mathbf{S}
        ~=~
        \int_{\partial V}\!
        rB_\phi\B\cdot\dd\mathbf{S}
\label{eq:torque-integrated}
\end{equation}
where $\dd\mathbf{S}$ is the vector area element directed
outward.
In the corresponding dimensional equation, $\tfrac{1}{2} r^2$ is
replaced by $\OmegaI r^2$, the absolute angular momentum
per unit mass
after neglecting contributions $O(\Ro)$.

The \RHS\
of (\ref{eq:torque-integrated})
represents the
total
Alfv\'enic torque
exerted
on the \cl.
The \LHS\
represents the
net
rate of absolute
angular momentum export by the
flow
coming in
through the top and out
through the
periphery,
as in \fig\ref{fig:section}.
For any such velocity field
the \LHS\ is
positive.
Therefore 
the
Alfv\'enic torque 
must
also be positive, \ie\ prograde.

The need for the field $\BI$ is now apparent.
Without it,
the balance described by (\ref{eq:torque-integrated})
would be impossible.
Instead,
the fluid within $V$,
and outside it as well,
would begin to be spun down by the flow.
In the Sun, this would cause the
slow rotation of the high-latitude \cz\
to spread down
into the radiative envelope,
in the Haynes--Spiegel--Zahn burrowing process
noted in \S\ref{sec:intro}.
We also note that the torque balance
described
by (\ref{eq:torque-integrated})
is very different from the torque balance
described
by \citet{RudigerKitchatinov97}
and \citet{KitchatinovRudiger06}
in which
there is no Coriolis effect,
the Lorentz torque being balanced
instead
by a viscous torque.
In fact
for realistic solar parameter values
it will be seen that, as already indicated,
viscous torques are entirely negligible
--- perhaps counterintuitively for shear layers as thin as the
\cl\ and the \hsl.

The balance expressed by (\ref{eq:torque-integrated})
must
apply also throughout the
interior apple-core region
that surrounds the rotation axis, magnetically linking
the \cl\ at the north pole to that at the south.
The \muchoke, both in the Sun's
helium-rich inner core
and in the
\HSL, if present, is enough to suppress MMCs in
the apple-core region and
make the \LHS\ of (\ref{eq:torque-integrated})
negligible,
when the volume of integration $V$
is taken as the interior apple-core region.
Any Alfv\'enic torque exerted on the bottom of one \cl\
must therefore be balanced by an equal and opposite
Alfv\'enic
torque
exerted on the bottom of the other
\cl.
If we assume
that the two \cl s
are mirror-symmetric about the equatorial plane,
then it follows that
the torque, and therefore $B_{\ii\phi}$, must vanish
on the bottom of each \cl.
The prograde Alfv\'enic torque on
each \cl\ must therefore
come
wholly
from its sideways connection to lower latitudes
via the \tc,
which is consistent with the global picture suggested in
\fig\ref{fig:sphere}.

\vspace{-0.3cm}

\section{The \san\
solutions}
\label{sec:self-similar}

\Eq(\ref{eq:T-W}) describes how the Coriolis and
Lorentz force-curls act to tilt the stratification
surfaces
within the \cl.
Now
if the stable stratification is sufficiently strong,
then the tilting will be only slight.
The
formal
asymptotic limit to describe this is
$\alphat,\,\alphamu\to\infty$
with $\bu$ and $\B$
finite
so as to preserve the steady-state torque balance
(\ref{eq:torque-integrated})
and
a steady-state
balance in the induction equation
(\ref{eq:B-induc}).
In that limit,
(\ref{eq:T-W}) implies that
${\partial  \pottemp}/{\partial r}\to0$ and
${\partial\mu}/{\partial r}\to0$,
with both sides of (\ref{eq:T-W}) finite.
The
stratification surfaces
become perfectly flat:
$\pottemp \to \pottemp(z)$ and,
in the \hsl,
$\mu \to \mu(z)$.

In the limit of perfect flatness
thus enforced by (\ref{eq:T-W}), the remaining
equations can be
reduced to
a system of coupled ordinary differential equations.
With
$\pottemp = \pottemp(z)$ and $\mu = \mu(z)$,
(\ref{eq:T}) and (\ref{eq:mu})
imply for steady flow that
\vspace{-0.1cm}
\begin{equation}
  u_z
  \;=\;
  \frac{\kappa}{\eta}\frac{\dd}{\dd z}\ln\frac{\dd \pottemp}{\dd z}
  \;=\;
  \frac{\chi}{\eta}\frac{\dd}{\dd z}\ln\frac{\dd\mu}{\dd z}~.
  \label{eq:compatibility}
\end{equation}
Therefore
$u_z$ is also a function of $z$ alone,
$u_z = u_z(z)$. \
From (\ref{eq:divu})
it then
follows that
$u_r$ is $r$ times a function of $z$ alone,
on the assumption of regularity at the pole $r=0$.
We say that
the poloidal velocity field is ``horizontally self-similar''.
The induction equation (\ref{eq:B-induc})
then permits a steady
poloidal magnetic field that is horizontally self-similar
in the same sense.
The resulting equations are
\vspace{-0.19cm}
\begin{align}
\phantom{\Big|}
r\sliver u_r
&=
\B\cdot\bm{\nabla}(rB_\phi)
  \label{eq:torque_duplicate} \\
  0
&= \frac{1}{r}\frac{\partial(ru_r)}{\partial r}
  + \frac{\dd u_z}{\dd z}
  \label{eq:divu-1d} \\
  r\bu\cdot\bm{\nabla}\frac{B_\phi}{r}
&= r\B\cdot\bm{\nabla}\frac{u_\phi}{r}
  + \left(\nabla^2-\frac{1}{r^2}\right)B_\phi
  \label{eq:indp-1d} \\
  u_z\frac{\dd B_z}{\dd z}
&= B_z\frac{\dd u_z}{\dd z}
  + \frac{\dd^2B_z}{\dd z^2}
  \label{eq:indz-1d} \\
  0
&= \frac{1}{r}\frac{\partial(rB_r)}{\partial r}
  + \frac{\dd B_z}{\dd z}
  \label{eq:divB-1d}
~.
\vspace{-0.32cm}
\end{align}
The azimuthal component of (\ref{eq:momentum})
is replaced by
(\ref{eq:torque})
and
multiplied by $r$
to give (\ref{eq:torque_duplicate}),
expressing the balance between Coriolis and Lorentz
torques
as before.
\Eqs(\ref{eq:divu-1d})--(\ref{eq:divB-1d})
correspond to (\ref{eq:divu})--(\ref{eq:divB});
\eqs(\ref{eq:T}) and~(\ref{eq:mu})
have no further
role,
beyond their connection to the downwelling expressed by
(\ref{eq:compatibility}).
\Eq(\ref{eq:indp-1d}) permits more general toroidal
magnetic
and differential-rotation
fields
than
were considered in WM07.

For large but finite $\alphat$ and $\alphamu$,
departures from perfect flatness arise
as small corrections.
To describe these corrections it is necessary to bring back
\eqs(\ref{eq:T}), (\ref{eq:mu}) and (\ref{eq:T-W}); 
the order of magnitude of the corrections is
analysed in \S\S\ref{sec:CL-scalings}--\ref{sec:helium}.
The whole picture will be independently checked by
the numerical solutions, which automatically contain
the departures from flatness governed by
(\ref{eq:T}), (\ref{eq:mu}) and (\ref{eq:T-W})
since the
full set of equations is used, with finite values of
$\alphat$ and $\alphamu$, for instance to produce the
almost-flat numerical solution
plotted in \fig\ref{fig:section}.

Returning now to the limit of perfect flatness, we focus on
\eqs(\ref{eq:torque_duplicate})--(\ref{eq:divB-1d}).
These equations
admit a family of solutions in which
the function
$u_z(z)$
is arbitrary except for certain restrictions
on its asymptotic behaviour as $z\to\pm\infty$.
In particular,
we require
$u_z(z)\to-1$ as $z\to+\infty$
and $u_z(z)\to0$ as $z\to-\infty$.
These statements
will shortly be made
more
precise.
As discussed in \S 7,
the arbitrariness
in $u_z(z)$
is needed to permit matching to the
\cl's surroundings.

Solutions can be most conveniently constructed
by taking advantage of the arbitrariness to
specify a suitable $u_z(z)$ at the start.
With
$u_z(z)$ specified,
we can
then
find $B_z(z)$
numerically
by solving the vertical component
(\ref{eq:indz-1d})
of the induction equation
as a linear
ordinary differential equation,
assuming
that
the time-averaged field
$B_z$ vanishes far above the confinement
layer ($z \to +\infty$) and
that it
matches on to
the
imposed
interior dipolar
magnetic field structure beneath
($z \to -\infty$).
The interior dipole
has the same horizontally self-similar
structure as the \cl,
with components satisfying
$B_{\ii r}/r=$ constant,
and $B_{\ii z}$ a linear function of $z$
consistent with
(\ref{eq:divB-1d}).
Even though the balance in (\ref{eq:indz-1d})
is not simple advective--diffusive,
we find that $B_z$ decays upward like $\exp(-z)$.

The radial components of $\bu$ and $\B$ can be found
directly from their vertical components,
by
using (\ref{eq:divu-1d}) and
(\ref{eq:divB-1d}) and assuming
regularity at the pole $r=0$:
\vspace{-0.05cm}
\begin{align}
  u_r
&\,=\:
  -\Sliver\dfrac{r}{2}\Sliver
   \dfrac{\dd u_z}{\dd z}
\;,
\label{eq:divusim}
\\
  B_r
&\,=\:
  -\Sliver\dfrac{r}{2}\Sliver
   \dfrac{\dd B_z}{\dd z}
\;.
\label{eq:divBsim}
\end{align}
So
once we have $B_z(z)$
we can calculate $B_r$ from (\ref{eq:divBsim}), and
then
the toroidal field
$B_\phi$ from
(\ref{eq:torque_duplicate})
by using (\ref{eq:divusim})
and taking advantage of
the hyperbolic character of the operator
$\B\cdot\bm{\nabla}$. \
By calculating
$B_\phi$
in this way,
we ensure that the Lorentz torque
balances the Coriolis torque along
each
magnetic field line.
Requiring that
$B_\phi(r,\,z)\to 0$ for all $r$ as $z\to-\infty$
(recall the end of \S\ref{sec:model-equations})
leads to
the following, unique solution
of (\ref{eq:torque_duplicate}):
\begin{equation}
~~B_\phi
~=~
B_z
\int_{\!\antisliver-\antisliver\infty}^{\,z}
   \Antisliver\frac{u_r }{B_z^2}\,
\sliver \dd z
\,.
\label{eq:hatBphi}
\end{equation}
For any $u_r$ profile that decays
exponentially
as $z\to-\infty$,
this solution for
$B_\phi$, and with it the
Maxwell stress and Alfv\'enic torque,
will
also decay exponentially
as $z\to-\infty$.
The expression (\ref{eq:hatBphi})
then shows that $B_\phi$
has the same horizontally self-similar
functional form
as $u_r$ and $B_r$,
namely
$r$ times a function of
$z$ alone.

To ensure that $\B_\phi$ decays aloft,
as $z \to +\infty$,
it is sufficient to assume that
\begin{equation}
  u_r = O(\exp(-\twoandabit z))
\;\;\; \mbox{as} \;\;\;
  z \to +\infty
\label{eq:decay}
\end{equation}
for constant $\twoandabit > 1$,
implying
that
$u_z(z) \sim -1 + O(\exp(-\twoandabit z))$. \
The assumption that $\twoandabit > 1$
ensures that we get
solutions with qualitatively reasonable behaviour aloft.
Otherwise we leave the value of $\twoandabit$ arbitrary.
Once again this is an arbitrariness whose resolution will
depend on matching to the
\cl's surroundings, in this case to conditions aloft.
As already indicated,
the conditions aloft
probably involve turbulent flow,
for which we do not yet have quantitative models.
So
here we restrict ourselves to surveying the possible range of
behaviours for $\twoandabit > 1$.

The three cases
$\twoandabit > 2$, 
$\twoandabit = 2$, 
and $2 > \twoandabit > 1$
need separate consideration.
When $\twoandabit > 2$,
the only case considered in WM07,
the integral in (\ref{eq:hatBphi})
converges
to a constant plus $O(\exp(-(\twoandabit\!-\!2)z))$
as $z \to +\infty$. \
That in turn means that
$B_\phi$ decays upward like $\exp(-z)$.
When $\twoandabit = 2$, the integral in (\ref{eq:hatBphi})
asymptotes
to a linear function of $z$, and
$B_\phi$ decays upward like $z\exp(-z)$.
When $2 > \twoandabit > 1$, the integral in (\ref{eq:hatBphi})
increases upward like
$\exp((2\!-\!\twoandabit)z)$, but
$B_\phi$ still decays upward,
like $\exp(-(\twoandabit\!-\!1)z)$.

In all these
cases
it is clear from (\ref{eq:hatBphi}) that
$B_\phi$
does not have exactly the same
$z$-dependence
aloft
as does $B_z$.
This is
contrary to what
might might have been
expected from a naive appeal to
advective--diffusive balance,
with advection by constant downwelling $u_z=-1$.
Having the same $z$-dependences
would make
the \RHS\ of (\ref{eq:torque_duplicate}) vanish.
Hence
it is
the more or less subtle
\emph{departures} from advective--diffusive balance,
including the contribution to $B_\phi$ from
the twisting of field lines
by the differential rotation~$u_\phi$,
that
allow the
\RHS\ of
(\ref{eq:torque_duplicate})
not to vanish and
thereby to provide a Lorentz torque to support the
flow $u_r$
at all altitudes.

The $u_\phi$ field that does the twisting
can be calculated next, from
(\ref{eq:indp-1d}) and the condition that the
interior rotates \solid ly,
$u_\phi\to0$ as $z\to-\infty$.
Again this calculation depends on the hyperbolic character of
$\B\cdot\bm{\nabla}$. \
When $B_\phi$ is
given by (\ref{eq:hatBphi})
we have, uniquely,
\begin{equation}
~~u_\phi
~=~
\int_{\!\antisliver-\antisliver\infty}^{\,z}\!\!
\left(
   u_z\sliver\frac{\partial B_\phi}{\partial z}
   -
   \frac{\partial^2B_\phi}{\partial z^2}
\right)
\frac{\dd z}{B_z }
\label{eq:hatuphi}
\end{equation}
showing that
$u_\phi$ is also
$r$ times a function of
$z$ alone.  That is, the differential rotation is
what astrophysicists call
``shellular solid
  rotation''.
In the three cases
$\twoandabit > 2$, 
$\twoandabit = 2$, 
and $2 > \twoandabit > 1$,
the behaviours of $u_\phi$ as $z \to +\infty$
are respectively
$u_\phi \sim$ constant,
$u_\phi \sim \pm z$,
and $u_\phi \sim \pm\exp((2\!-\!\twoandabit)z)$.

Cases with negative shear aloft, $\partial u_\phi/\partial z < 0$
---
especially the
last case, with exponen\-ti\-al\-ly-increasing
negative shear aloft ---
are
suggestive of a possible way to match upwards
to the observed, much stronger
negative
shear in the bulk of the \tc.
By using
(\ref{eq:hatBphi}) to eliminate $B_\phi$ from
$\partial/\partial z$ of
(\ref{eq:hatuphi}),
then (\ref{eq:indz-1d}) to eliminate $\dd^2 B_z/\dd z^2$
and (\ref{eq:divusim}) to eliminate $\dd u_z/\dd z$,
we find
\begin{equation}
~~\frac{\partial u_\phi}{\partial z}
~=~ 
\frac{u_zu_r}{B_z^2}
-
\frac{1}{B_z^2}
\frac{\partial u_r}{\partial z}
-
\frac{2u_r}{r}
\int_{\!\antisliver-\antisliver\infty}^{\,z}
   \Antisliver\frac{u_r }{B_z^2}\,
\sliver \dd z
\;\approx\;
(\twoandabit\!-\!1)
\frac{u_r}{B_z^2}
\quad\mbox{as~~} z\to\infty
\,.
\label{eq:uphi-shear}
\end{equation}
The asymptotic behaviour on the right
comes from the first two terms in
the exact expression.  The third term
involving the integral
is smaller by a factor $O(\exp(-\twoandabit z))$.
Asymptotically, therefore,
the sign of the shear $\partial u_\phi/\partial z$ aloft
is the same as the sign of $u_r$ aloft.
We can therefore find
solutions that match on to
the strong
negative
\tc\ shear
provided that there is an exponentially weak poleward
mass flux above the confinement layer.
However, a more precise description of such matching
must await future work,
for reasons already mentioned.
We do not yet have quantitative models of the precise conditions
aloft, which are likely to be affected by
small-scale
MHD turbulence
\citep[\eg][\& refs.]{Spruit02,GilmanCally07,ParfreyMenou07}.
Aspects of this are touched on again in \S\ref{sec:comparison},
the main point for present purposes being that the
structure aloft is sensitive to conditions aloft 
whereas, as is clear from 
(\ref{eq:hatBphi}) and (\ref{eq:hatuphi}), the 
rest of the \cl\ is
insensitive to conditions aloft,
as we have verified by varying $\twoandabit$.

\begin{figure}
  \centering
  \psfrag{uz}[cc][cr]{$u_z$}
\psfrag{ur}[cc][cc]{$\dfrac{u_r}{r}$}
\psfrag{up}[cc][cc]{$\dfrac{u_\phi}{r}$}
\psfrag{Bz}[cc][cr]{$B_z$}
\psfrag{Br}[cc][cr]{$\dfrac{B_r}{r}$}
\psfrag{Bp}[cc][cr]{$\dfrac{B_\phi}{r}$}
\psfrag{z}[cc][cc]{$z$}
\psfrag{Velocity Field}[cc][cc]{Velocity Field}
\psfrag{Magnetic Field}[cc][cc]{Magnetic Field}
  \includegraphics[width=0.25\textwidth]{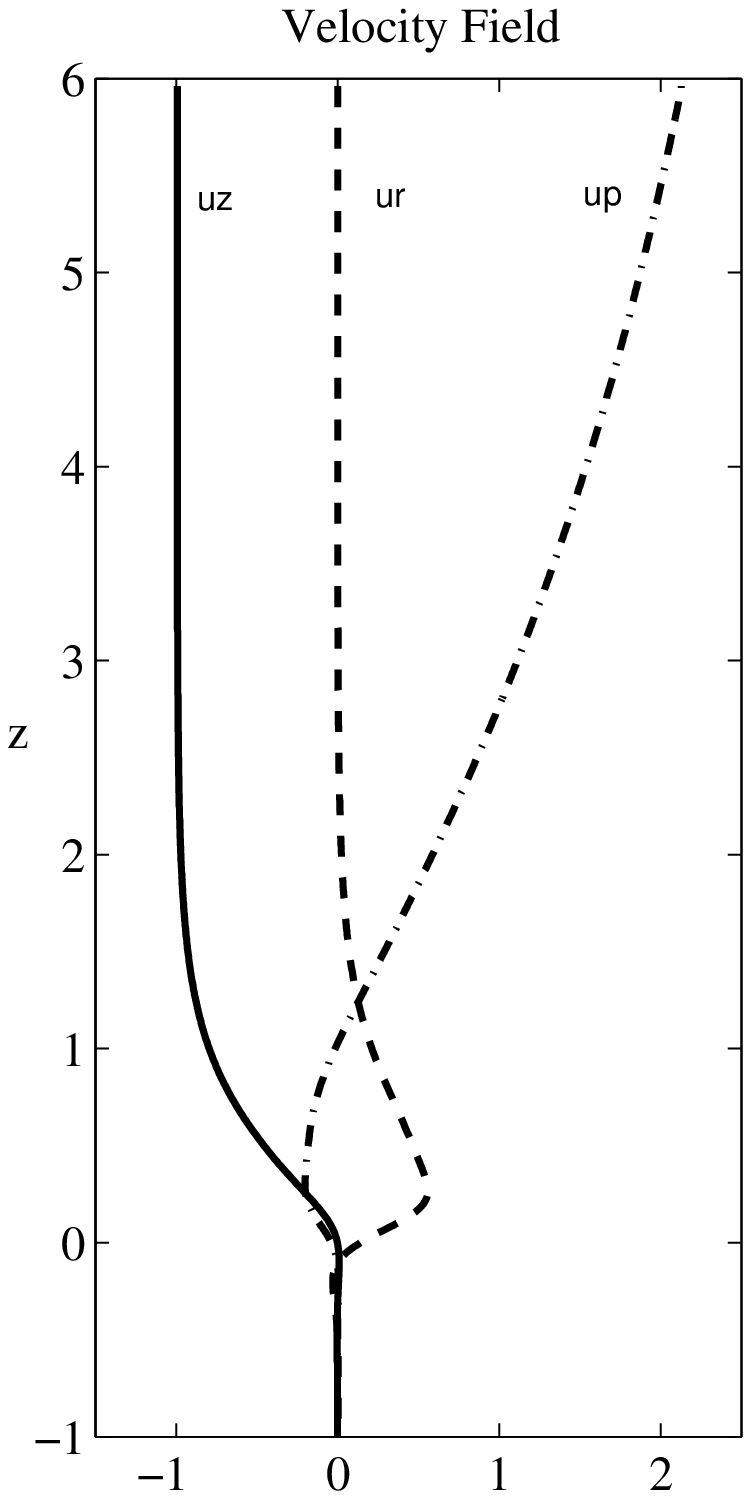}
  \includegraphics[width=0.25\textwidth]{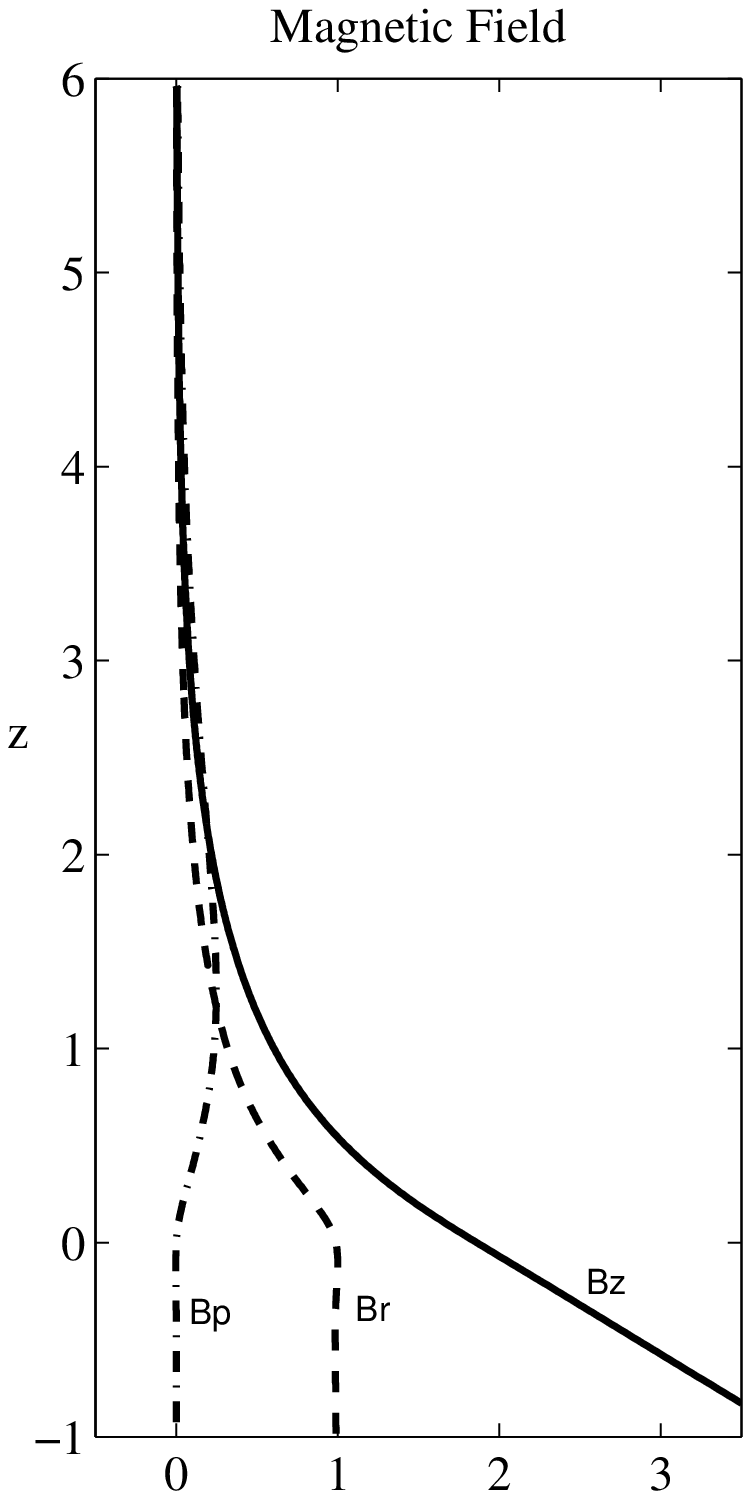}
  \caption{\footnotesize
 Vertical profiles from a
    \san\ solution of the \chl\
    equations in the strong-stratification
    limit $\alphat,\,\alphamu \to \infty$.
    The downwelling profile $u_z(z)$,
    solid curve on the
    left, was chosen to match the downwelling profile from
    the numerical solution shown in \fig\ref{fig:section}.
    For numerical reasons,
    small adjustments were
    made to this profile in the ``slippery''
    upper region
    $z > 1.5\CLthick$;
    see \S\S\ref{sec:boundary-conditions}--\ref{sec:comparison}.\;
    In (\ref{eq:decay}) the decay constant $\twoandabit = 2.24$,
    and the $u_\phi$ profile therefore approaches a constant
    like $\exp(-0.24z)$.
The parameter $\Lambda$,
(\ref{eq:Lambda}) below,
takes the value $\Lambda \approx 3.5$.
\label{fig:sim-profiles}}
\end{figure}

Purely for illustration we show
one of the solutions in
\fig\ref{fig:sim-profiles},
somewhat arbitrarily choosing $\twoandabit=2.24$.
In this case
the interior field $\BI$
is taken such that
$B_r/r = 1$.
The downwelling profile $u_z(z)$
was adapted from the numerical solution shown in
\fig\ref{fig:section},
in the manner
described in \S\ref{sec:comparison}.

\begin{figure}
  \centering
  \includegraphics[width=0.479\textwidth]{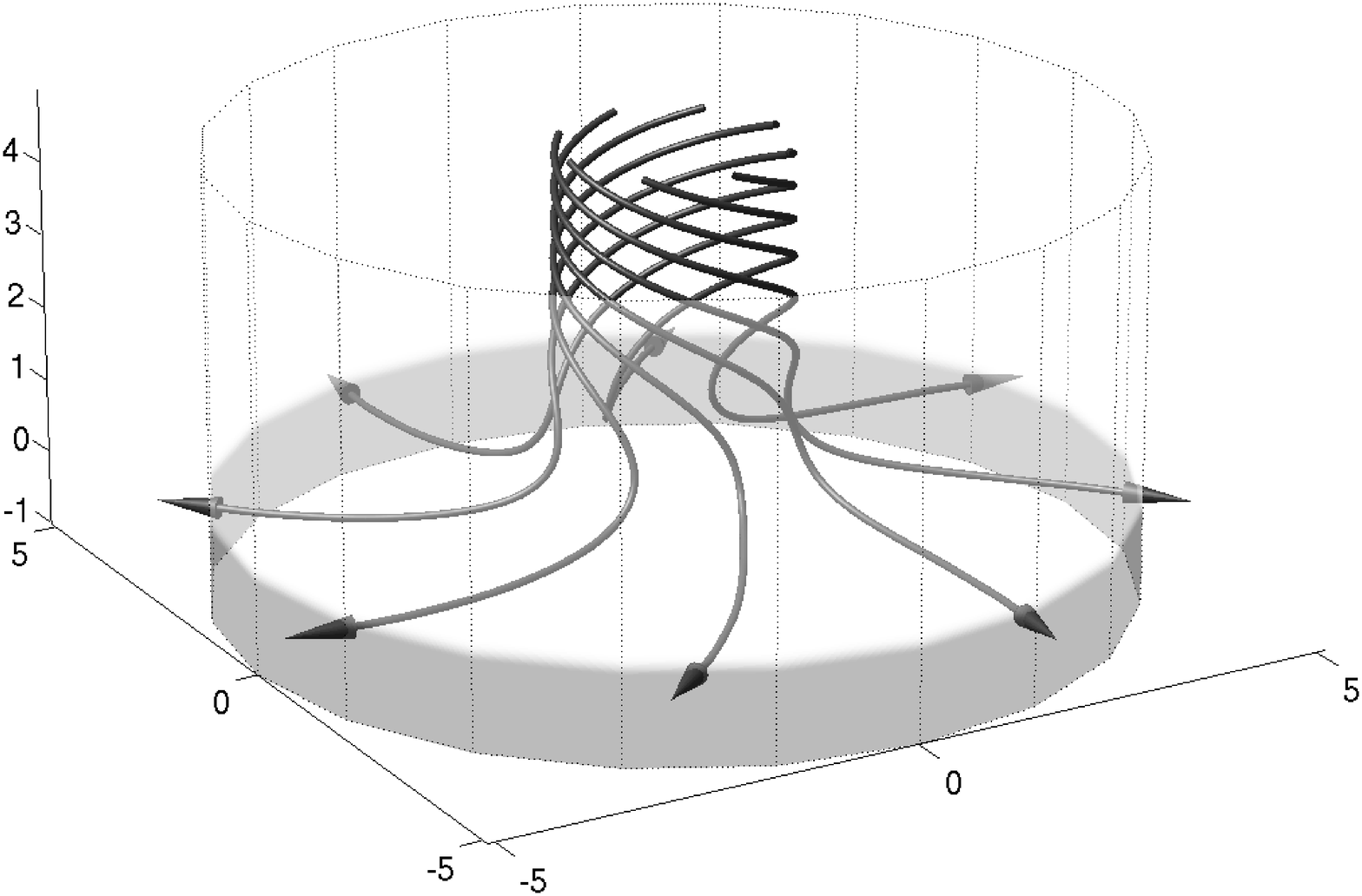}
  \includegraphics[width=0.479\textwidth]{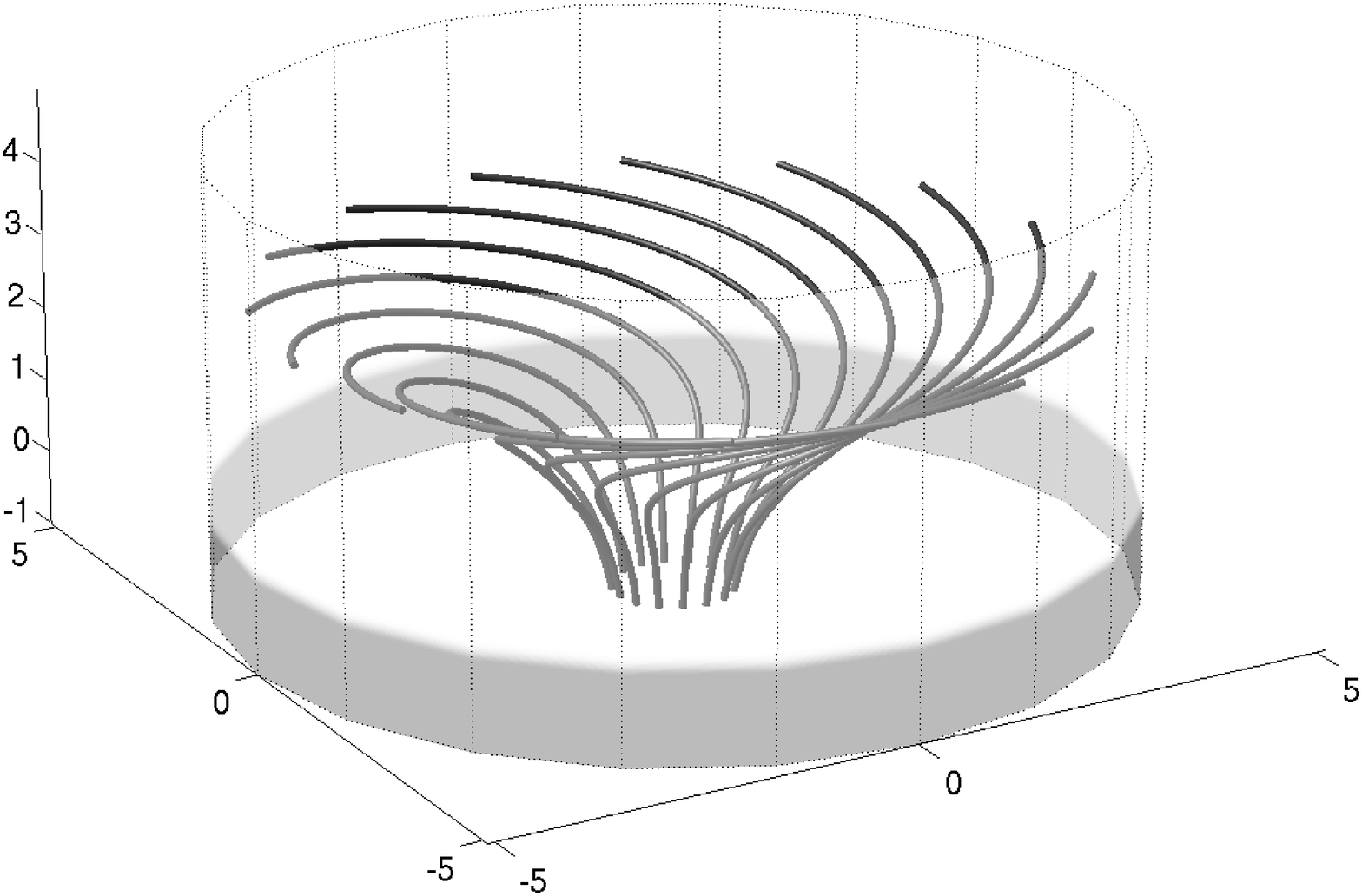}
  \caption{\footnotesize
    Streamlines (left) and magnetic field lines (right)
    from
    the \san\ solution
    whose vertical profiles are displayed in \fig\ref{fig:sim-profiles}.
    The peripheral shading locates
    the \HSL\ and \hsl.\label{fig:similar}}
\end{figure}

Some
three-dimensional streamlines and magnetic field lines
corresponding to the
solution
in
\fig\ref{fig:sim-profiles}
are plotted in \fig\ref{fig:similar},
visualizing
how the
prograde Lorentz
torque on the right of (\ref{eq:torque_duplicate}),
associated with field-line curvature,
balances the retrograde Coriolis
torque on the left of (\ref{eq:torque_duplicate})
and satisfies the overall torque balance
(\ref{eq:torque-integrated}).

\section{Confinement-layer scalings}
\label{sec:CL-scalings}

We are interested in
the departures from
perfect flatness
associated with large but finite
stratification.
We may regard
those departures as the
$O(\smallparam)$
correction terms in an asymptotic expansion
whose leading, $O(1)$ term is a \san\ solution,
where the small parameter $\smallparam$
is inversely proportional to the
stratification
and where the limit $\smallparam\to 0$ is taken within
a cylindrical domain of fixed dimensionless size
$r=
\rperiph$.

It will be found that the departures from flatness
behave like
$O(\smallparam r^2)$
near the pole.  So
instead
of saying
that the departures are small within the fixed domain
$r\leqslant\rperiph$ we may say,
perhaps more usefully,
that the \san\ solutions are valid
as leading-order approximations
as long as we are
well
within some dimensionless colatitudinal distance of the pole,
$\rt$ say,
that is
large
in comparison with unity.
For
a fair range of parameter values
the dimensional counterpart of $\rt$
turns out to
be quite large numerically,
of the order of hundreds of megametres, or
tens of degrees of colatitude.
Of course for the solutions to apply
we must also be within a region of
approximately uniform downwelling.
In this
section we
use scaling arguments to arrive at an
appropriate definition of
$\rt$  for the bulk of the \cl,
where the stratification
is purely thermal.  The next section
presents the corresponding analysis for the
\hsl.

Consider, then,
the scaling regime
in the bulk of the \cl.
Because
the
photon mean free path
makes
the thermal
diffusivity
$\kappa$
relatively large,
with
$\kappa/\eta \sim 3\times10^4$,
the \chl\ flow only weakly
perturbs the background thermal stratification.
Consistently with (\ref{eq:NTdef}) and (\ref{eq:Tdef})
we define
a dimensionless
thermal anomaly $\pottemp'$ such that
$\pottemp = z + \pottemp'\Sliver+\!$ constant,
and such that $\pottemp'\!\to0$ beneath the \cl\
at the pole.
Then, at the pole, and by implication
sufficiently close to it,
the leading-order balance in the
dimensionless
steady-state thermal equation
(\ref{eq:T})
involves only the vertical component $u_z$
of $\bu$,
\vspace{-0.2cm}
\begin{equation}
  u_z = \frac{\kappa}{\eta}\nabla^2\pottemp'
\,.
\label{eq:flat-T}
\end{equation}
On the \RHS\ we have
$
\nabla^2\pottemp'
\sim
\partial^2\pottemp'/\partial z^2
\sim
\pottemp'
$\Sliver\Sliver\Sliver
since in the \cl\ $\partial/\partial z \sim 1$,
and since $\pottemp'\to0$ beneath.  Hence
with $|u_z|\sim 1$ we have
\begin{equation}
\pottemp'
\;\sim\;
\eta/\kappa
\;\approx\;
3\times10^{-5}
\ll 1
\;.
\label{eq:total-pottemp-prime}
\end{equation}
In the bulk of the \cl\ we may estimate the
departure from flatness, equivalently
the $r$-dependence of $\pottemp'$,
from (\ref{eq:T-W})
with $\alphamu$ neglected
and
$\alphat$ considered large.  Since
the leading-order solution
is a \san\ solution,
the remaining terms in (\ref{eq:T-W})
all take the form
$r$ times a function of $z$ alone,
to leading order,
from which
we may deduce that
$\partial\pottemp'/\partial r$
is $r$ times a function of $z$ alone
and hence that
\begin{equation}
\pottemp'
\,=\:
\aaat
\,+\,
\bbbt\sliver r^2
\,,
\label{eq:central-parabola}
\end{equation}
to leading order,
where $\aaat$ and $\bbbt$ are
functions of $z$ alone
and where $\aaat$
has the small magnitude given by
(\ref{eq:total-pottemp-prime}), \
$\aaat\sim\eta/\kappa$. \
With
$\partial/\partial z \sim 1$
the condition for
(\ref{eq:central-parabola})
to be compatible with
uniform downwelling, $u_z = u_z(z)$ in
(\ref{eq:flat-T}),
can now be seen to be
$r\ll|\aaat/\bbbt|^{1/2}$\!.\:
We may therefore take
$\rt\sim|\aaat/\bbbt|^{1/2}$\!\:
or, alternatively,
$\smallparam\sim|\bbbt\sliver\rperiph^2/\aaat|$.\:
The
magnitude of $\bbbt$
is governed by
the remaining terms in (\ref{eq:T-W}).

Perhaps surprisingly,
the magnitudes
of $\bbbt$ and $\rt$
can be simply related to a single magnitude, that of
the vertical component $B_z$ of the magnetic field.
This is because
the velocity and magnetic fields of the
leading-order,
\san\ solutions
have components that are all
simply related to $B_z$, thanks to the horizontally self-similar
  structure.
Let $\BZO$ be the
dimensional magnitude of
$B_z$
at the \base\ of the \cl,
in units of Alfv\'en speed.
Then the corresponding
dimensionless magnitude of
$B_z$
is $\Lambda^{1/2}$,
where
\begin{equation}
  \Lambda
  \;\definedas\;
  \frac{\BZO^2}{2\OmegaI\eta}
  \label{eq:Lambda}
\;,
\end{equation}
an
Elsasser number based on $\BZO$.  
We assume that
$\Lambda^{1/2}$ characterizes
the order of magnitude of $B_z$ in
the bulk of the \cl,
and
that $\partial/\partial z \sim 1$
as before.\footnote{These
scaling assumptions are consistent with the example
solution shown in
\fig\ref{fig:sim-profiles},
for which we imposed $B_r/r = 1$ below the \cl,
implying that $B_z$ is of order unity within and just below
the \cl.
Taking
$z=0$
as the bottom of the \cl\
in \fig\ref{fig:sim-profiles},
we read off
$B_z=\Lambda^{1/2}\approx
1.9$.
Consistently, the
numerical output
gives
$\Lambda\approx 3.5$.
}
Then the
leading-order, \san\ solutions
satisfy
the dimensionless scaling
relations
\begin{eqnarray}
  u_z &\sim& 1
\label{eq:Lambda-scalings-a}\\
  u_r &\sim& r
\label{eq:Lambda-scalings-b}\\
  u_\phi &\sim& r
  \Lambda^{-1}
\label{eq:Lambda-scalings-c}\\
  B_z &\sim& \Lambda^{1/2}
\label{eq:Lambda-scalings-d}\\
  B_r &\sim& r
  \Lambda^{1/2}
\label{eq:Lambda-scalings-e}\\
  B_\phi &\sim& r
  \Lambda^{-1/2}
\label{eq:Lambda-scalings-f}
\end{eqnarray}
and these relations are expected to apply
at least
for $r\ll\rt$,
and probably also for
$r\lesssim\rt$ as a
guide
to orders of magnitude.
They
can
be derived alternatively
from scaling arguments applied directly
to (\ref{eq:torque})
along with the steady-state versions of
(\ref{eq:divu})--(\ref{eq:divB}).
We note that, since the horizontal components of $\B$ increase with
$r$, the \emph{total} magnetic field strength $|\B|$
can greatly
exceed $\BZO$ for $r
\gg 1$.

For increasing values of $\Lambda$
the field lines become stiffer, so that both they and the velocity
streamlines spiral less tightly, as noted in WM07.
The
magnitude of $\Lambda$
is
not well
constrained by observations,
depending as it does
on the magnitude of
the interior field at
the top of the
radiative envelope.
Fortunately, however,
it will be found
that the \chl\ regime described here can accommodate a
considerable range of $\Lambda$ values.

Now
in (\ref{eq:T-W}), with
the term in $\alphamu$ neglected,
the term in $\alphat$ cannot exceed
the largest of the other terms in order of magnitude.
From (\ref{eq:central-parabola})
we have
$\partial \pottemp'/\partial r
\sim \bbbt r$
so that
the term in $\alphat$ has magnitude
$\sim\alphat\bbbt r$,
at least
for $r\ll\rt$.
The remaining terms in (\ref{eq:T-W})
have magnitudes either
$\sim r\Lambda^{-1}$
(the terms in $u_\phi$ and $B_\phi^2$)
or
$\sim r\Lambda$
(the last term).
Those magnitudes
follow from
the horizontally self-similar structure of the
\san\ solutions,
along with
$\partial/\partial z \sim 1$
and the magnitudes
(\ref{eq:Lambda-scalings-a})--(\ref{eq:Lambda-scalings-f}).
We may therefore define the typical magnitude of
$\Sliver\bbbt\Sliver$ to be
$
\alphat^{-1}\max(\Lambda,\Lambda^{-1})
$. \
Correspondingly,
with $\aaat \,\sim\, \eta/\kappa$
we may define $\rt^2$, the typical magnitude of
$\aaat/\bbbt$,
to be
\begin{equation}
\rt^2
\,=\: 
\frac{\alphat\eta}
     {\kappa}
\min(\Lambda,\Lambda^{-1})
\,=\:
\frac{N_\pottemp^2\CLthick^2}{2\OmegaI\kappa}
\sliver\min(\Lambda,\Lambda^{-1})
\;,
\label{eq:cl-flatness}
\end{equation}
since
$\alphat = {N_\pottemp^2\CLthick^2}/{2\OmegaI\eta}$.
For realistic
$N_\pottemp\approx0.8\times10^{-3}$\sm,
for downwelling $U \sim
10^{-5}$\cms,
and for $\Lambda \sim 1$,
we have
$\CLthick\sim 0.4$\Mm\ and
$\rt \sim 4\times10^3$,
corresponding to
quite a large dimensional colatitudinal distance
$\rt\CLthick$
$\sim 1600$\Mm.
\footnote{Even
when
the tilting is significant,
such as to become incompatible with (\ref{eq:flat-T}),
the slopes
of the \tsss\ are
still geometrically small
--- far smaller than the geometrical aspect ratio
$\rt^{-1}$\antisliver.
Indeed, even on a global scale
we expect the
stratification surfaces
to
depart from the horizontal by only
``a very tiny fraction of a megametre''
from pole to equator \citep[][end of \S8.5]{McIntyre07},
based on
observational constraints on shear in the \tc.
Here
of course
``horizontal''
means
tangential to the heliopotentials,  \ie\
to the sum of
the centrifugal
and
gravitational
potentials.
}
The validity of
the foregoing scale analysis,
and that of the next section,
will be independently checked
by the numerical solutions
in \S\ref{sec:num-solutions}.

Of course our cylindrical model with
its assumption of uniform downwelling
will itself cease to apply, almost certainly,
well inside
such large
distances
from the pole.
One reason
for this limitation
is that the
cylindrical coordinates, regarded as
distorted spherical coordinates,
become increasingly inaccurate at large distances
from the pole.
However, that is not the most serious limitation.
Using true spherical coordinates
would not qualitatively
alter the
dynamics of the \cl.
It is the
assumption of uniform downwelling
that places the more serious limitation
on the range of applicability of our
model.
At some colatitude the downwelling
from
the real \tc\
must give way to upwelling,
as required by mass conservation.
The \chl\ regime cannot then apply even qualitatively.
Instead,
the
interior
magnetic field lines are
free to advect and
diffuse
upward
until they encounter the
\mfp\
associated with the
convective overshoot layer,
as assumed in \fig\ref{fig:sphere}.
This has wider
implications to be discussed in
\S\ref{sec:conclusions},
including implications for lithium burning.

The range of interior field strengths
accommodated by
the \chl\ regime
is determined by (\ref{eq:cl-flatness}).
For uniform downwelling
$U$,
the condition
for
the regime to apply
quantitatively
within, say,
10\degg\ colatitude or 90\Mm\
of the poles is
$\rt\CLthick \gg 90$\Mm.
We assume qualitative applicability for
$\rt\CLthick \gtrsim 90$\Mm.
We can use (\ref{eq:cl-flatness})
together with
realistic $N_\pottemp$
and diffusivity values
to write this
last
condition as
\begin{equation}
\max(\Lambda,\Lambda^{-1})
\lesssim
 3\times10^{2}\left(\frac{U}{10^{-5}\mbox{\cms}}\right)^{-4}
        \!\!.
\label{eq:Lambda-range}
\end{equation}
So
for $U \sim 10^{-5}$\cms\
we expect
the
regime to apply
qualitatively
over a range
of more than four
decimal
orders of magnitude in
$\Lambda$.
The corresponding range of field strengths,
being
proportional to $\Lambda^{1/2}$, covers more than
two decimal orders of magnitude.
To relate this to global-scale interior field strengths,
we estimate
$|B_r|$
from  (\ref{eq:Lambda-scalings-e})
at a nominal
30\degg\ colatitude,
with
$\CLthick\approx0.4$\Mm\ so that
$r \approx 650$.
Converting to dimensional units, we see that
the range of
field strengths is roughly
3\G\
$\lesssim
|B_r|
\lesssim
10^3$\G,
near the top of the radiative envelope
at 30\degg\ colatitude.
It is noteworthy that
these values
lie substantially above the threshold,
more like $10^{-2}$\G,
for the field strength required
to enforce the Ferraro constraint in the interior
over the Sun's lifetime
\citep[\eg][]{MestelWeiss87,CharbonneauMacGregor93}.
For smaller values of $U$,
a
still
wider range of interior
field strengths
becomes possible.

\section{The \hsl}
\label{sec:helium}

In this section we show that
the flow through the \hsl\ has the character
of flow through a porous medium.
We also show that, over a
large range of
$N_\mu^2$ and $\alphamu$ values,
the tilting of the compositional stratification surfaces in
the sublayer is even less significant than that of the thermal
stratification surfaces in the confinement layer.

The \hsl\ marks the transition between the compositionally
well-ven\-tilated \cl\ and the nearly \mbox{imperm}\-eable,
compositionally
stratified \HSL.
Therefore,
we expect the dimensional
sublayer
thickness scale,
$\SLthick$ say,
to be determined
by a balance between advection and diffusion of helium.
In (\ref{eq:mu})
the advection
operator $\bu\cdot\bm{\nabla}$
scales like the strain rate
$\sim U/\CLthick$
just above
and within
the sublayer,
because
$u_z\to0$
just beneath.
The strain rate $U/\CLthick$ must
therefore be comparable to
the helium diffusion rate $\chi/\SLthick^2$. \
Since
$U/\CLthick = \eta/\CLthick^2$,
\begin{align}
\SLthick &\sim (\chi/\eta)^{1/2}\CLthick
  \;\approx
\seventh\sliver\CLthick
\label{eq:sublayer-thick}
\end{align}
for realistic solar parameters.

Although the sublayer thickness scale $\SLthick$ is small in
magnitude relative to $\CLthick$,
it nevertheless
greatly exceeds the Ekman thickness scale
$\Ekthick \definedas
\left(
{\nu}/{2\OmegaI}
\right)^{1/2}$. \
Specifically,
\begin{flalign}
        \frac{\SLthick^2}{\Ekthick^2}
&\;\sim\;
     \frac{\chi\CLthick^2}{\eta}\frac{2\OmegaI}{\nu} 
\;\sim\;
     \frac{\chi\CLthick}{U}\frac{2\OmegaI}{\nu}
\nonumber \\
&\;\sim\;
     \frac{\chi}{\nu}\Ro^{-1}
\;\gg\; 1,
\label{eq:no-Ek-criterion}
\end{flalign}
because $\chi/\nu \approx 0.3$ while
$\Ro^{-1} \gg 1$, typically by many decimal orders of
magnitude; recall (\ref{eq:EkRo}).
The relations (\ref{eq:sublayer-thick}) and (\ref{eq:no-Ek-criterion})
suggest that
the dynamics of the \hsl\
should be well
described by
the asymptotic regime
\begin{equation}
\Ekthick \ll \SLthick \ll \CLthick
\;.
\label{eq:sublayer-regime}
\end{equation}
We
assume
(\ref{eq:sublayer-regime}) throughout this section.

Under (\ref{eq:sublayer-regime})
the magnetic diffusion rate $\eta/\SLthick^2$
in the sublayer
greatly exceeds the helium diffusion rate
$\chi/\SLthick^2$,
by a factor $\eta/\chi \sim (\CLthick/\SLthick)^2$.
The flow within the sublayer
can therefore
induce
only a small
perturbation
$\B - \BI = \B'$, say,
to
the interior field $\BI$.
In
\figs\ref{fig:section}
and~\ref{fig:similar},
the field lines are hardly deflected
as they cross the sublayer.
We may
therefore analyse the sublayer as a
perturbation to the state with
$\bu=0$ and $\B = \BI$,
where
$\BI$
has the
simple
dipolar structure
already assumed,
with components satisfying
$B_{\ii\phi} = 0$,
$B_{\ii r}/r=$ constant,
and $B_{\ii z}$ a linear function of $z$
consistent with
$\bm{\nabla}\Antisliver\cdot\antisliver\BI=0$.

Any such
$\BI$
has $\bm{\nabla}\antisliver\times\antisliver\BI=0$
and
Lorentz force
$(\bm{\nabla}\antisliver\times\antisliver\BI)\antisliver\times\antisliver\BI=0$.
Using this we
show in Appendix~\ref{sec:darcy_scalings} that,
in the asymptotic regime given by (\ref{eq:sublayer-regime}),
the steady-state
induction equation
becomes
simply
\vspace{-0.3cm}
\begin{equation}
0
\;=\;
B_{\ii z}\frac{\partial}{\partial z}\bu
+ \frac{\partial^2}{\partial z^2}\B'
\,,
\label{eq:SL-induc2}
\vspace{0.1cm}
\end{equation}
in the dimensionless variables of
\S\ref{sec:model-equations}.
The momentum balance (\ref{eq:Tmubalance})
becomes
\begin{flalign}
  \mathbf{e}_z\times\bu
\;=\;
-&\bm{\nabla} \tilde{p}
  - \alphamu\sliver
  \mu\sliver\mathbf{e}_z
+ B_{\ii z}\frac{\partial}{\partial z}\B'
\,. \label{eq:SL-momentum2}
\end{flalign}
Here the thermal-buoyancy term
$\alphat\sliver\pottemp\sliver\mathbf{e}_z$
has been absorbed into a
modified pressure gradient
$\bm{\nabla}\tilde p$, which also
incorporates
a gradient contribution to the Lorentz force;
see Appendix~\ref{sec:darcy_scalings}, below
(\ref{eq:SL-momentum1}).
Because $\SLthick\ll\CLthick$
and because, as verified shortly,
the sublayer will prove to be
sufficiently flat,
we may take $B_{\ii z}$
to be constant throughout the sublayer.
It is convenient to equate the dimensional value of this
constant to $\BZO$
in the definition
(\ref{eq:Lambda}) of the Elsasser number $\Lambda$.
Then the dimensionless magnitude of $B_{\ii z}$
in the sublayer
is precisely
$\Lambda^{1/2}$.
We can
now integrate (\ref{eq:SL-induc2}) to give
\begin{equation}
  0
~=~
  \Lambda^{\shalf}
  \Sliver\bu
\;+\;
  \frac{\partial\B'}{\partial z}
\phantom{xxx}
\label{eq:integd-sublayer-induc}
\end{equation}
since both $\B'$ and $\bu$ vanish beneath the sublayer.
Using (\ref{eq:integd-sublayer-induc})
we write
(\ref{eq:SL-momentum2}) as
\begin{align}
  \mathbf{e}_z\times\bu
\;=\; -\bm{\nabla} \tilde p
&  - \alphamu\mu\mathbf{e}_z
- \Lambda\bu
  \,.
\label{eq:Tmubalance-sublayer}
\end{align}
The term $-\Lambda\bu$ has the form
of a Darcy or Rayleigh drag,
showing that
the sublayer behaves like a porous medium
on the timescale set by
the strain flow.
The impermeability of the \HSL\
together with
the sublayer's flatness and thinness act to
keep the flow nearly horizontal,
compelling it to push past, and slightly deflect,
the
magnetic field lines spanning the sublayer
at angles
steep by comparison with sublayer aspect ratios.
So
the Lorentz force
from the deflected field lines
resists the sublayer flow in the manner of a Darcy
friction; similar behaviour occurs in Hartmann layers
\citep[\eg][]{Debnath73},
although in such cases
viscosity
also contributes to the
balance of forces.  Within
our
sublayer,
by contrast,
viscosity
is wholly
negligible provided that
\begin{equation}
\Lambda \gg \Ekthick^2/\SLthick^2\,,
\end{equation}
that is, provided that the Darcy friction from the field lines
dominates the fluid friction from
viscosity.
This condition
is easily satisfied, in virtue of (\ref{eq:no-Ek-criterion}).

In this Darcy regime,
(\ref{eq:torque}) and (\ref{eq:T-W}) simplify to
\begin{flalign}
 u_r
&~=~
 -\Lambda u_\phi
  \phantom{\raisebox{-0.15cm}{j}}
&
\label{eq:torque-sublayer}
\\
\mbox{and}\hspace{4.4cm}
\frac{\partial u_\phi}{\partial z}
&~=~
- \alphamu\frac{\partial\mu}{\partial r}
+ \Lambda\frac{\partial u_r}{\partial z}
\;.
\label{eq:T-W-sublayer}
\end{flalign}
Together with
(\ref{eq:divu}) and (\ref{eq:mu}),
\eqs(\ref{eq:torque-sublayer}) and (\ref{eq:T-W-sublayer})
describe the sub\-lay\-er dynamics
to an order of accuracy that includes the
first corrections to
perfect flatness.

We now use scaling arguments, paralleling those
in \S\ref{sec:CL-scalings},
to verify
that the corrections can indeed be taken as small
and
the sublayer treated as flat.
As before,
we expect that
each term in (\ref{eq:T-W-sublayer}) is proportional to $r$ and that
\begin{equation}
\mu
\,=\;
\aaamu
\,+\,
\bbbmu r^2
\label{eq:central-parabola-mu}
\end{equation}
to leading order
close to the pole,
where $r$ is again
dimensionless and
where $\aaamu$ and $\bbbmu$ are
dimensionless functions of $z$ alone.
The matching to the \HSL\ beneath implies that
$\dd\aaamu/\dd z \sim 1$ and that
$\aaamu \sim \SLthick/\CLthick$
in the sublayer,
if we
take the constant value of
$\mu$
above the sublayer
to be zero.
The condition for validity of flat,
horizontally self-similar
sublayer solutions
is
$r^2 \ll \rmu^2$,
say, where $\rmu$ denotes a
typical magnitude of \,$|\aaamu/\bbbmu|^{1/2}$.

Within the sublayer, the pattern of dimensionless scalings
(\ref{eq:Lambda-scalings-a})--(\ref{eq:Lambda-scalings-f})
is replaced by
\begin{eqnarray}
  u_z &\sim& \SLthick/\CLthick
\label{eq:Lambda-scalings-mu-a}\\
  u_r &\sim& r
\label{eq:Lambda-scalings-mu-b}\\
  u_\phi &\sim& r\sliver\Lambda^{-1}
\label{eq:Lambda-scalings-mu-c}\\
  B_z' &\sim& \Lambda^{1/2}\Sliver(\SLthick/\CLthick)^2
\label{eq:Lambda-scalings-mu-d}\\
  B_r' &\sim&
r\sliver\Lambda^{1/2}\Sliver\SLthick/\CLthick
\label{eq:Lambda-scalings-mu-e}\\
  B_\phi' &\sim&
r\sliver\Lambda^{-1/2}\sliver\SLthick/\CLthick
\label{eq:Lambda-scalings-mu-f}
\end{eqnarray}
for all
$r \ll \rmu,\,\rt$.
These dimensionless order-of-magnitude relations follow
from the matching to the \cl,
again noting its horizontally self-similar structure,
and
from $\bm{\nabla}\cdot\B'=\bm{\nabla}\cdot\bu=0$ together with
(\ref{eq:integd-sublayer-induc})
and (\ref{eq:torque-sublayer}),
with
$\partial/\partial z\sim\CLthick/\SLthick\gg1$.
Again, further detail is given in Appendix~\ref{sec:darcy_scalings}.

The horizontal velocity components
in (\ref{eq:Lambda-scalings-mu-b})--(\ref{eq:Lambda-scalings-mu-c})
inherit their magnitudes directly
from those in the overlying \cl.
For this reason,
the
vertical shear
in the sublayer is larger than that in the \cl\ by a
factor $\CLthick/\SLthick$.
This shows
how, in the limit $\SLthick\to0$,
preserving (\ref{eq:sublayer-regime}),
the sublayer regime goes over into the
slip regime
analysed in WM07.
The slip regime has infinite shear
at the top of the \HSL,
with finite
discontinuities in both horizontal velocity components.

We can now estimate $\rmu$
and hence the flatness of the sublayer,
from (\ref{eq:T-W-sublayer}),
on the same basis as before, namely consistency with
(\ref{eq:mu}) and
uniform downwelling.
From (\ref{eq:central-parabola-mu}) we have
$\partial\mu/\partial r\sim\bbbmu r$
for all $r\ll\rmu\sliver$,
so that in
(\ref{eq:T-W-sublayer})
the $\alphamu$ term
has magnitude
$\sim\alphamu\bbbmu r$. \
Using (\ref{eq:Lambda-scalings-mu-a})--(\ref{eq:Lambda-scalings-mu-f})
we find that
the other terms
in (\ref{eq:T-W-sublayer})
have magnitudes
$\sim r\sliver(\CLthick/\SLthick)\Lambda^{-1}$
(the term in $u_\phi$)
and
$\sim r\sliver(\CLthick/\SLthick)\Lambda$
(the term in $u_r$).
The typical magnitude of
$\alphamu\bbbmu$ can therefore be taken to be
$
(\CLthick/\SLthick)\max(\Lambda,\Lambda^{-1})
$. \
Correspondingly,
since $\aaamu \,\sim\, \SLthick/\CLthick\sliver$,
we may define $\rmu^2$,
the typical magnitude of $\aaamu/\bbbmu$, to be
$\alphamu(\SLthick/\CLthick)^2\min(\Lambda,\Lambda^{-1})
=\alphamu(\chi/\eta)\min(\Lambda,\Lambda^{-1})$.
Comparing this with (\ref{eq:cl-flatness})
and recalling that
$\alphamu = {N_\mu^2\CLthick^2}/{2\OmegaI\eta}$
we see that
\vspace{-0.2cm}
\begin{equation}
\rmu^2
\;=~
\frac{\alphamu}{\alphat}\sliver
\frac{\kappa\chi}{\eta^2}\Sliver
\rt^2
\;=~
\frac{N_\mu^2}{N_\pottemp^2}\sliver
\frac{\kappa\chi}{\eta^2}\Sliver
\rt^2
\;.
\label{eq:cl-flatness-mu-new}
\end{equation}
Since
$\kappa\chi/\eta^2\approx 0.75\times10^3$,
we see that
$\rmu^2\:\gtrsim\:\rt^2$
for a large range of $N_\mu^2$ values, from today's
value $\sim N_\pottemp^2$ down to
almost three decimal orders of magnitude less.
This means that, in most cases,
our flatness assumptions hold
even more strongly for the sublayer than for the \cl,
in the sense of compatibility with uniform downwelling
$u_z=u_z(z)$ in (\ref{eq:mu}).

It is worth going beyond scale analysis
to say more about the vertical structure
of the sublayer,
especially in its
lower extremity
or ``\st'',
wherein we expect
$|\bu|$
to decay
exponentially with depth.
Within this \st,
the \HSL\
suffers only small
perturbations
to its otherwise uniform
compositional
stratification
$\partial\mu/\partial z = -1$. \
We denote the perturbation to $\mu$ by $\mu'$.
In the steady state,
\eq(\ref{eq:mu})
may then be approximated as
\vspace{-0.3cm}
\begin{equation}
-u_z
\;=\;
\frac{\chi}{\eta}\Sliver
\frac{\partial^2\mu'}{\partial z^2}
\,,
\end{equation}
which can be combined with (\ref{eq:divu}),
(\ref{eq:torque-sublayer}), and (\ref{eq:T-W-sublayer})
with $\mu$ replaced by $\mu'$,
to yield a single equation for
$\mu'$,
\begin{equation}
       \alphamu
       (\eta/\chi)
       \nabla_H^2\mu'
        \;=\;
        \left(
          \Lambda + \Lambda^{-1}
        \right)\frac{\partial^4\mu'}{\partial z^4}
\;,
\label{eq:porous}
\end{equation}
where
$\nabla_H^2
\definedas
r^{-1}\partial(r\sliver\partial/\partial r)/\partial r
=
\nabla^2\antisliver - \sliver
{\partial^2}/{\partial z^2}$. \
It is now clear
that the leading-order scale analysis
given above applies only to the main part of the sublayer
and not to the \st.
\Eq(\ref{eq:porous})
tells us that the
vertical scale,
$\subtailthick$ say,
for the
\st\
must depend on the horizontal scale
in a manner reminiscent of the heuristic boundary-layer
analysis given in GM98.

For instance
if we assume that the actual horizontal scale is
the scale $\rt$ set by
the \cl, so that
$\alphamu \nabla_H^2
\sim
\alphamu \rt^{-2}$,
then a straightforward scale analysis of
(\ref{eq:porous})
shows that,
in terms of the definitions
(\ref{eq:cl-flatness}) and
(\ref{eq:cl-flatness-mu-new}),
\begin{flalign}
{\subtailthick}
&\;\sim\;
\left(
   \frac
   {\rt}
   {\rmu}
\right)^{\!\shalf}\!\!
\SLthick
\;,
\label{eq:subtailthick}
\end{flalign}
which is
generally
smaller than $\SLthick$.
Even at this scale, however, viscosity remains negligible
provided
that the Darcy friction from the field lines
dominates the fluid friction from microscopic viscosity.
That is,
viscosity remains negligible
provided that
\begin{flalign}
&&
  \Lambda
\;\gg\;
\Ek\Sliver\nabla^2
&\,\sim~
\Ek\Sliver(\CLthick/\subtailthick)^2
,&
\label{eq:subtail-visc-a}\\
&\mbox{equivalently}
&\subtailthick^2
&\;\gg~
\Ekthick^2/\Lambda
\,.&
\label{eq:subtail-visc-b}
\end{flalign}
For realistic solar parameters, (\ref{eq:subtail-visc-a})
is easily satisfied
because $\Ek\sim0.4\times10^{-8}$.
A more detailed
analysis
\citep{Wood10}
verifies
all these properties of the \st.
The exponentially weak
flow within
the \st\
is invisible
in \fig\ref{fig:section},
and barely visible
in~\fig\ref{fig:sim-profiles}.

The actual horizontal scale may or may not be set by the
value of $\rt$ since,
along with boundary conditions for (\ref{eq:porous}),
it will depend
on the size of the downwelling region and
indeed on how the entire
global picture sketched in \fig\ref{fig:sphere}
fits together, a point to which we return in
\S\ref{sec:conclusions}.

\section{Boundary conditions for the numerical solutions}
\label{sec:boundary-conditions}

To go beyond the self-similar,
perfectly flat solutions described
in \S\ref{sec:self-similar}
we must
compute solutions numerically.  To this end
a numerical code
has been written
to solve
the axisymmetric version of
(\ref{eq:momentum})--(\ref{eq:mu})
in
a cylindrical domain
of radius $r=\rperiph$,
                   say.
The scheme on which the code is based
is summarized
in
Appendix~\ref{sec:numerics_details}.
For reasons explained there and in
Appendix~\ref{sec:magnetostrophic},
it proves necessary to use the full time-dependent
equations
(\ref{eq:momentum})--(\ref{eq:mu}),
rather than assuming \mb.

To solve the equations numerically in the
cylindrical domain,
we need to
specify
boundary conditions.
This inevitably involves artificial choices.
The only way to avoid making such choices would be
to fit the polar caps into the complete,
and highly complicated,
global picture.
That remains a challenge for the future,
requiring
the quantification
of turbulent processes in
the \tc\ and \cz.
Such quantification would have to include
realistic descriptions of
turbulent
\mfp\ and of
turbulent magnetic
diffusion in the bulk of the \tc.
Also crucial is the turbulent
gyroscopic pumping of
polar downwelling and of
the complementary upwelling
in lower latitudes,
together with the effects on
the global-scale pattern of heat flow
and the resulting feedback on
$N_\pottemp$ distributions.

As already noted,
in the present work
we are imposing
a dipolar magnetic field structure underneath
the \cl,
and a uniform
downwelling of magnitude $U$ from a field-free region aloft.
Field-free refers
to time-averaged
fields.
In the example shown in \fig\ref{fig:section},
the numerical domain was defined by
$0\leqslant r \leqslant 5$,
\ie\ $\rperiph=5$,
and
$-1\leqslant z \leqslant 6$,
one dimensionless unit
taller than shown
in the figure.
We imposed $u_z = -1$ at $z = 6$
and $B_r/r = 1$ at $z = -1$.

As shown in \S\ref{sec:self-similar}, the
bulk of the
\cl\ is relatively insensitive to conditions
within the
field-free region
aloft,
and in particular to the vertical shear
$\partial u_\phi/\partial z$ aloft.
There,
the vertical shear is tied to the temperature
distribution via \eq(\ref{eq:justT-W}),
and hence to the global-scale heat flow.
To avoid having to solve the
complete global-scale problem
we simply
imposed $\pottemp = \mbox{const.}$
and $\partial u_\phi/\partial z = 0$ at $z = 6$,
which is consistent with
the imposed
uniform downwelling and also
ensures that no viscous torque is exerted on the top
of the domain.

At the periphery
of the domain,
the
artificial
cylindrical surface
$r=\rperiph=5$,
the numerical algorithm requires us
to
impose
three vertical profiles,
including the vertical profile of
Maxwell stress.
The stress profile
represents the
field lines' connection to
lower latitudes
and the Alfv\'enic torque exerted therefrom.
We also need to
impose
thermal and compositional stratification
profiles
$\pottemp(z)$ and $\mu(z)$
at the periphery,
in a manner consistent with scalings in the \cl\ and \hsl\
(\S\S\ref{sec:CL-scalings},~\ref{sec:helium}).
In this way we
artificially
fix the altitude of the \hsl.
We
thereby
influence
the velocity field as well, since it
is tightly
linked to the two stratifications
by \eqs(\ref{eq:T}) and (\ref{eq:mu}).
The remaining peripheral boundary conditions are used to
promote
smoothness of the fields, and hence to minimise
spurious
boundary effects in the steady state.
In particular, we impose
\vspace{-0.4cm}
\begin{flalign}
  &&\frac{\partial}{\partial r}\Antisliver
  \left(
    \frac{u_r}{r}
  \right)
  &= 0\,,& \\
  &&\frac{\partial}{\partial r}\Antisliver
  \left(
    \frac{u_\phi}{r}
  \right)
  &= 0\,,& \\
  &\mbox{and}&\frac{\partial B_z}{\partial r}\:
  &= 0&
\end{flalign}
at $r=  \rperiph = 5$.

At the bottom of the domain,
the horizontal surface
$z=-1$,
we impose
the conditions (\ref{eq:NTdef}) and (\ref{eq:Nmudef}),
equivalently
$\partial \pottemp/\partial z = 1$
and $\partial\mu/\partial z = -1$.
We also impose $u_\phi=0$, \ie\
that the interior
is in \solid\ rotation,
with dimensional angular velocity $\OmegaI$,
as
required by the global picture.
As discussed at the end of \S\ref{sec:model-equations}
the global picture also requires that
$B_\phi = 0$, \ie\ that there is no Alfv\'enic torque,
on the bottom of the domain.
This last condition cannot be directly imposed, however.
Instead, it must be approached via
iterative adjustments
to
the
Maxwell stress profile
at the periphery
$r=\rperiph$.
The iteration procedure
and its rationale
are described
in \S\ref{sec:num-solutions}.

We can now see more clearly why $u_z(z)$ could be specified
arbitrarily
when constructing the \san\ solutions in
\S\ref{sec:self-similar}.
As
already mentioned,
for the numerical solution
we need to
specify three vertical profiles at the periphery
$r=\rperiph$,
which are taken to be
$\pottemp(z)$, $\mu(z)$, and $B_\phi(z)$. \
For the \san\ solutions
of \S\ref{sec:self-similar},\;
$\pottemp(z)$ and $\mu(z)$ could not
be specified independently. \
Rather, they were both determined
by (\ref{eq:compatibility}),
up to boundary conditions,
as soon as
$u_z(z)$ was specified.
We could still have specified $B_\phi(z)$,
but gave up that freedom in order to ensure the vanishing
of the Alfv\'enic torque as $z\to-\infty$,
thus determining
$B_\phi(z)$
via
the expression (\ref{eq:hatBphi}).
Also allowed by the \san\ framework
was the freedom to specify $u_\phi(z)$
at $r=\rperiph$,
which we similarly gave up in order to
ensure \solid\ rotation
as $z\to-\infty$,
thus determining $u_\phi(z)$
via (\ref{eq:hatuphi}).

More generally,
within the \san\ framework, the
peripheral and bottom
profiles
of $B_\phi$
contain
equivalent
information,
and similarly for $u_\phi$. \
This is because of the
Alfv\'enic coupling along the field lines
linking the
periphery to the bottom
of the domain,
expressed
by the $\B\cdot\bm{\nabla}$ operator in
\eqs(\ref{eq:torque_duplicate}) and~(\ref{eq:indp-1d}).
There is no such precise equivalence within the numerical
framework.
The time-dependence,
in the equations solved numerically,
replaces
static Alfv\'enic coupling by
       Alfv\'enic
wave propagation, requiring
one peripheral and one bottom
profile to be specified, which we take to be
$B_\phi(z)$ and $u_\phi(r)$
respectively.
This is analogous to the need for boundary conditions
at both ends
of a stretched string in motion.

\section{The numerical solutions}
\label{sec:num-solutions}

Computing
limitations preclude
a perfect match to
the real Sun's parameter values.
They also require a slight modification to
(\ref{eq:momentum})--(\ref{eq:mu}),
explained in Appendix~\ref{sec:numerics_details},
in which artificial horizontal diffusivities $\nu_H$, $\chi_H$
are introduced.
These maintain numerical stability
while allowing
small enough $\nu$ and $\chi$ in the important
vertical diffusion terms.

From the
scale analyses
in \S\S\ref{sec:CL-scalings},~\ref{sec:helium}
we may identify
the conditions most
essential
to reaching a qualitatively
similar
parameter regime
---
that is, qualitatively similar
to a regime with a perfect parameter match
to the real Sun.
Those essential conditions are:
\begin{enumerate}
  \item The Rossby number $\Ro$ should
be small
in comparison with unity,
so that the steady state is
close to
magnetostrophic.

  \item The thermal diffusivity $\kappa$
should
be large in comparison with
the magnetic diffusivity $\eta$, so that the \chl\ flow
only weakly perturbs the background thermal stratification.

  \item\label{itemlabel:c}
The \cl\
and \hsl\
should
both
be reasonably flat,
to the extent that
$\rt$ and $\rmu$ are both distinctly greater than 1.
With (\ref{eq:cl-flatness-mu-new}) in mind,
we also take $\rmu > \rt$.

  \item The helium diffusivity $\chi$
should be
small in comparison with the magnetic diffusivity $\eta$,
so that the helium sublayer is thinner than, and
therefore distinct from, the magnetic confinement layer.

  \item The viscosity $\nu$ should be small enough that
an Ekman layer does not form at the
top of the \HSL,
so that
the flow is everywhere inviscid, even in the \hsl.
With small $\Ro$ this condition is easily satisfied
in the numerical model
as well as in the real Sun,
because of the factor $\Ro^{-1}$ in
(\ref{eq:no-Ek-criterion}).

\end{enumerate}

\begin{figure}
  \centering
  \includegraphics[width=0.479\textwidth]{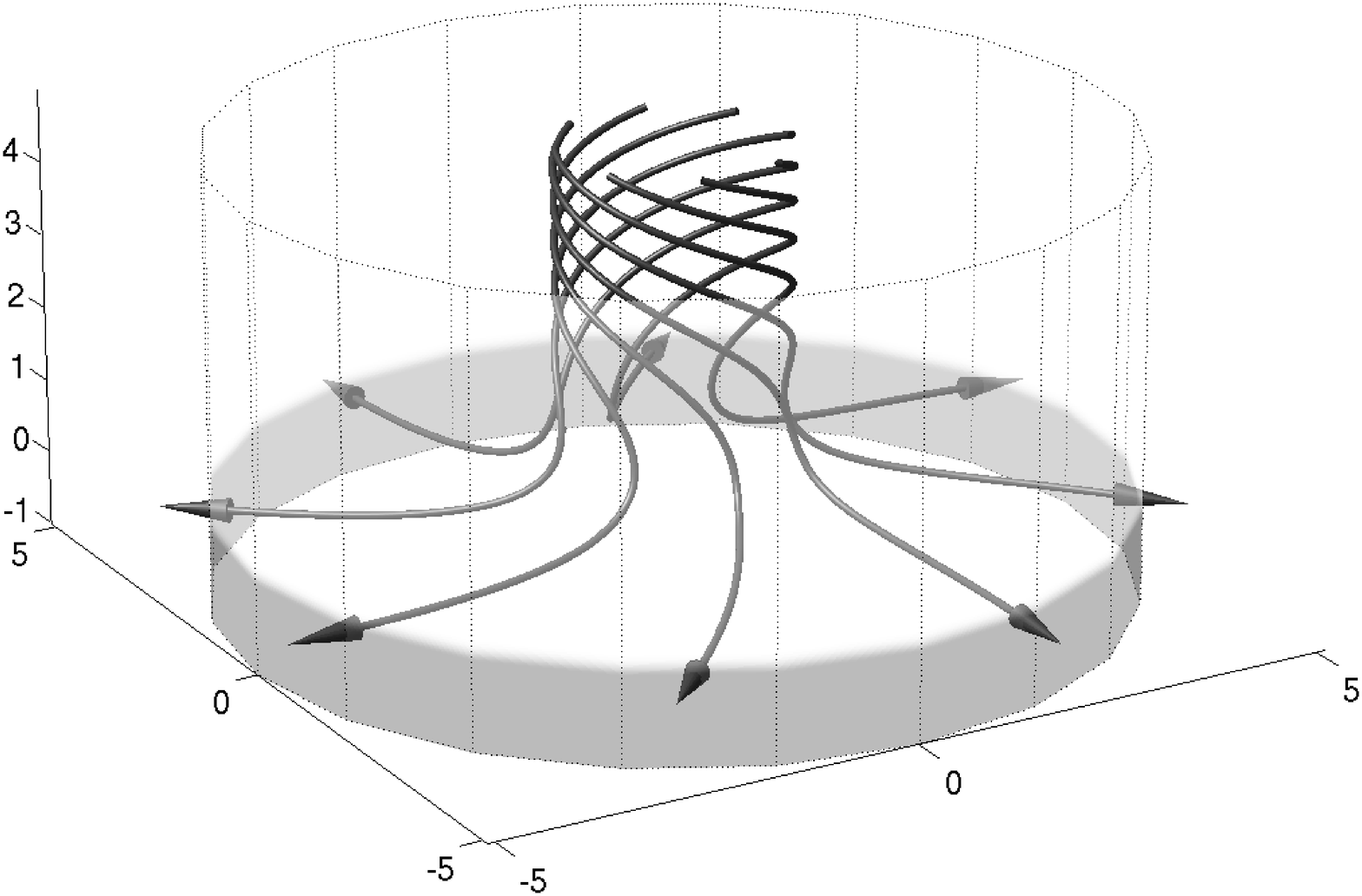}
  \includegraphics[width=0.479\textwidth]{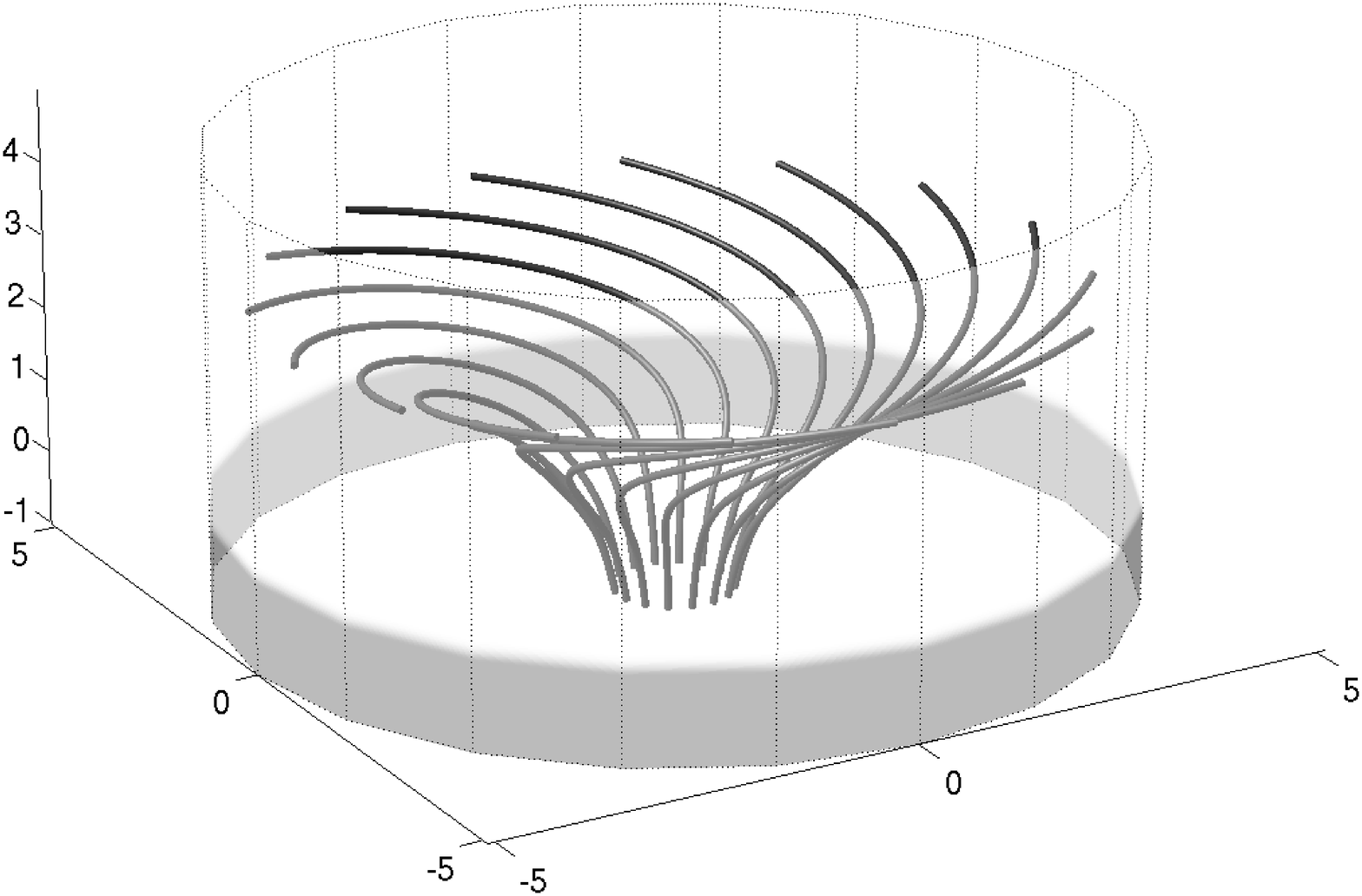}
  \caption{\footnotesize A numerical solution of the
  \chl\ equations with
  $\rt^2\approx14$
  and Elsasser number $\Lambda \approx 3.5$.
  Other parameter values are given in the second-last column of
   Table~\ref{tab:num-params}.  From the same solution as
  shown in
  \fig\ref{fig:section}.
Both here and in \fig\ref{fig:section}, a top layer
$5\leqslant z \leqslant 6$ has been omitted from the plots.
\Figs\ref{fig:section}, \ref{fig:full}, and
\ref{fig:num-profiles}
are all views of the same numerical solution.
  }
\label{fig:full}
\end{figure}

Leaving $\nu_H$ and $\chi_H$ aside for the moment
(see Appendix~\ref{sec:numerics_details})
we can
characterize the system
by
seven
dimensionless parameters,
including the Elsasser number $\Lambda$, which enters through the
boundary conditions.
Table~\ref{tab:num-params}
presents the
other
six
dimensionless parameters,
with
nominal solar values
alongside the
values used
for
the
numerical
solution
presented here
(\figs\ref{fig:section}, \ref{fig:full}, and \ref{fig:num-profiles}).
The last column
echoes aspects of the
qualitative
parameter
conditions
just stated.
The nominal
solar values assume $U = 10^{-5}$\cms.

\begin{table}
\begin{minipage}{\textwidth}
\begin{center}
\begin{tabular}{@{}c@{\hspace{1cm}}c@{\hspace{1cm}}c@{\hspace{1cm}}c@{}}
Dimensionless & Nominal & Value for & Condition \\
    parameter   & solar value &  numerical solution   &            \\
  $\Ro$         & $~5\times10^{-8}$ & $10^{-2}$        & $\ll 1$                 \\
  $\kappa/\eta$ & $3\times10^4$     & $10^2$           & $\gg 1$                 \\
  $\alphat(\eta/\kappa)$ & $1.4\times10^7\;$ & $50$ &
                $> \max(\Lambda,\Lambda^{-1})$\footnote{
  This condition
  corresponds to
  $\rt^2>1$, \\ with
  $\rt^2 = \alphat(\eta/\kappa)\min(\Lambda,\Lambda^{-1})$
    as
    defined
    in
    (\ref{eq:cl-flatness}).} \\
  $(\rmu/\rt)^2$ & $3\times10^2$
& $2$ & $> 1$ \\
  $\chi/\eta$   & $\:2\times10^{-2}$  & $~~2\times10^{-2}$ & $\ll 1  $                 \\
  $\nu/\chi$    & $3~$               & $1$              & $\ll 1/\Ro$
\end{tabular}
\end{center}
\end{minipage}
\caption{Parameter values and conditions; see text.}
\label{tab:num-params}
\end{table}

In order to allow the stratification surfaces to develop
a significant tilt,
the horizontal size of the numerical domain, $\rperiph$,
was chosen to be of the same order as $\rt$;
specifically, we chose $\rperiph=5$.
However, the precise value of $\rt$, as determined by
(\ref{eq:cl-flatness}),
depends on the precise value of $\Lambda$,
which
is set indirectly
via the boundary condition for $B_r$. \
For
the case shown in
\figs\ref{fig:section}, \ref{fig:full}, and \ref{fig:num-profiles}
this boundary condition was $B_r/r = 1$
at $z = -1$,
and
we find
from the solution
that
$\Lambda\definedas\left.B_{\ii z}^2\right|_{z=0}\approx 3.5$.
With the parameter values in the second-last column of
Table~\ref{tab:num-params},
this in turn means that $\rt^2 \approx 50/3.5 \approx 14$,
so in fact $\rt$ is slightly smaller than $\rperiph$.
Nevertheless
the numerical solution
is
remarkably
close to being flat,
even close to the periphery of the domain $r = \rperiph$.
Some effects of
departures from flatness can be seen near the periphery in,
for instance, \fig\ref{fig:strophic}
below, but they do not qualitatively alter the nature of the flow.
The \chl\ dynamics can therefore apply, at least qualitatively,
even at colatitudes for which the \san\ solutions of \S\ref{sec:self-similar}
are not strictly valid.
We should therefore regard (\ref{eq:Lambda-range})
as a conservative estimate of the range of interior field strengths
for which the \chl\ regime applies.

The vertical profiles of $\pottemp$, $\mu$ and $B_\phi$ at the periphery of
the computational domain were initially taken from a
\san\
solution.
The resulting
steady-state meridional flow,
and the distribution of
Coriolis torque
it exerts
on each field line,
cannot be
precisely
known in advance.
So the steady state found,
with this choice of the peripheral $B_\phi$ profile,
will generally
be such that
some
of the
balancing
Alfv\'enic torque is exerted
on the bottom of the computational
domain, rather than on the periphery $r = \rperiph$.

As mentioned in the previous section,
we can
eliminate this
bottom torque through
iterative
adjustments to the torque
at the periphery.
Experience with
the numerical solutions reveals that
such
adjustments
are propagated
along the magnetic field lines without,
in most cases of interest,
greatly
perturbing the poloidal velocity and magnetic field components.
So we can reduce the bottom torque
simply by mapping
or transferring
that torque
from the bottom
to the periphery of the domain,
along the field
lines.
The bottom torque
is thereby reduced, though not
all the way to zero since the other fields adjust,
slightly reshaping the field lines.
The process can then be repeated, further reducing the
bottom torque.
After a
sufficient number of iterations,
the bottom torque
can be
reduced to
negligible values,
with practically all the torque transferred to the periphery.

\Fig\ref{fig:full} shows plots of the steady-state streamlines
and magnetic field lines from
the numerical
solution whose
parameter values are listed in Table~\ref{tab:num-params},
and whose meridional cross-section was presented in
\fig\ref{fig:section}.
The bottom torque is negligibly small.
\Fig\ref{fig:num-profiles} shows the vertical profiles of
$u_z$, $u_\phi/r$, $B_z$ and $B_\phi/r$
on the rotation axis,
from the same numerical solution.

\section{Upper-domain ``slipperiness''}
\label{sec:comparison}

\begin{figure}
  \centering
\psfrag{uz}[cc][cr]{$u_z$}
\psfrag{ur}[cc][cc]{$\dfrac{u_r}{r}$}
\psfrag{up}[cc][cc]{$\dfrac{u_\phi}{r}$}
\psfrag{Bz}[cc][cr]{$B_z$}
\psfrag{Br}[cc][cr]{$\dfrac{B_r}{r}$}
\psfrag{Bp}[cc][cr]{$\dfrac{B_\phi}{r}$}
\psfrag{z}[cc][cc]{$z$}
\psfrag{Velocity Field}[cc][cc]{Velocity Field}
\psfrag{Magnetic Field}[cc][cc]{Magnetic Field}
  \includegraphics[width=0.25\columnwidth]{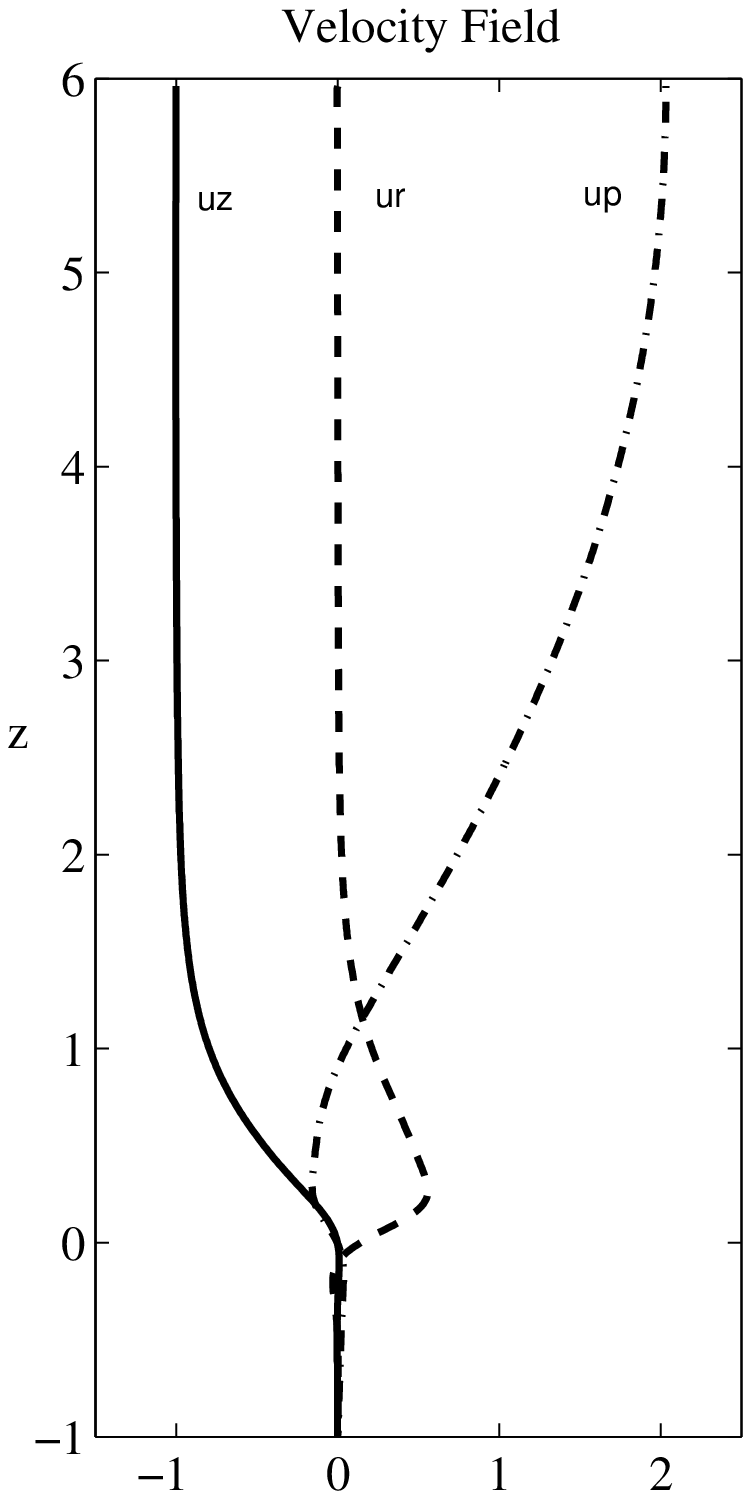}
  \includegraphics[width=0.25\columnwidth]{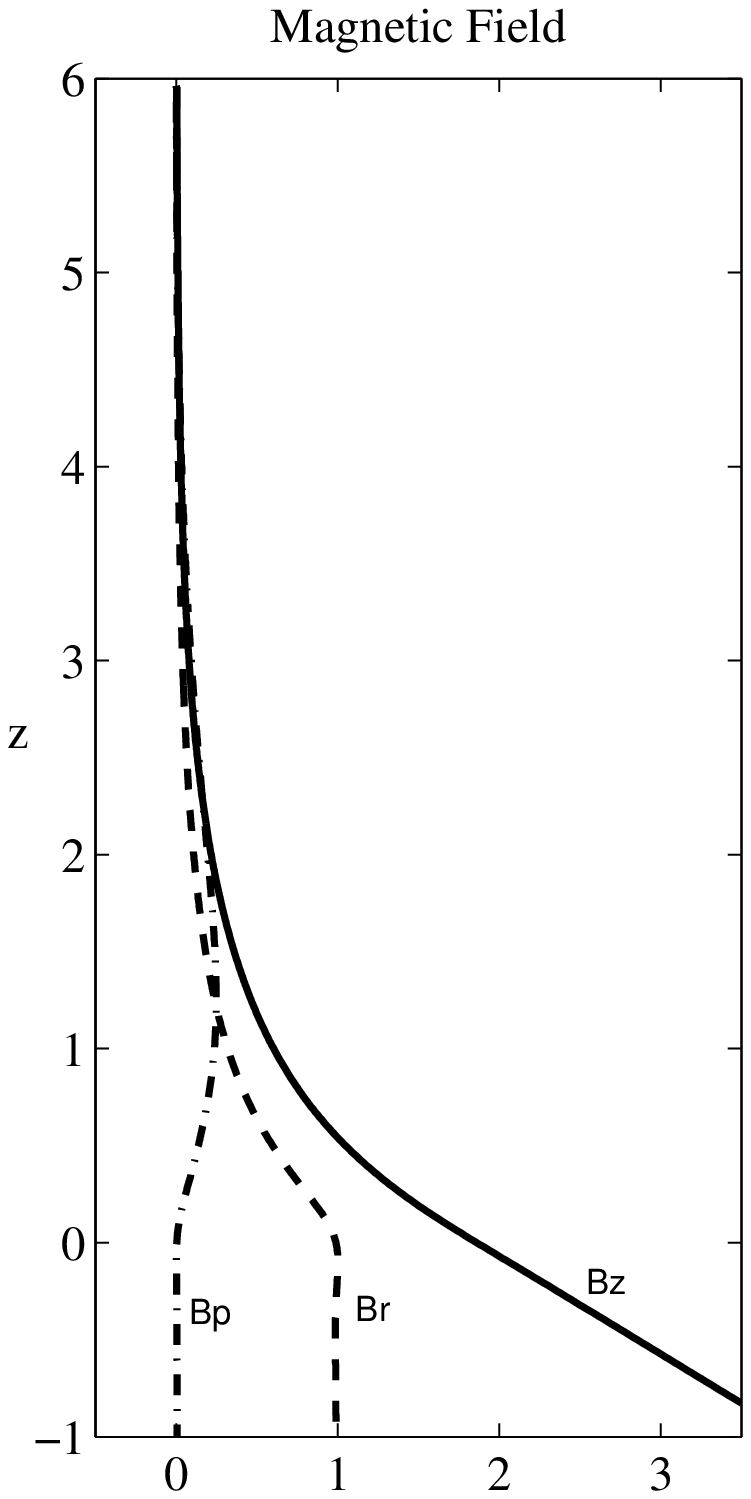}
  \caption{\footnotesize
  Vertical profiles from a
  numerical solution of the
  \chl\ equations with
  $\rt^2\approx14$
  and Elsasser number $\Lambda \approx 3.5$.
  Other parameter values are given in the second-last column of
   Table~\ref{tab:num-params}.  From the same solution as
  shown in
  \fig\ref{fig:section}.
  }
  \label{fig:num-profiles}
\end{figure}

On the rotation axis, where
the profiles in \fig\ref{fig:num-profiles} were taken,
the stratification surfaces
are
flat for
any finite
$\alphat$ and $\alphamu$.
If the
numerical
solution were in perfect \mb\ then
we could
use its
$u_z(z)$ profile to
calculate
the other field components
on the axis
by the procedure
for constructing \san\ solutions
described in \S\ref{sec:self-similar}.
But the numerical solutions are not in perfect balance,
especially toward the upper part of the domain,
where the Lorentz and Coriolis forces become small and the
artificial
viscous forces relatively more significant,
along with
numerical truncation errors
and
other small effects.
In particular, the numerical
$u_z(z)$ and
$u_r(z)$ profiles will
not conform
precisely
to the decay law
(\ref{eq:decay}) as $z$ increases.
So the \san\ solution obtained by this process
cannot
perfectly match the numerical solution, even
on
the rotation axis.
Indeed, such a
\san\ solution
will often exhibit wild deviations
from the numerical solution toward the upper part
of the domain.
There,
the delicate balance of terms gives
the dynamics a certain ``slipperiness'',
as
already
evidenced by
the upper-domain sensitivity
of $u_\phi$ and $\partial u_\phi/\partial z$
to
values of the decay constant
$\twoandabit$
in
(\ref{eq:decay}).

To enable
a meaningful comparison between the numerical and
\san\ solutions, we are
therefore
compelled to make
small
adjustments to
$u_z(z)$
in the upper
domain,
to make
$u_z(z)$ and
$u_r(z)$
conform to (\ref{eq:decay}),
before using
them to compute a self-similar
solution.
In the case shown here
the decay constant $\gamma$ was chosen,
purely for illustration,
to be $2.24$.
The required adjustment to $u_z(z)$ is
then
very small indeed;
the solid $u_z$ curves on the left
of \figs\ref{fig:sim-profiles} and \ref{fig:num-profiles}
are practically indistinguishable
from each other.

As already mentioned, conditions aloft in the real Sun
may well involve small-scale MHD turbulence.  The resulting
departures from \mb\ may well produce asymptotic behaviour aloft
that disagrees with all the solutions obtained here, whether
\san\ or numerical.
A realistic matching to conditions aloft remains a challenge
for future modelling work.

\section{Confinement layers with no \HSL}
\label{sec:noHSL}
Although the presence of the \HSL\ influences the structure
of the \cl\ in today's Sun,
it is not actually essential to magnetic confinement.
The qualitative picture sketched in
\fig\ref{fig:sphere}
might therefore apply
also to
the early Sun, before the \HSL\ developed.

The analysis presented in \S\ref{sec:self-similar}
holds good for any suitable profile of $u_z(z)$,
provided only that the
thermal stratification is sufficiently
strong.
Compositional stratification enters only indirectly,
via the shape of $u_z(z)$
in its lower part representing the
\hsl\ and subtail.
As pointed out below (\ref{eq:subtailthick})--(\ref{eq:subtail-visc-b}),
the shape depends in turn on how the \cl\ and sublayer
fit into the global picture sketched in \fig\ref{fig:sphere}.

For the early Sun, with no compositional stratification,
we have a similar indeterminacy in the shape of $u_z(z)$.
However, we may note that with no \HSL\
the scaling analysis
in
\S\ref{sec:CL-scalings} then applies not only within the \cl,
but also in the region immediately beneath.
An argument
similar
to that
leading to the estimate
$\subtailthick$
of the subtail scale in
(\ref{eq:subtailthick})
predicts that
$|\bu|$ decays
with depth on a lengthscale
$\sim\CLthick$, rather than $\subtailthick$.

\begin{figure}
  \centering
  \psfrag{uz}[cc][cr]{$u_z$}
  \psfrag{ur}[cc][cc]{$\dfrac{u_r}{r}$}
  \psfrag{up}[cr][cl]{$\dfrac{u_\phi}{10r}$}
  \psfrag{Bz}[cc][cr]{$B_z$}
  \psfrag{Br}[cc][cr]{$\dfrac{B_r}{r}$}
  \psfrag{Bp}[cc][cr]{$\dfrac{B_\phi}{r}$}
  \psfrag{z}[cc][cc]{$z$}
  \psfrag{Velocity Field}[cc][cc]{Velocity Field}
  \psfrag{Magnetic Field}[cc][cc]{Magnetic Field}
  \includegraphics[width=0.25\textwidth]{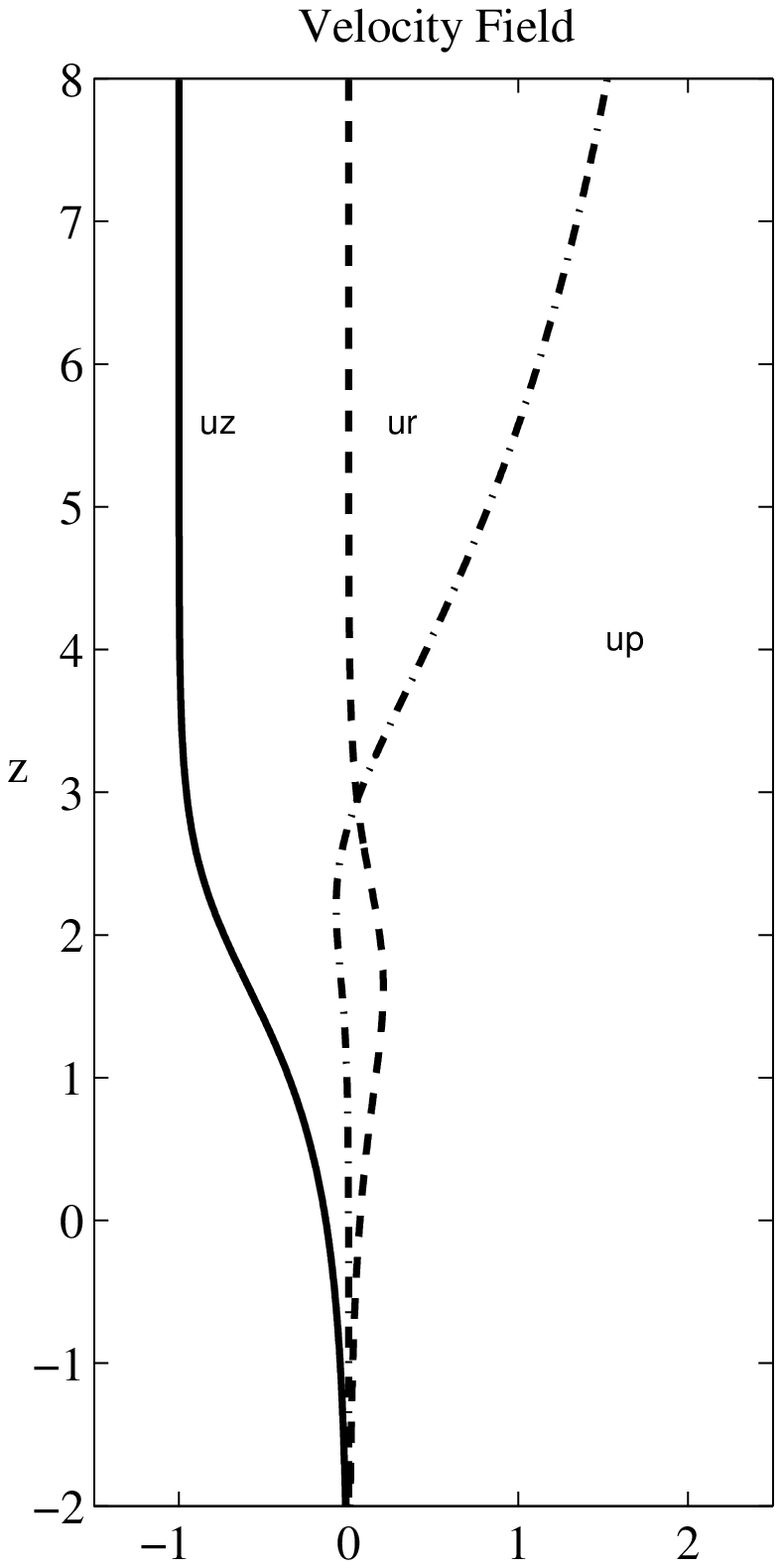}
  \includegraphics[width=0.25\textwidth]{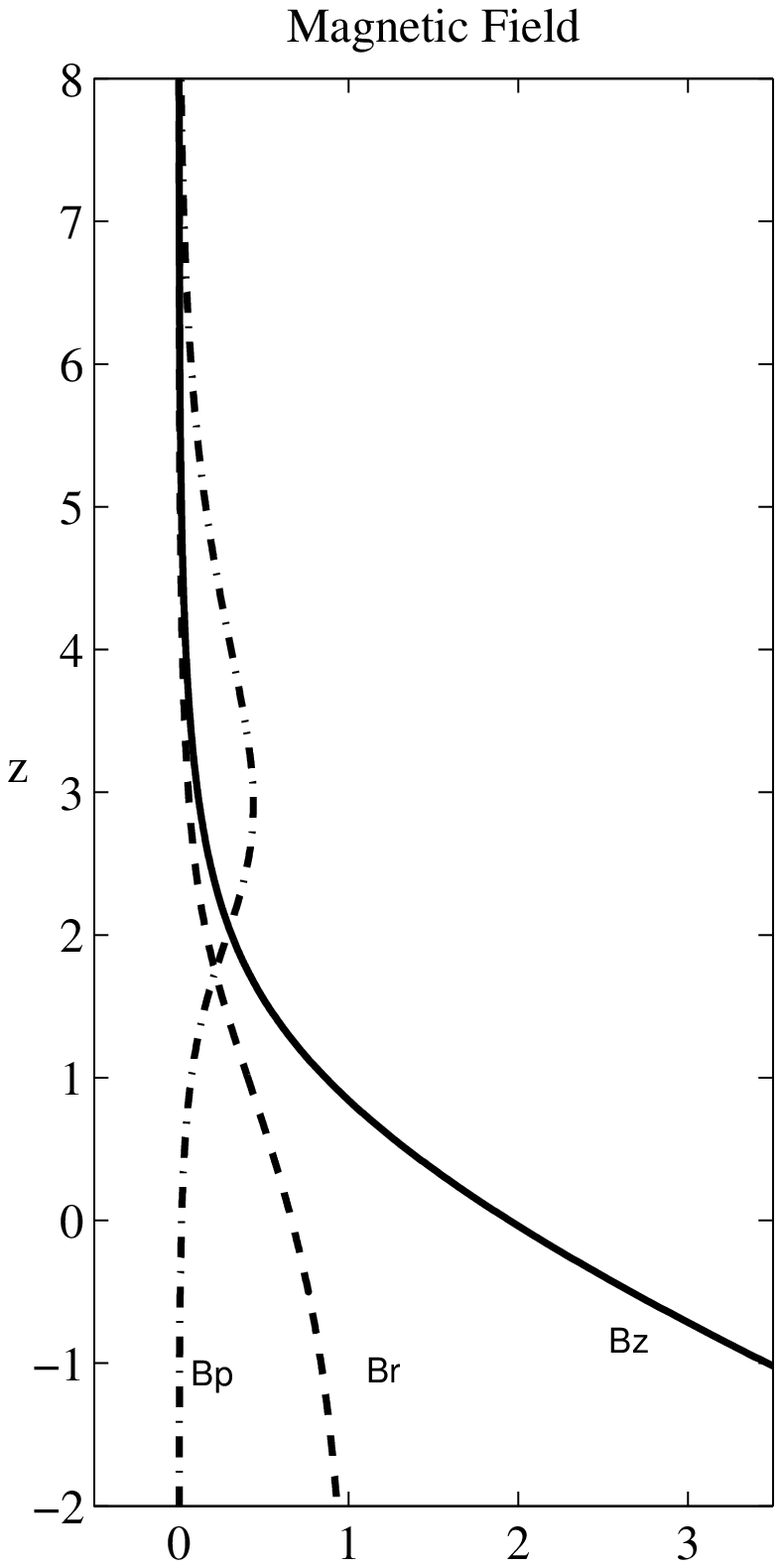}
  \caption{\footnotesize An \san\ solution of the
  \chl\ equations with no \HSL.
  The downwelling was chosen such that $\gamma = 2.24$.
  The $u_\phi$ profile has been rescaled by a factor of 10.
  The \mbox{$z$-origin} is arbitrary.
  }
  \label{fig:noHSL}
\end{figure}

\begin{figure}
  \centering
  \psfrag{U}[l][]{$\mathbf{u}$}
  \psfrag{B}[tl][B]{$\mathbf{B}$}
  \psfrag{W}[][t]{$_{\phantom{\ii}}\OmegaI$}
  \includegraphics[width=10cm]{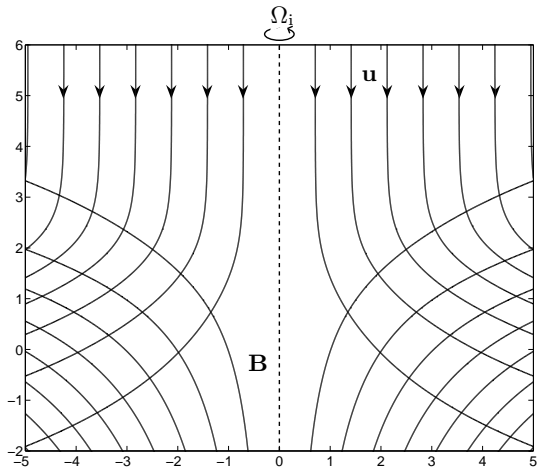}
  \caption{\footnotesize The magnetic confinement layer
  near the
  north pole
  in a model for the early Sun,
  with no \HSL.
  The plot is from the same \san\ solution
as that of \fig\ref{fig:noHSL}.
\label{fig:noHSL-section}}
\end{figure}

Analytical solutions with no \HSL\ can be calculated
in the same way as
in \S\ref{sec:self-similar}.
\Fig\ref{fig:noHSL} presents such a solution.
In this case, the downwelling profile was chosen to be
$u_z(z) = -(1+\exp(-\gamma (z-2)))^{-1/\gamma}$,
so that $u_z = O(\exp(z))$ as $z\to-\infty$
and $u_z = -1+O(\exp(-\gamma z))$ as $z\to+\infty$.
We have taken $\gamma=2.24$,
again purely for illustration,
to allow a more direct comparison
with \figs\ref{fig:sim-profiles}
and \ref{fig:num-profiles}.
\Fig\ref{fig:noHSL-section} shows a vertical cross-section
through the
solution presented in \fig\ref{fig:noHSL}.

\section{Conclusions and future directions}
\label{sec:conclusions}

We cannot yet claim to have a complete
\tc\ theory.  Indeed, the \cl\ and \hsl\ form only two pieces
of a complicated
jigsaw puzzle.
Other
aspects of that jigsaw include
the way in which the \cl\
matches upward to the
relatively large
negative shear in the bulk of the \tc,
and
the way in which the baroclinic
temperature anomalies induced by the
\tc's MMC
fit into the perturbed global-scale heat flow.
In particular, without putting the whole jigsaw together we
cannot quantitatively predict the
thickness of the \tc.
Nor can we predict the precise shapes of the
vertical profiles
of $\B$ and $\bu$ in the \cl.
Those profile shapes
depend
on matching to conditions not only aloft
but also equatorward,
where
stratification surfaces and
field lines extend into colatitudes outside the polar
downwelling regions.

However,
the results obtained here give us
the first fully consistent
model
of
polar field confinement, as such,
together with insight into
how it could work in today's Sun.
The results cover a wide range of
possible
downwelling values
and interior field strengths
(end of
\S\ref{sec:CL-scalings}).
We have also shown,
in \S\ref{sec:noHSL},
how confinement could have worked in the early Sun.
The dynamics is similar apart from the
slightly deeper penetration of the MMC
in
the absence of the \HSL\ and \hsl.
We can use the resulting
insights,
alongside our well-established understanding of
the gyroscopic pumping of MMCs,
to say something
new
about the early Sun and the solar
lithium-burning problem.

Standard solar-evolution models predict
surface lithium abundances higher than observed
by a factor $\sim10^2$
\citep[\eg][]{Vauclair-etal78}.
The reason is that the standard
models mix material down to the bottom of the \cz\ but
no further.  To destroy lithium,
material from the \cz\ must be mixed or
circulated to somewhat greater depths and
therefore to somewhat higher temperatures,
beyond those at the
bottom of today's \tc.
However, there is no evidence of depletion of the
\cz's beryllium, which is destroyed at only moderately
higher temperatures than lithium.
Further discussion and references
may be found in
\citet{JCDGoughThompson92},
and in \citet[][chap.\ 6]{Wood10}.
Here
we argue that a quantitative version of the scenario
sketched in \fig\ref{fig:sphere} has promise as a way of
circulating material to the required depth in the early Sun,
and no further,
thus
making sense
of the high beryllium
as well as the low lithium
abundance.

As already mentioned
in \S\ref{sec:intro},
the downwelling MMC in the polar \tc\ that makes field
confinement possible
can be regarded as due to a
gyroscopically-pumped
MMC trying
to burrow
downward, but
held in check
by its encounter with the interior field
$\BI$
and
with the \HSL, if present.
If, in a \te, we were to switch off the interior field $\BI$,
then the downwelling would spread or burrow
to ever-increasing depths.
The
timescale
for such burrowing
is inversely proportional to
$\OmegaI^2$
\citep[\eg][\eq(8.15)ff.]{McIntyre07};
one may think of
rotational stiffness as
strengthening the burrowing tendency.

Now,
because the early Sun rotated much faster than today,
not only
would there have been no \HSL\
but, also,
the burrowing tendency would have been
much
stronger than today,
tending to push the bottom of the \tc\
downward.
This reopens the possibility
conjectured in
{GM98} that there might have been a ventilated
``polar pit''
in which most
of the
\cz's
lithium,
though not too much of its beryllium,
was burnt
during
the first gigayear or
two
of the Sun's main-sequence evolution.

To take this further we again need to consider
the way in which the
\cl\ fits into the global picture.  It is
arguable that the
bottom of the
entire polar downwelling region is depressed relative
to its surroundings, forming
not so much a ``pit''
as a
shallow ``frying pan'',
too shallow to burn lithium in today's Sun
but possibly
just
deep enough in the early Sun.

\begin{figure}
  \centering
  \includegraphics[width=7cm]{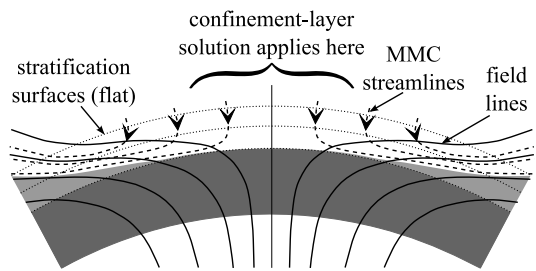}
  \caption{\footnotesize
    Schematic drawing of
    the magnetic confinement layer and its immediate
    surroundings at the \base\ of the high-latitude \tc.
    Close to the pole the
    interior
    magnetic field (solid lines)
    is confined by the downwelling MMC (dashed streamlines).
    The vertical scale
    has been
    greatly
    exaggerated.\label{fig:surroundings}}
\end{figure}
Here we need to distinguish
the shape of the ventilated region from the shapes
of the
stratification surfaces, which latter must remain relatively
flat, meaning close to the horizontal.
\Fig\ref{fig:surroundings}
sketches the way in which the \cl\ might fit into its
surroundings near the bottom of the polar \tc.
The stratification surfaces are shown dotted.
At the periphery of the polar downwelling region,
the field lines (solid)
spiral
outward and upward
from the \cl\ on their way
to lower latitudes.  They will tend to
splay out,
as well as
slanting upward, as they
emerge from
the downwelling region.
The MMC will similarly slant upward, flowing approximately along the
field lines
(dashed streamlines).  This is because the
splaying-out increases the magnetic Reynolds number beyond the
order-unity values characteristic of the \cl.
Further out, the field lines must continue to rise through the
\tc\ until they encounter the
\cz's overshoot layer,
where they
are held horizontal
by turbulent \mfp\
as suggested in \fig\ref{fig:sphere}.
On the way we must expect
turbulent
eddy fluxes to become increasingly important,
decoupling the MMC's upwelling
streamlines
from the
time-averaged field lines and leaving the upwelling free to
spread over a wide
range of latitudes, constrained only by mass conservation
and
global-scale heat flow.

Such a picture
applies equally well to today's Sun and to the early Sun,
the main difference being that the ventilated polar region
(unshaded in \fig\ref{fig:surroundings})
is likely to have been pushed deeper in the early Sun with its
much faster rotation, stronger
burrowing tendency, and global-scale
$|\BI|$ values only modestly larger.
The ventilated polar regions
could well have been deeper
by many tens of megametres,
as required to burn
lithium.\footnote{This
deepening is additional
to the deepening of the \cz\ itself, in the early Sun relative to
today's Sun,
amounting to
several more tens of megametres
according to standard solar models
\citep[e.g.][]{Ciacio-etal97}.
}
The early Sun would have started to form
its \HSL\ just below these
ventilated,
lithium-destroying polar
regions,
\ie\ just below the polar
\cl s.
Then, with the gradual diminution of $\OmegaI$
through solar-wind braking,
the \cl s, marking the bottom of the ventilated
regions,
would have retreated
upward,
and the top of the \HSL\
would have followed them upward
as new helium
strata formed.

Within
the
peripheral lightly-shaded
region in \fig\ref{fig:surroundings},
into which the MMC does not penetrate, we suggest that
ventilation is weak or nonexistent and that shear will
be limited by the Ferraro constraint.
The darker shading
represents
the top part of
today's \HSL.

As the
lightly-shaded
region expands upward and outward
beyond the immediate surroundings
sketched in \fig\ref{fig:surroundings},
through the \tc\ toward
the overshoot layer,
we may surmise that
small-scale MHD
instabilities will kick in
\citep[\eg][\& refs.]{Spruit02,GilmanCally07,ParfreyMenou07},
breaking the Ferraro constraint and
blurring
the distinction between the
shaded and unshaded
regions as turbulent eddy fluxes increase.
So a
larger-scale picture of the
``lithium frying pan''
would show its upward-sloping lower boundary
becoming increasingly porous and indistinct
at greater colatitudes.
 
The
global \tc\ model that would be
needed to test, and to begin to quantify,
the foregoing speculations
would have to describe

\smallskip

\begin{enumerate}
  \item the precise
        way in which turbulent stresses in the convection
        zone and
        \tc\ gyroscopically
        pump
        the polar downwelling needed to confine
        $\BI$ in polar latitudes;
  \item the global-scale distribution of temperature
        and heat flow
        that fits in with the
        MMCs;
  \item the turbulent \mfp\
        by convective overshoot
        assumed to
        confine
        $\BI$ in
        \extrapolar\
        latitudes;
  \item the extent to which the winding-up of the
        time-averaged
        toroidal field
        in \extrapolar\ latitudes (\fig\ref{fig:sphere})
        is limited by turbulent eddy fluxes;
  \item the reaction of the overlying turbulent layers to all
        of the above,
especially the deficit in the \cz's
differential rotation governing the torques exerted from
above,
for instance via
feedback on the
strength of gyroscopic pumping of the MMC.
\end{enumerate}

\smallskip

Progress on these formidable problems will
depend on finding suitable ways to model the turbulent
processes.

\acknowledgments
We thank
N. H. Brummell,
J. Christensen-Dalsgaard,
W. D\"appen,
S. Degl'Innocenti,
P. Garaud,
D. O. Gough,
D. Jault,
C. A. Jones,
D. Lecoanet,
L. Mestel,
M. Miesch,
M. R. E. Proctor,
T. M. Rogers,
S. N. Shore,
M. J. Thompson,
S. M. Tobias,
N. O. Weiss,
L. A. Willson,
and several
referees
for helpful and stimulating comments and discussions.
TSW was supported by a UK Science and Technology Facilities
Council Research Studentship,
and MEM's early thinking on the \tc\ problem was supported by
a SERC/EPSRC Senior Research Fellowship.

\appendix

\section{HELIUM SUBLAYER SCALINGS}

\label{sec:darcy_scalings}

For perturbations $\B'$ to a background magnetic field $\BI$,
the Lorentz force may be written as
\begin{equation}
\mathbf{F}
=
(\bm{\nabla}\times(\BI+\B'))\times(\BI+\B')
=
-\bm{\nabla}(\tfrac{1}{2}|\BI|^2+\BI\cdot\B'+\tfrac{1}{2}|\B'|^2)
+ (\BI+\B')\cdot\bm{\nabla}(\BI+\B')
\end{equation}
If, as here,
the background field $\BI$ is curl-free, then
all the terms quadratic in $\BI$ vanish and
we have
\begin{equation}
\mathbf{F}
=
-\bm{\nabla}(\BI\cdot\B'+\tfrac{1}{2}|\B'|^2)
+ \B'\cdot\bm{\nabla}\BI
+ (\BI+\B')\cdot\bm{\nabla}\B'
\label{eq:perturbed-Lorentz}
       \,.
\end{equation}
If $\BI$ and $\B'$ are axisymmetric,
and have the scalings given by
(\ref{eq:Lambda-scalings-mu-d})--(\ref{eq:Lambda-scalings-mu-f}),
with $\partial/\partial z\sim\CLthick/\SLthick\gg1$,
then all components of $\B'\cdot\bm{\nabla}\B'$
are smaller than the corresponding components of
$\BI\cdot\bm{\nabla}\B'$ by factors $(\SLthick/\CLthick)^2$,
with one exception.  The $r$ component of
$\B'\cdot\bm{\nabla}\B'$ includes a term $B\sliver'^{\Sliver2}_\phi/r$,
the divergence of the hoop stress.
Relative to the $r$ component of $\BI\cdot\bm{\nabla}\B'$
this term is of order $(\SLthick/\CLthick)^2/\Lambda^2$.
However, the hoop-stress
term itself
is smaller than the $r$ component
of the Coriolis force by a factor $(\SLthick/\CLthick)^2$,
even for small $\Lambda$,
because of (\ref{eq:Lambda-scalings-mu-c}).
We also find that,
because $B_{\ii \phi} = 0$,
the poloidal components of
$\bm{\nabla}(\BI\cdot\B')$ exceed in magnitude those of
$\bm{\nabla}(\tfrac{1}{2}|\B'|^2)$ by a factor $(\CLthick/\SLthick)^2$.
Their
azimuthal components are both zero
since we consider only axisymmetric fields here.
We may therefore neglect all
terms quadratic in $\B'$,
in the asymptotic regime (\ref{eq:sublayer-regime}),
so that (\ref{eq:Tmubalance}) becomes
\begin{align}
  \mathbf{e}_z\times\bu
\;=\; -\bm{\nabla} \pp
  + \alphat\sliver
  \pottemp\mathbf{e}_z
  - \alphamu\sliver
  \mu\sliver\mathbf{e}_z
  -\bm{\nabla}(\BI\cdot\B')
+ \B'\cdot\bm{\nabla}\BI
+ \BI\cdot\bm{\nabla}\B'
\;.
\label{eq:SL-momentum1}
\end{align}
The flow through the sublayer produces only a
small perturbation to
the otherwise
uniform thermal stratification.
From (\ref{eq:T}) and (\ref{eq:mu})
we see that, within the sublayer,
variations in $\pottemp$ are smaller than variations in $\mu$
by a factor $\chi/\kappa \ll 1$.
The thermal stratification in the sublayer may therefore be
treated as horizontally uniform,
allowing
the thermal buoyancy term
$\alphat\sliver \pottemp\mathbf{e}_z$ in (\ref{eq:SL-momentum1})
to be incorporated into the
pressure field
along with the
$-\bm{\nabla}(\BI\cdot\B')$ term.
We denote the modified pressure field as $\tilde{p}$.

The perturbed steady-state induction equation
is
\begin{equation}
0 = (\BI+\B')\cdot\bm{\nabla}\bu - \bu\cdot\bm{\nabla}(\BI+\B')
+ \nabla^2\B'
\label{eq:perturbed-induc}
\;,
\end{equation}
again on the assumption that $\BI$ is curl-free.
It is readily verified
from the scalings
(\ref{eq:Lambda-scalings-mu-a})--(\ref{eq:Lambda-scalings-mu-f})
and $\partial/\partial z\sim\CLthick/\SLthick\gg1$
that each component of
$\bu\cdot\bm{\nabla}\B'$ is of the same order as the
corresponding component of $\B'\cdot\bm{\nabla}\bu$,
but smaller than $\BI\cdot\bm{\nabla}\bu$ by a factor
$(\SLthick/\CLthick)^2$.
Furthermore, provided that the horizontal scales $\rt$ and
$\rmu$ are both $\gg1$ we may also make the boundary-layer
approximation,
$\nabla^2\approx\partial^2/\partial z^2$.
Thus, in the asymptotic regime (\ref{eq:sublayer-regime}),
we may simplify
(\ref{eq:perturbed-induc})
to
\begin{equation}
0 = \BI\cdot\bm{\nabla}\bu - \bu\cdot\bm{\nabla}\BI
+ \frac{\partial^2}{\partial z^2}\B'
\;.
\label{eq:SL-induc1}
\end{equation}

If the field $\BI$ were uniform, and directed along the
axis of rotation, then
(\ref{eq:SL-momentum1}) and (\ref{eq:SL-induc1})
would reduce immediately to
(\ref{eq:SL-momentum2}) and (\ref{eq:SL-induc2})
respectively.
Since $\BI$ is axisymmetric and smooth,
this reduction still holds as a first approximation
within some neighbourhood of the axis.
It is sufficient
to show that this neighbourhood
includes the entire sublayer
within a radius $r\sim\rmu$.
We first
note that, since $\bm{\nabla}\cdot\BI = 0$, we have
$B_{\ii r}/r =
-\tfrac{1}{2}\partial B_{\ii z}/\partial z$
as in (\ref{eq:divBsim}).
In the sublayer we have $\partial B_{\ii z}/\partial z \sim B_{\ii z}$
(dimensionally, $\partial B_{\ii z}/\partial z \sim B_{\ii z}/\CLthick$)
as a consequence of the matching to the \cl,
as can be seen, for instance, from the solid curve in
the right-hand panel of \fig\ref{fig:sim-profiles}.
Now applying the scalings
(\ref{eq:Lambda-scalings-mu-a})--(\ref{eq:Lambda-scalings-mu-f})
we find that,
in (\ref{eq:SL-momentum1}),
each component of $\B'\cdot\bm{\nabla}\BI$ is smaller
than the corresponding component of $\BI\cdot\bm{\nabla}\B'$
by a factor $\SLthick/\CLthick$, and that,
in (\ref{eq:SL-induc1}),
each component of $\bu\cdot\bm{\nabla}\BI$ is smaller
than the corresponding component of $\BI\cdot\bm{\nabla}\bu$
by the same factor $\SLthick/\CLthick$.
Moreover, even at colatitudes
$r \sim \rmu$
the contributions
$B_{\ii z}\partial\B'/\partial z$
and $B_{\ii z}\partial\bu/\partial z$
dominate all other contributions to
$\BI\cdot\bm{\nabla}\B'$ and $\BI\cdot\bm{\nabla}\bu$
by factors of at least $\CLthick/\SLthick$.
So
(\ref{eq:SL-momentum1}) and (\ref{eq:SL-induc1})
do indeed reduce to
(\ref{eq:SL-momentum2}) and (\ref{eq:SL-induc2}).

For $r \lesssim \rmu$,
tilting of the compositional isopleths produces
variations in the
dimensionless
altitude of the sublayer
no greater than $O(\SLthick/\CLthick)$,
so that $B_{\ii z}$ may be assumed constant
within the sublayer
as assumed in the derivation
of (\ref{eq:torque-sublayer}) and (\ref{eq:T-W-sublayer}).
Finally, we note that
the foregoing picture
applies not
only to steady states but also to
time-dependent states with any timescale,
such as $\CLthick/U=\CLthick^2/\eta$,
that is long in comparison with
the timescale \
$\SLthick^2/\eta$ \
for magnetic diffusion across the sublayer.
On any such
timescale
the sublayer  therefore
behaves like a porous medium.

\section{THE NUMERICAL SCHEME}
\label{sec:numerics_details}

We wish to solve
a suitable version of
\eqs(\ref{eq:momentum})--(\ref{eq:mu})
in axisymmetric cylindrical
polar coordinates.
We introduce streamfunctions $\Psi$ and $\APHI$, \ie\ azimuthal
vector-potential components,
for the poloidal velocity and magnetic fields,
such that
\begin{eqnarray}
  u_z = \frac{1}{r}\frac{\partial(r\Psi)}{\partial r}
  \;\; &\mbox{and}& \;\;
  u_r = -\frac{\partial\Psi}{\partial z}\;, \\
  B_z = \frac{1}{r}\frac{\partial(r\APHI)}{\partial r}
  \;\; &\mbox{and}& \;\;
  B_r = -\frac{\partial\APHI}{\partial z}\;,
\end{eqnarray}
guaranteeing that the fields are divergence-free;
$r\Psi$ is sometimes called
the Stokes streamfunction.
The azimuthal vorticity $\omega_\phi$
and
electric current
density
$J_\phi = (\nabla\times\B)_\phi$
are
related to $\Psi$ and $\APHI$ by
\begin{flalign}
&
& \omega_\phi &= - \left(\nabla^2-r^{-2}\right)\Psi&
  \label{eq:invert} \\
&\mbox{and}
& J_\phi &= - \left(\nabla^2-r^{-2}\right)\APHI
  \;.&
\end{flalign}
As already explained,
and further discussed
below, we
introduce anisotropic viscosity and helium
diffusivity
with
dimensionless
horizontal components $\nu_H$ and $\chi_H$. \
So \eqs(\ref{eq:momentum})--(\ref{eq:mu})
are replaced by
\begin{flalign}
  \Ro\,\frac{1}{r}\frac{\DD(ru_\phi)}{\DD t} - \frac{\partial\Psi}{\partial z}
  &=
  \frac{1}{r}\B\cdot\bm{\nabla}(rB_\phi)
  + \Ek\left[
    \frac{\partial^2}{\partial z^2}+\nu_H\left(\nabla_H^2-\frac{1}{r^2}\right)
  \right]u_\phi \label{eq:num-mom-phi} \\
  \Ro\left[
    r\frac{\DD(\omega_\phi/r)}{\DD t}
    \right.+&\left.
    \frac{\partial(ru_\phi,u_\phi/r)}{\partial(z,r)}
  \right]
  - \frac{\partial u_\phi}{\partial z} \nonumber \\
  &=
  - \alphat\frac{\partial \pottemp}{\partial r} + \alphamu\frac{\partial\mu}{\partial r}
  + r\B\cdot\bm{\nabla}(J_\phi/r) + \frac{\partial(rB_\phi,B_\phi/r)}{\partial(z,r)} \nonumber \\
  &\phantom{xx}
  + \Ek\left[
    \frac{\partial^2}{\partial z^2}+\nu_H\left(\nabla_H^2-\frac{1}{r^2}\right)
  \right]\omega_\phi \label{eq:num-vort-phi} \\
  r\frac{\DD(B_\phi/r)}{\DD t} &= r\B\cdot\bm{\nabla}(u_\phi/r)
  + \left(\nabla^2-\frac{1}{r^2}\right)B_\phi
  \label{eq:num-ind-phi} \\
  \frac{1}{r}\frac{\DD(r\APHI)}{\DD t} &= \left(\nabla^2-\frac{1}{r^2}\right)\APHI \\
  \frac{\DD \pottemp}{\DD t} &= \frac{\kappa}{\eta}\nabla^2\pottemp \\
  \frac{\DD\mu}{\DD t} &= \frac{\chi}{\eta}\left[
    \frac{\partial^2}{\partial z^2}+\chi_H\nabla_H^2
  \right]\mu
\;.
\label{eq:num-mu}
\end{flalign}
We solve these equations
using a simple finite-difference scheme on an Eulerian grid
regularly spaced in $r$ and $z$
at intervals $\dr$ and $\dz$.
The outer boundary of the computational domain is at $r=\rperiph$.
The inner boundary is at $r = 2\dr$, \ie\
two grid
intervals from the coordinate singularity at the
rotation axis.
Because of the directionality of operators like
$\bu\cdot\bm{\nabla}$ and $\B\cdot\bm{\nabla}$,
the spatial derivatives are
calculated using two-point,
one-sided (first-order)
finite differences whose
directions are chosen to ensure numerical stability at
the grid scale.
For reasons of symmetry and good behaviour near the
coordinate singularity,
the finite differencing
is done
by locally approximating
the fields
$\Psi/r$, $u_\phi/r$, $\omega_\phi/r$,
$\APHI/r$, $B_\phi/r$, $\pottemp$, and $\mu$
as functions that are linear in $z$ and in $r^2$
over a single grid interval.
This ensures that the error is $O(\dr)$ even
for small $r$.
Field values for $r < 2\dr$ are
obtained by extrapolation from
$r=3\dr$ and $r=2\dr$,
again assuming linear functional dependence on $r^2$.

With the parameter values given in Table~\ref{tab:num-params},
the dimensionless
\hhsl\ and Ekman-layer thicknesses
are $\SLthick/\CLthick = (\chi/\eta)^{1/2} \approx 0.14$
and $\Ekthick/\CLthick = \Ek^{1/2} \approx 0.01$
respectively.
We have chosen a vertical grid
interval $\dz = 0.01$,
dimensionally $0.01\CLthick$, which is
small enough to
resolve the helium sublayer accurately.
This $\dz$ is
too large
to resolve any Ekman layers.
However,
Ekman layers are
prevented from becoming significant
by careful choice of
the
code representing the
boundary conditions.
By allowing slip velocities and
making viscous stresses negligible at the boundaries,
we have been able to
keep Ekman-layer formation so weak as to
play
no significant role in the dynamics.
Uniform rotation
is
imposed at
and just beneath
the bottom of the
domain, via the $\mathbf{B}\cdot\bm{\nabla}$ term in the
azimuthal component of the induction equation
(\ref{eq:num-ind-phi}).

An explicit Eulerian timestepping scheme
is used to evolve the system.  The timestep $\dt$
must be small enough to
resolve thermal diffusion at the grid scale (which is the
fastest process at this scale and therefore determines
the Courant--Friedrichs--Lewy condition).
So
from Table~\ref{tab:num-params},
$\dt \lesssim (\eta/\kappa)(\dz)^2 =
10^{-2}\times(0.01)^2 = 10^{-6}$,
dimensionally
$10^{-6}\CLthick/U$ or
$10^{-4}(2\OmegaI)^{-1}$.

We use a \san\ \chl\ solution,
of the kind described in \S\ref{sec:self-similar},
as the initial condition.
The system
typically takes several
domain-scale
magnetic diffusion times
to reach a steady state.
As explained in \S\ref{sec:num-solutions},
multiple iterations of the
peripheral $B_\phi(z)$
profile are
then
required to achieve a steady state
with vanishing $B_\phi(r)$
at the bottom.
To make the computation feasible, in a domain wide enough to
accommodate noticeable
tilting of the stratification surfaces,
we have used
$\rperiph = 5$,
dimensionally $5\CLthick$,
and a horizontal
grid
interval $\dr = 0.1$,
dimensionally $0.1\CLthick$,
larger than the vertical
grid
interval $\dz$ by a
factor of 10.  For numerical stability,
the dimensionless
horizontal viscosity $\nu_H$ and helium
diffusivity $\chi_H$
must then be chosen so that the diffusive terms in
(\ref{eq:num-mom-phi}), (\ref{eq:num-vort-phi})
and (\ref{eq:num-mu})
dominate the advective terms at the gridscale.
In practice,
we
found it
sufficient
to take
$\nu_H = \chi_H = 10$,
\ie\ to make them a factor 10 greater
than the corresponding vertical diffusivities,
as
suggested by
advective--diffusive
scaling
when $\dr/\dz=10$.
We have verified,
in smaller computational domains,
that the
coarser horizontal
resolution and
anisotropic
diffusivities
do not qualitatively affect the steady state of the system.

At each timestep,
the azimuthal vorticity
$\omega_\phi$ is updated and
the streamfunction $\Psi$ then computed
from (\ref{eq:invert})
by inverting
the operator  $\nabla^2 - r^{-2}$,
approximated
using centred differences.
The inversion is performed iteratively,
using a
successive-overrelaxation method
described in \citet{Press-etal86}.
During the
early evolution, when the dynamics is dominated
by timescales
not much longer than
the timestep $\dt$, many such
iterations are required,
at each timestep,
to achieve convergence.
At later times
the same degree of convergence
can be achieved with far fewer iterations.
Since we are interested only in the ultimate steady state,
we can tolerate a larger error in the inversion
during the system's transient evolution.
Further details of the numerical code are spelt out in
\citet{Wood10}.

\begin{figure}
  \centering
  \includegraphics[height=6cm]{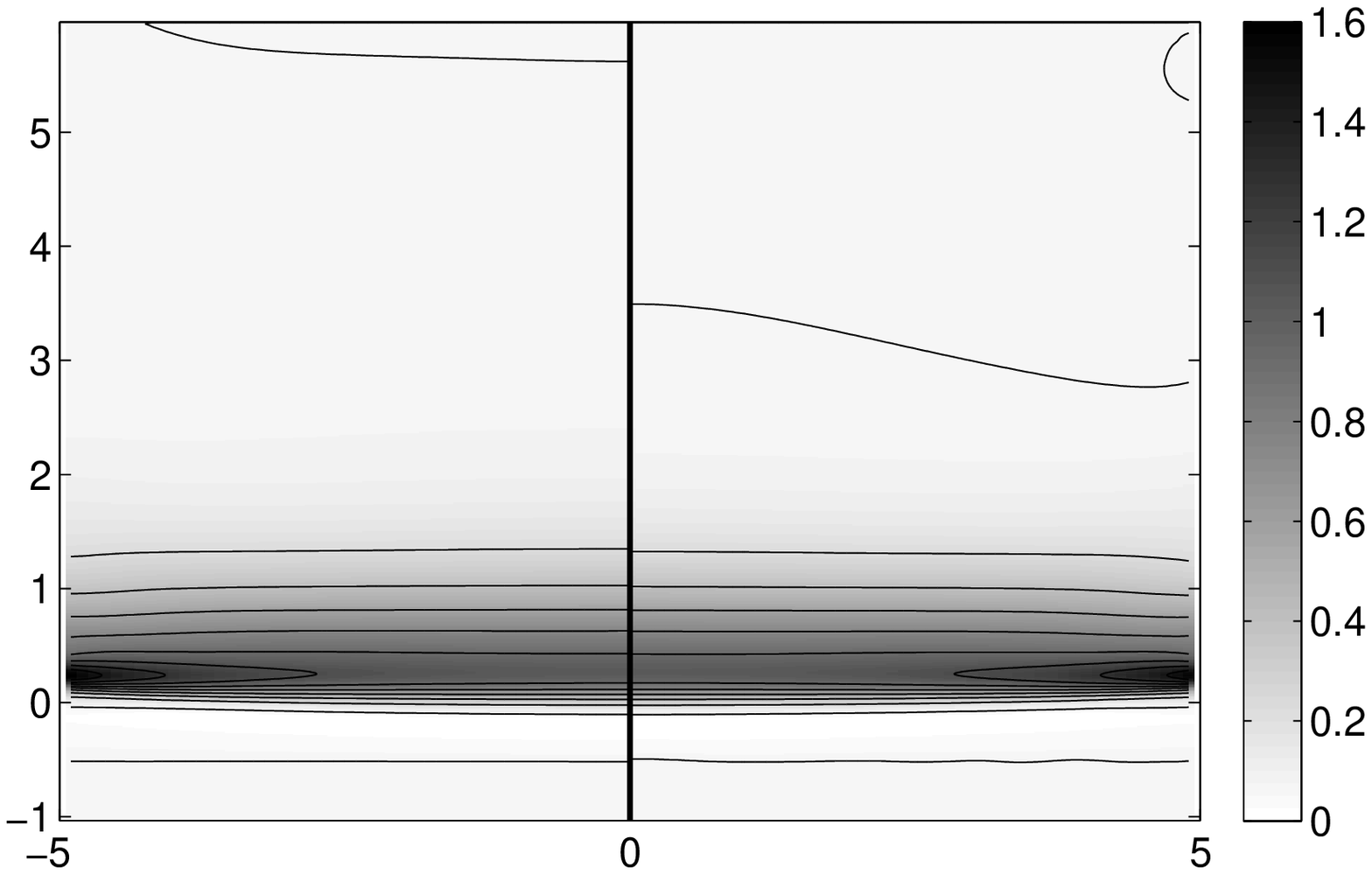}
  \caption{\footnotesize
    The left and right panels show contours of the \LHS\
    and \RHS\ of
    (\ref{eq:numerical-balance}) respectively,
    in a vertical cross-section
    for
    the numerical solution of
\figs\ref{fig:section}, \ref{fig:full}, and
\ref{fig:num-profiles}.
    The bulk of the \cl\ is close to \mb.
    The largest fractional
    departures from balance
    occur in the
    regions above and below the \cl\ where both sides of
    (\ref{eq:numerical-balance}) are numerically small.
    Unlike that in \fig\ref{fig:section},
    the cross-sections shown here span the
    entire numerical domain including the
    top part
    $5 \leqslant z \leqslant 6$.
  }
  \label{fig:strophic}
\end{figure}
As anticipated from
the small values of $\Ro$ and $\Ek$,
the steady state is found to be close to \mb:
the motion is scarcely distinguishable from one in which
the balance conditions
(\ref{eq:torque}) and (\ref{eq:T-W}) hold exactly.
Indeed, in (\ref{eq:T-W}) the terms in $\alphat$ and $\alphamu$
dominate so strongly that, in the case of
\fig{\ref{fig:section} for instance,
the tilting is barely visible, as has been verified from
plots, not shown, of the thermal as well as the compositional
stratification
surfaces.
This of course
is no more than
was
anticipated from the scaling arguments
of \S\S\ref{sec:CL-scalings},\;\ref{sec:helium}. \
So the balance (\ref{eq:T-W})
holds in an almost trivial sense, with the
remaining
terms
scarcely able to produce any noticeable effect.
The
azimuthal
momentum balance
described by (\ref{eq:torque})
is less trivial,
but here also
we find that the \LHS\ and \RHS\ are
close to being equal, in the present notation
\begin{equation}
  -\frac{1}{r}\frac{\partial\Psi}{\partial z}
  \approx
  \frac{1}{r^2}\B\cdot\bm{\nabla}(rB_\phi)
\,.
  \label{eq:numerical-balance}
\end{equation}
\Fig\ref{fig:strophic} shows the
\LHS\ and \RHS\ of
(\ref{eq:numerical-balance})
for the case of
\fig\ref{fig:section}.
The balance is quite accurate
despite
the
modest $\Ro$ value
chosen for the numerical solution,
$10^{-2}$ in Table~\ref{tab:num-params}.

It might be thought that
imposing \mb\
from the start,
as first suggested by \citet{Taylor63},
would filter out all the fast oscillations ---
including inertial or epicyclic
oscillations
as well as
Alfv\'en waves, gravity waves, and the various hybrid
types ---
and thereby allow larger timesteps to be used.
However,
the imposition of \mb\
leads to pathological behaviour at small
scales \citep{Walker-etal98}.
Far from eliminating or slowing the fast oscillations,
the imposition of balance exacerbates
the problem,
for reasons explained in Appendix~\ref{sec:magnetostrophic}.

\section{MAGNETOSTROPHIC BALANCE AND NUMERICAL ILL-CONDITIONEDNESS}
\label{sec:magnetostrophic}

Filtering out fast oscillations by imposing some kind of
balance is a
familiar,
and often effective,
device
in many other problems
involving stiff differential equations.  A well known example
is that of fluid flow in non-MHD fluid systems
with strong rotation (small $\Ro$) and stable stratification.
The standard ``quasi-geostrophic equations''
result from
imposing geostrophic or Coriolis--pressure-gradient
as well as hydrostatic
balance,
thereby filtering out
inertia and gravity waves as well as sound waves.
The filtering saves computational resources by
allowing relatively large time steps.

Such filtering turns out, however, to be ineffective
in the \chl\ problem.  Indeed
--- at first sight paradoxically ---
the imposition of \mb\
exacerbates the timestepping problem.
Far from eliminating fast oscillations, it introduces
spurious modes of oscillation that are even faster,
as shown by \citet{Walker-etal98}
in the context of
the terrestrial dynamo problem.
Following \citeauthor{Walker-etal98},
we show how this
pathology
can be understood through an idealized
analysis of the fast oscillations, first in the unfiltered and
then in the filtered equations.

The reason for the pathology is the interplay
between the Coriolis and
Lorentz forces.
Stratification $N^2$ is relatively unimportant,
as will be shown shortly.
We therefore start with
the linear theory of
MC (magneto--Coriolis) waves, \ie\
small plane-wave disturbances
to an
unstratified, incompressible fluid
with \solid\ rotation $\bm{\Omega}$
and a uniform magnetic field $\B$.
Neglecting viscosity and
magnetic diffusivity, we find the well-known
dispersion relation
\begin{equation}
  \omega^2 - (\B\cdot\mathbf{k})^2
  =
  \pm 2\bm{\Omega}\cdot\mathbf{k}\,\omega/|\mathbf{k}|
  \label{eq:MC}
  \,,
\end{equation}
where $\omega$ is the frequency and $\mathbf{k}$ is the wavevector,
both dimensional here.
If we take the limit of rapid rotation, $|\bm{\Omega}| \to \infty$,
then for most
choices of $\mathbf{k}$ the four roots of this
dispersion relation are asymptotically
\begin{flalign}
  &
  &\omega
  &\sim
  \pm \frac{2\bm{\Omega}\cdot\mathbf{k}}{|\mathbf{k}|}&
  \label{eq:fast}
  \\
  &\mbox{and} \;\;\;
  &\omega
  &\sim
  \pm
  \frac{(\B\cdot\mathbf{k})^2}{2\bm{\Omega}\cdot\mathbf{k}}|\mathbf{k}|
  \;.&
  \label{eq:slow}
\end{flalign}
The modes corresponding to (\ref{eq:fast})
are inertial
or epicyclic
waves --- in this context
sometimes called ``fast MC waves'' ---   and
those corresponding to (\ref{eq:slow})
are ``slow MC waves''.
By imposing \mb\
we
neglect relative fluid accelerations, which
corresponds to dropping the $\omega^2$ term from the \LHS\
of (\ref{eq:MC}).
The dispersion relation
then
becomes
\vspace{-0.25cm}
\begin{equation}
  \omega = \pm \frac{(\B\cdot\mathbf{k})^2}{2\bm{\Omega}\cdot\mathbf{k}}|\mathbf{k}|
  \label{eq:mb}
  \;.
\end{equation}
So imposing \mb\
eliminates the two
``fast'' branches
(\ref{eq:fast})
of the full
dispersion relation (\ref{eq:MC}).

However, not all modes of
the full dispersion relation
(\ref{eq:MC})
have the asymptotic behaviour described by
(\ref{eq:fast}) and (\ref{eq:slow}).
Even in the presence of rapid rotation,
there are always some
modes whose $\mathbf{k}$ values satisfy
\begin{equation}
|\B\cdot\mathbf{k}|
\gg
|2\bm{\Omega}\cdot\mathbf{k}|/|\mathbf{k}|
\label{eq:lorentz-gg-coriolis}
\end{equation}
by an arbitrarily large factor.
For instance we can fix the direction of
$\mathbf{k}$ and make $|\mathbf{k}|$ arbitrarily
large.
Such modes
behave like Alfv\'en waves,
with
$\omega \approx \pm \B\cdot\mathbf{k}$.
Imposing \mb\ removes the mechanism
for Alfv\'en wave propagation,
and must therefore alter the behaviour of these modes.
In fact their frequencies become
arbitrarily
higher than Alfv\'en wave frequencies.
This can be seen at once by inspection of
(\ref{eq:mb})
and (\ref{eq:lorentz-gg-coriolis}).
In summary,
even in a rapidly rotating system
\emph{some modes
of the full dispersion relation
always
feel the Lorentz force more strongly than the Coriolis
force},
\ie, they satisfy (\ref{eq:lorentz-gg-coriolis}),
\emph{and
these modes become
ill-behaved under the assumption of \mb}.
A numerical scheme that imposes \mb\
while retaining realistic (Laplacian)
magnetic
dissipation
will therefore be ill-conditioned.

If we introduce stratification $N^2$ then (\ref{eq:mb})
becomes
\begin{equation}
  \omega
  =
  \pm \frac{\B\cdot\mathbf{k}}{2\bm{\Omega}\cdot\mathbf{k}}\left[
    (\B\cdot\mathbf{k})^2|\mathbf{k}|^2
    + N^2|\mathbf{k}_H|^2
  \right]^{1/2}
  \;,
\end{equation}
 where
$\mathbf{k}_H$ is the horizontal
projection of $\mathbf{k}$.
Therefore the
stratification
(a)~makes little difference to the
large-$|\mathbf{k}|$ behaviour
but
(b)~always
increases the frequency of the ill-behaved modes and thereby,
if anything,
further exacerbates the problem.

\bibliography{tsw-thesis-refs}

\begin{thebibliography}{46}
\expandafter\ifx\csname natexlab\endcsname\relax\def\natexlab#1{#1}\fi

\bibitem[{Braithwaite} \& {Spruit}(2004)]{BraithwaiteSpruit04}
{\sc {Braithwaite}, J. \& {Spruit}, H.~C.} 2004 {A fossil origin for the
  magnetic field in A stars and white dwarfs}. {\em Nature\/} {\bf 431},
  819--821.

\bibitem[{Brun} \& {Zahn}(2006)]{BrunZahn06}
{\sc {Brun}, A.~S. \& {Zahn}, {J.-P.}} 2006 {Magnetic confinement of the solar
  tachocline}. {\em A\&A\/} {\bf 457}, 665--674.

\bibitem[{Charbonneau} \& {MacGregor}(1993)]{CharbonneauMacGregor93}
{\sc {Charbonneau}, P. \& {MacGregor}, K.~B.} 1993 {Angular Momentum Transport
  in Magnetized Stellar Radiative Zones. II. The Solar Spin-down}. {\em
  Astrophys. J.\/} {\bf 417}, 762.

\bibitem[{Charbonnel} \& {Talon}(2007)]{CharbonnelTalon07}
{\sc {Charbonnel}, C. \& {Talon}, S.} 2007 {On Internal Gravity Waves in
  Low-mass Stars}. In {\em Unsolved Problems in Stellar Physics: A Conference
  in Honor of Douglas Gough\/} (ed. {R.~J.~Stancliffe, J.~Dewi, G.~Houdek,
  R.~G.~Martin, \& C.~A.~Tout}), {\em American Institute of Physics Conference
  Series\/}, vol. 948, pp. 15--26.

\bibitem[{Christensen-Dalsgaard} {\em et~al.\/}(1992){Christensen-Dalsgaard},
  {Gough} \& {Thompson}]{JCDGoughThompson92}
{\sc {Christensen-Dalsgaard}, J., {Gough}, D.~O. \& {Thompson}, M.~J.} 1992 {On
  the rate of destruction of lithium in late-type main-sequence stars}. {\em
  A\&A\/} {\bf 264}, 518--528.

\bibitem[{Christensen-Dalsgaard} {\em et~al.\/}(1993){Christensen-Dalsgaard},
  {Proffitt} \& {Thompson}]{JCD-etal93}
{\sc {Christensen-Dalsgaard}, J., {Proffitt}, C.~R. \& {Thompson}, M.~J.} 1993
  {Effects of diffusion on solar models and their oscillation frequencies}.
  {\em Astrophys. J.\/} {\bf 403}, L75--L78.

\bibitem[{Christensen-Dalsgaard} \& {Thompson}(2007)]{JCDThompson07}
{\sc {Christensen-Dalsgaard}, J. \& {Thompson}, M.~J.} 2007 {Observational
  results and issues concerning the tachocline}. In {\em The Solar
  Tachocline\/} (ed. {D.~W.~Hughes, R.~Rosner, \& N.~O.~Weiss}), p.~53.
  Cambridge: University Press.

\bibitem[{Ciacio} {\em et~al.\/}(1997){Ciacio}, {degl'Innocenti} \&
  {Ricci}]{Ciacio-etal97}
{\sc {Ciacio}, F., {degl'Innocenti}, S. \& {Ricci}, B.} 1997 {Updating standard
  solar models}. {\em A\&AS\/} {\bf 123}, 449--454.

\bibitem[{Debnath}(1973)]{Debnath73}
{\sc {Debnath}, L.} 1973 {On Ekman and Hartmann boundary layers in a rotating
  fluid}. {\em Acta Mechanica\/} {\bf 18}, 333--341.

\bibitem[{Elliott}(1997)]{Elliott97}
{\sc {Elliott}, J.~R.} 1997 {Aspects of the solar tachocline}. {\em A\&A\/}
  {\bf 327}, 1222--1229.

\bibitem[{Elliott} \& {Gough}(1999)]{ElliottGough99}
{\sc {Elliott}, J.~R. \& {Gough}, D.~O.} 1999 {Calibration of the thickness of
  the solar tachocline}. {\em Astrophys. J.\/} {\bf 516}, 475--481.

\bibitem[{Ferraro}(1937)]{Ferraro37}
{\sc {Ferraro}, V.~C.~A.} 1937 {The non-uniform rotation of the Sun and its
  magnetic field}. {\em MNRAS\/} {\bf 97}, 458.

\bibitem[{Garaud} \& {Brummell}(2008)]{GaraudBrummell08}
{\sc {Garaud}, P. \& {Brummell}, N.~H.} 2008 {On the penetration of meridional
  circulation below the solar convection zone}. {\em Astrophys. J.\/} {\bf
  674}, 498--510.

\bibitem[{Garaud} \& {Garaud}(2008)]{GaraudGaraud08}
{\sc {Garaud}, P. \& {Garaud}, {J.-D.}} 2008 {Dynamics of the solar tachocline
  - II. The stratified case}. {\em MNRAS\/} {\bf 391}, 1239--1258.

\bibitem[{Gilman} \& {Cally}(2007)]{GilmanCally07}
{\sc {Gilman}, P.~A. \& {Cally}, P.~S.} 2007 {Global MHD instabilities of the
  tachocline}. In {\em The Solar Tachocline\/} (ed. {D.~W.~Hughes, R.~Rosner,
  \& N.~O.~Weiss}), p. 243. Cambridge: University Press.

\bibitem[{Gough}(2007)]{Gough07}
{\sc {Gough}, D.} 2007 {An introduction to the solar tachocline}. In {\em The
  Solar Tachocline\/} (ed. {D.~W.~Hughes, R.~Rosner, \& N.~O.~Weiss}), p.~3.
  Cambridge: University Press.

\bibitem[{Gough} \& {McIntyre}(1998)]{GM98}
{\sc {Gough}, D.~O. \& {McIntyre}, M.~E.} 1998 {Inevitability of a magnetic
  field in the Sun's radiative interior}. {\em Nature\/} {\bf 394}, 755--757.

\bibitem[{Haynes} {\em et~al.\/}(1991){Haynes}, {McIntyre}, {Shepherd}, {Marks}
  \& {Shine}]{Haynes-etal91}
{\sc {Haynes}, P.~H., {McIntyre}, M.~E., {Shepherd}, T.~G., {Marks}, C.~J. \&
  {Shine}, K.~P.} 1991 {On the `downward control' of extratropical diabatic
  circulations by eddy-induced mean zonal forces}. {\em J.~Atmos.~Sci.\/} {\bf
  48}, 651--680.

\bibitem[{Hughes} {\em et~al.\/}(2007){Hughes}, {Rosner} \&
  {Weiss}]{Hughesetal07}
{\sc {Hughes}, D.W., {Rosner}, R. \& {Weiss}, N.O.} 2007 {\em The Solar
  Tachocline\/}. Cambridge: University Press.

\bibitem[{Kitchatinov} \& {R{\"u}diger}(2006)]{KitchatinovRudiger06}
{\sc {Kitchatinov}, L.~L. \& {R{\"u}diger}, G.} 2006 {Magnetic field
  confinement by meridional flow and the solar tachocline}. {\em A\&A\/} {\bf
  453}, 329--333.

\bibitem[Kleeorin {\em et~al.\/}(1997)Kleeorin, Rogachevskii, Ruzmaikin, Soward
  \& Starchenko]{Kleeorinetal97}
{\sc Kleeorin, N., Rogachevskii, I., Ruzmaikin, A., Soward, A.~M. \&
  Starchenko, S.} 1997 Axisymmetric flow between differentially rotating
  spheres in a dipole magnetic field. {\em J. Fluid Mech.\/} {\bf 344},
  213--244.

\bibitem[{MacGregor} \& {Charbonneau}(1999)]{MacGregorCharbonneau99}
{\sc {MacGregor}, K.~B. \& {Charbonneau}, P.} 1999 {Angular momentum transport
  in magnetized stellar radiative zones. IV. Ferraro's theorem and the solar
  tachocline}. {\em Astrophys. J.\/} {\bf 519}, 911--917.

\bibitem[{McIntyre}(1994)]{McIntyre94}
{\sc {McIntyre}, M.} 1994 {The quasi-biennial oscillation (QBO) : some points
  about the terrestrial QBO and the possibility of related phenomena in the
  solar interior}. In {\em The Solar Engine and its Influence on Terrestrial
  Atmosphere and Climate\/} (ed. {E.~Nesme-Ribes}), p. 293. Reprint available
  at www.atm.damtp.cam.ac.uk/people/mem/tachocline.

\bibitem[{McIntyre}(2003)]{McIntyre03}
{\sc {McIntyre}, M.~E.} 2003 {Solar tachocline dynamics: eddy viscosity,
  anti-friction, or something in between?} In {\em Stellar astrophysical fluid
  dynamics\/} (ed. {Thompson, M.~J.~\& Christensen-Dalsgaard, J.}), pp.
  111--130. Cambridge: University Press.

\bibitem[{McIntyre}(2007)]{McIntyre07}
{\sc {McIntyre}, M.~E.} 2007 {Magnetic confinement and the sharp tachopause}.
  In {\em The Solar Tachocline\/} (ed. {D.~W.~Hughes, R.~Rosner, \&
  N.~O.~Weiss}), p. 183. Cambridge: University Press, corrected version
  available at \texttt{\footnotesize
  www.atm.damtp.cam.ac.uk/people/mem/tachocline}.

\bibitem[{Mestel}(1953)]{Mestel53}
{\sc {Mestel}, L.} 1953 {Rotation and stellar evolution}. {\em MNRAS\/} {\bf
  113}, 716.

\bibitem[{Mestel}(2011)]{Mestel11}
{\sc {Mestel}, L.} 2011 {\em {Stellar Magnetism}\/}, 2nd edn. Oxford:
  University Press.

\bibitem[{Mestel} \& {Moss}(1986)]{MestelMoss86}
{\sc {Mestel}, L. \& {Moss}, D.~L.} 1986 {On mixing by the Eddington-Sweet
  circulation}. {\em MNRAS\/} {\bf 221}, 25--51.

\bibitem[{Mestel} \& {Weiss}(1987)]{MestelWeiss87}
{\sc {Mestel}, L. \& {Weiss}, N.~O.} 1987 {Magnetic fields and non-uniform
  rotation in stellar radiative zones}. {\em MNRAS\/} {\bf 226}, 123--135.

\bibitem[{Parfrey} \& {Menou}(2007)]{ParfreyMenou07}
{\sc {Parfrey}, K.~P. \& {Menou}, K.} 2007 {The origin of solar activity in the
  tachocline}. {\em Astrophys. J.\/} {\bf 667}, L207--L210.

\bibitem[Plumb \& McEwan(1978)]{PlumbMcEwan78}
{\sc Plumb, R.~A. \& McEwan, A.~D.} 1978 The instability of a forced standing
  wave in a viscous stratified fluid: a laboratory analogue of the
  quasi-biennial oscillation. {\em J. Atmos. Sci.\/} {\bf 35}, 1827--1839.

\bibitem[{Press} {\em et~al.\/}(1986){Press}, {Flannery}, {Teukolsky} \&
  {Vetterling}]{Press-etal86}
{\sc {Press}, W.~H., {Flannery}, B.~P., {Teukolsky}, S.~A. \& {Vetterling},
  W.~T.} 1986 {\em {Numerical recipes. The art of scientific computing}\/}.
  Cambridge: University Press.

\bibitem[{Rogers} \& {Glatzmaier}(2006)]{RogersGlatzmaier06}
{\sc {Rogers}, T.~M. \& {Glatzmaier}, G.~A.} 2006 {Angular Momentum Transport
  by Gravity Waves in the Solar Interior}. {\em Astrophys. J.\/} {\bf 653},
  756--764.

\bibitem[{R\"udiger} \& {Kitchatinov}(1997)]{RudigerKitchatinov97}
{\sc {R\"udiger}, G. \& {Kitchatinov}, L.~L.} 1997 {The slender solar
  tachocline: a magnetic model}. {\em Astron. Nachr.\/} {\bf 318}, 273.

\bibitem[{Schatzman}(1993)]{Schatzman93}
{\sc {Schatzman}, E.} 1993 {Transport of angular momentum and diffusion by the
  action of internal waves}. {\em A\&A\/} {\bf 279}, 431--446.

\bibitem[{Schou} {\em et~al.\/}(1998){Schou}, {Antia}, {Basu}, {Bogart},
  {Bush}, {Chitre}, {Christensen-Dalsgaard}, {di Mauro}, {Dziembowski},
  {Eff-Darwich}, {Gough}, {Haber}, {Hoeksema}, {Howe}, {Korzennik},
  {Kosovichev}, {Larsen}, {Pijpers}, {Scherrer}, {Sekii}, {Tarbell}, {Title},
  {Thompson} \& {Toomre}]{Schou-etal98}
{\sc {Schou}, J., {Antia}, H.~M., {Basu}, S., {Bogart}, R.~S., {Bush}, R.~I.,
  {Chitre}, S.~M., {Christensen-Dalsgaard}, J., {di Mauro}, M.~P.,
  {Dziembowski}, W.~A., {Eff-Darwich}, A., {Gough}, D.~O., {Haber}, D.~A.,
  {Hoeksema}, J.~T., {Howe}, R., {Korzennik}, S.~G., {Kosovichev}, A.~G.,
  {Larsen}, R.~M., {Pijpers}, F.~P., {Scherrer}, P.~H., {Sekii}, T., {Tarbell},
  T.~D., {Title}, A.~M., {Thompson}, M.~J. \& {Toomre}, J.} 1998 {Helioseismic
  studies of differential rotation in the solar envelope by the Solar
  Oscillations Investigation using the Michelson Doppler Imager}. {\em
  Astrophys. J.\/} {\bf 505}, 390--417.

\bibitem[{Spiegel} \& {Zahn}(1992)]{SpiegelZahn92}
{\sc {Spiegel}, E.~A. \& {Zahn}, {J.-P.}} 1992 {The solar tachocline}. {\em
  A\&A\/} {\bf 265}, 106--114.

\bibitem[{Spruit}(2002)]{Spruit02}
{\sc {Spruit}, H.~C.} 2002 {Dynamo action by differential rotation in a stably
  stratified stellar interior}. {\em A\&A\/} {\bf 381}, 923--932.

\bibitem[{Taylor}(1963)]{Taylor63}
{\sc {Taylor}, J.~B.} 1963 {The magneto-hydrodynamics of a rotating fluid and
  the earth's dynamo problem}. {\em Proc. R. Soc. Lond. A\/} {\bf 274},
  274--283.

\bibitem[{Tobias} {\em et~al.\/}(2001){Tobias}, {Brummell}, {Clune} \&
  {Toomre}]{Tobias-etal01}
{\sc {Tobias}, S.~M., {Brummell}, N.~H., {Clune}, T.~L. \& {Toomre}, J.} 2001
  {Transport and storage of magnetic field by overshooting turbulent
  compressible convection}. {\em Astrophys. J.\/} {\bf 549}, 1183--1203.

\bibitem[{Vauclair} {\em et~al.\/}(1978){Vauclair}, {Vauclair}, {Schatzman} \&
  {Michaud}]{Vauclair-etal78}
{\sc {Vauclair}, S., {Vauclair}, G., {Schatzman}, E. \& {Michaud}, G.} 1978
  {Hydrodynamical instabilities in the envelopes of main-sequence stars -
  Constraints implied by the lithium, beryllium, and boron observations}. {\em
  Astrophys. J.\/} {\bf 223}, 567--582.

\bibitem[{Walker} {\em et~al.\/}(1998){Walker}, {Barenghi} \&
  {Jones}]{Walker-etal98}
{\sc {Walker}, M.~R., {Barenghi}, C.~F. \& {Jones}, C.~A.} 1998 {A note on
  dynamo action at asymptotically small Ekman number}. {\em Geophys. Astrophys.
  Fluid Dyn.\/} {\bf 88}, 261--275.

\bibitem[{Weiss} {\em et~al.\/}(2004){Weiss}, {Thomas}, {Brummell} \&
  {Tobias}]{Weiss-etal04}
{\sc {Weiss}, N.~O., {Thomas}, J.~H., {Brummell}, N.~H. \& {Tobias}, S.~M.}
  2004 {The origin of penumbral structure in sunspots: downward pumping of
  magnetic flux}. {\em Astrophys. J.\/} {\bf 600}, 1073--1090.

\bibitem[{Wood}(2010)]{Wood10}
{\sc {Wood}, T.~S.} 2010 {\em {The Solar Tachocline: A Self-Consistent Model of
  Magnetic Confinement}\/}. Cambridge, University Library, Superintendent of
  Manuscripts: PhD thesis.

\bibitem[{Wood} \& {McIntyre}(2007)]{WM07}
{\sc {Wood}, T.~S. \& {McIntyre}, M.~E.} 2007 {Confinement of the Sun's
  interior magnetic field: some exact boundary-layer solutions}. In {\em
  Unsolved Problems in Stellar Physics: A Conference in Honor of Douglas
  Gough\/} (ed. {R.~J.~Stancliffe, J.~Dewi, G.~Houdek, R.~G.~Martin, \&
  C.~A.~Tout}), {\em American Institute of Physics Conference Series\/}, vol.
  948, pp. 303--308. Corrigendum at \texttt{\footnotesize
  www.atm.damtp.cam.ac.uk/people/mem/tachocline}.

\bibitem[{Zahn} {\em et~al.\/}(1997){Zahn}, {Talon} \& {Matias}]{Zahn-etal97}
{\sc {Zahn}, {J.-P.}, {Talon}, S. \& {Matias}, J.} 1997 {Angular momentum
  transport by internal waves in the solar interior}. {\em A\&A\/} {\bf 322},
  320--328.

\end{thebibliography}

\clearpage

\end{document}